\documentclass[preprint,journal]{vgtc}            

\onlineid{1708}

\vgtccategory{Research}

\vgtcpapertype{algorithm/technique}

\title{Uncertainty-Aware Deep Neural Representations for Visual Analysis of Vector Field Data}

\author{
  Atul Kumar,
  Siddharth Garg, and 
  Soumya Dutta
}
\authorfooter{
  \item
    Atul Kumar and Soumya Dutta are with the department of Computer Science and Engineering, and Siddharth Garg is with the Electrical Engineering department at IIT Kanpur, India. Soumya Dutta is the corresponding author. Contact him at soumyad@cse.iitk.ac.in.
}

\abstract{%
  The widespread use of Deep Neural Networks (DNNs) has recently resulted in their application to challenging scientific visualization tasks. While advanced DNNs demonstrate impressive generalization abilities, understanding factors like prediction quality, confidence, robustness, and uncertainty is crucial. These insights aid application scientists in making informed decisions. However, DNNs lack inherent mechanisms to measure prediction uncertainty, prompting the creation of distinct frameworks for constructing robust uncertainty-aware models tailored to various visualization tasks. In this work, we develop uncertainty-aware implicit neural representations to model steady-state vector fields effectively. We comprehensively evaluate the efficacy of two principled deep uncertainty estimation techniques: (1) Deep Ensemble and (2) Monte Carlo Dropout, aimed at enabling uncertainty-informed visual analysis of features within steady vector field data. Our detailed exploration using several vector data sets indicate that uncertainty-aware models generate informative visualization results of vector field features. Furthermore, incorporating prediction uncertainty improves the resilience and interpretability of our DNN model, rendering it applicable for the analysis of non-trivial vector field data sets.}


\graphicspath{{figs/}{figures/}{pictures/}{images/}{./}} 

\usepackage{tabu}                      
\usepackage{booktabs}                  
\usepackage{lipsum}                    
\usepackage{mwe}                       

\usepackage{mathptmx}                  

\usepackage{color}
\usepackage{subcaption}
\usepackage{amsmath}
\usepackage{amsfonts}
\usepackage{graphicx}  
\usepackage{amssymb}
\usepackage{multirow}
\usepackage[ruled,vlined]{algorithm2e}
\usepackage{mathrsfs}

\DeclareUnicodeCharacter{2212}{\ensuremath{-}}

\newcommand{\rmark}[1]{#1}

\newenvironment{tight_enumerate}{
\begin{enumerate}
  \setlength{\itemsep}{0pt}
  \setlength{\parskip}{4.5pt}
}{\end{enumerate}}

\begin{document}


\firstsection{Introduction}
\maketitle
The indisputable success of deep neural networks (DNNs)~\cite{lebh15} has resulted in numerous applications of it in the scientific visualization domain~\cite{dl4scivis}. Analysis of intricate vector fields using DNNs has shown promising results with applications such as generating super-resolution flow fields~\cite{SSRVFD, TSRVFD}, reconstructing flow fields from streamlines~\cite{flowdl_1, flowdl_2}, flow map reconstruction~\cite{neuralflowmap}, and predicting flow lines~\cite{particle_trace_NN}. \rmark{While these DNN-based approaches produce state-of-the-art results, learning a direct neural representation of the vector data is yet to be explored. Furthermore, the existing models do not quantify what it does not know or how confidently the predictions are generated.} Such missing knowledge, if estimated and conveyed to the experts, can significantly help them to make informed decisions about their data~\cite{Bonneau2014,gagh16}. \rmark{Consequently, DNN-based flow data visual analytics will become more trustworthy when they can assess their prediction uncertainty. However, the literature survey reveals that a direct vector data modeling approach augmented with uncertainty estimation capability using DNNs is still missing -- a gap that this work attempts to bridge.}

\rmark{Among various neural architectures used to analyze flow data, we study the efficacy of implicit neural representations (INRs) for directly learning vector fields, i.e., a INR will predict the vector components at any queried location in the spatial domain.} The choice of INR over other neural architectures is motivated by the recent success of INRs in producing state-of-the-art results for scientific data ~\cite{levine_neural_compression, coordnet, stsrinr, neuralflowmap}. \rmark{Since DNNs inherently do not provide prediction uncertainty, we employ deep uncertainty quantification techniques to obtain uncertainty estimates along with the predicted vectors from our INR model so that the downstream flow field analysis tasks can effectively leverage such information. Hence, we further focus on designing uncertainty-informed flow data visualization techniques to intuitively and effectively communicate the prediction uncertainty to domain experts.} 

Various factors specific can influence the choice of deep uncertainty estimation methods. While the deep learning community has developed several uncertainty quantification methods, our focus is on selecting methods that can be easily integrated into visualization models with minimal architecture modifications, facilitating smooth adoption of such methods for uncertainty-aware visual analysis. Literature survey indicates that Deep Ensembles often excel in producing accurate predictions and estimating uncertainty by leveraging the variation among ensemble member predictions~\cite{gustafsson2020evaluating, beluch2018power, huzw23}. 
However, Deep Ensembles demand prohibitively excessive training time and resource utilization since multiple ensemble members need to be trained. In contrast,  single-model-based uncertainty estimation methods are computationally cheap and are  gaining importance in the research community~\cite{single_model_uncert, deep_evidential_uncert, gagh16}. Among these methods, Monte Carlo Dropout (MCDropout) stands out as a suitable choice for our purposes since it only requires the addition of dropout layers to an existing model and theoretically equivalent to approximate inferencing in deep Gaussian processes~\cite{dala13,gagh16}. 
\rmark{Therefore, in this work, we compare and contrast} the characteristics of uncertainty estimates generated by Deep Ensemble and MCDropout methods for enabling uncertainty-aware visual analytics of steady-state flow fields.

We analyze the effectiveness and advantages of uncertainty-aware implicit neural representations (INRs) of steady-state flow fields coupled with two deep uncertainty estimation techniques: (1) Deep Ensemble and (2) MCDropout method. \rmark{We conduct a comprehensive assessment of the predictive accuracy of such uncertainty-aware INRs by (a) evaluating the quality of reconstructed vector fields when the INR learns the vector representation directly and (b) assessing the effectiveness in analyzing and visualizing  flow features through streamline and critical point visualizations. For each task, we devise uncertainty visualization methods to convey uncertainty information to the experts.} We compare and contrast the characteristics of uncertainties derived from these methods and illustrate how such information can assist domain scientists in making informed decisions and developing confidence in the model outputs. We validate our techniques using several 2D and 3D vector field data sets to showcase the efficacy of our approach in flow field analysis and visualization. Hence, our contributions are twofold:

\begin{tight_enumerate}
	\item \rmark{We propose developing uncertainty-aware INRs for vector field data to enhance the robust analysis of intricate flow features, emphasizing the significance of visualizing prediction uncertainty to improve how domain scientists interpret model predictions.}
	\item We extensively compare and contrast between two principled deep uncertainty estimation techniques: (1) Deep Ensemble, and (2) MCDropout, demonstrating their suitability for conducting uncertainty-informed flow feature analysis and visualization.
\end{tight_enumerate}

\section{Research Background and Uncertainty Estimation in Deep Neural Networks}

\subsection{Deep Learning in Scientific Data Visualization}
Deep learning has found diverse applications in scientific visualization. Lu et al.\cite{levine_neural_compression} and Weiss et al.~\cite{fvsrn} have introduced methods aimed at generating concise neural representations of scientific data. Hong et al.~\cite{DNNVolVis}, He et al.~\cite{insitunet}, and Berger et al.~\cite{GANvolren} have investigated the visualization of scalar field data, while Weiss et al. have explored isosurface visualization~\cite{weiss2019isosuperres} and volume visualization~\cite{9264699}. 
Spatiotemporal super-resolution volume generation has emerged as another research focus~\cite{SSR-TVD, TSR-TVD, Wurstersuperreso, superresoDL}. Novel models for domain knowledge-aware latent space generation have been proposed~\cite{idlat}. Moreover, DNNs have been employed as substitutes for generating visualizations and exploring parameter spaces in ensemble data~\cite{insitunet, gnnsurrogate, vdl_surrogate}. Analysis of flow field features using CNNs was proposed in~\cite{CNNflowdataDL}. Reconstruction of vector data from 3D streamlines was developed by Han et al.~\cite{flowdl_1}. More recently, Berenjkoub et al. highlighted how vortex boundaries could be extracted using CNNs~\cite{vortexDL}. In a recent work, Dutta et al.~\cite{Dutta_TVCG_DL_Uncet} demonstrated how visualizing prediction uncertainty  can provide insights about the model's robustness for view synthesizing tasks. For a comprehensive overview of flow data visualization using deep learning methods, please refer to~\cite{dl4flowvis} and for a broader set of applications of deep learning applications in scientific visualization please refer to~\cite{dl4scivis}.

\subsection{Uncertainty Visualization}
Pang et al. provide one of the earliest summaries of uncertainty visualization techniques~\cite{pang1997approaches}. Potter et al. focus on visualizing spatial probability distributions, preceded by a taxonomy of uncertainty visualization methods~\cite{potter2008, Potter2012}. Brodlie et al.~\cite{Brodlie2012} introduce visualization techniques enhanced with tools for uncertainty estimation. Liu et al. employ flickering as a method to depict uncertainty in volume data~\cite{Liu2012}, while Athawale et al. delve deeper into uncertainty visualization in volume rendering using non-parametric models~\cite{AthawaleAJE}. Uncertainty visualization techniques tailored for isocontouring methods have received significant attention in research. Pöthkow et al. devise a method to compute the level crossing probability between adjacent points, which is further refined to calculate the probability for each cell~\cite{51187755, Pothkow2011}. Whitaker et al.~\cite{6634129} explore uncertainty visualization in ensembles of contours. Streamline uncertainty in ensemble field~\cite{streamlinevariability} and pathline uncertainty in time-varying field~\cite{jimmypathlineuncert} is also explored. Otto et al.~\cite{uncert_flow_otto} study uncertainty in 2D vector fields. Bonneau et al. conduct a comprehensive survey of various uncertainty visualization techniques~\cite{Bonneau2014}. Recently, Gillmann et al. provide a summary of uncertainty visualization methods geared towards image processing~\cite{uncertimageproc}. 

\subsection{Uncertainty Estimation in Deep Neural Networks}

The uncertainty~\cite{blck15, deepEnsembles} of a DNN can be categorized into two broad types -- \emph{data or aleatoric uncertainty} and \emph{model or epistemic uncertainty}. Data uncertainty is attributed to errors and noise during data acquisition. Model (epistemic) uncertainty can arise for different reasons. Firstly, the DNNs produce a compressed representation of large-scale data. Such compression often results in prediction error and associated uncertainty. Secondly, DNNs are often fine-tuned carefully using learning rate variation, regularization, etc. Different decisions for such configurations lead to different learned model representations, and analysis using such models can lead to uncertainties. The data uncertainty can be addressed by improving data collection method and by improving the data quality. In contrast, the epistemic uncertainty is inherent to the model and needs to be studied in detail.

\subsubsection{Techniques for Modeling Uncertainty in DNNs}

\textbf{Deterministic methods.} Deterministic models can be equipped with uncertainty estimation capabilities by explicitly training a network to quantify uncertainties~\cite{evidential.neural.networks}. 

\textbf{Bayesian Methods.} Bayesian neural networks~\cite{BNNSurvey} employ prior distributions on the model parameters of DNNs to compute epistemic   uncertainty~\cite{deepEnsembles}. Training such networks require stochastic gradient MCMC~\cite{mafb17}, and variational inference~\cite{hahn19} methods. 

\textbf{Test-time Augmentation.} Test time augmentation methods perform data augmentation during the inference time and then estimate the uncertainty from the variability in the predicted results~\cite{Ayhan.2018, wang2019aleatoric}.

\textbf{Deep Evidential Regression.} In deep evidential regression, the network learns parameters as well as hyperparameters to the corresponding evidential distributions~\cite{deep_evidential_uncert}. Then finally, the evidential distributions are used to estimate model uncertainty.

\textbf{Uncertainty via Stochastic Data Centering.} Here, the authors show that when an ensemble of DNNs is trained with data sets that are shifted by a constant bias, model uncertainty becomes the variability across such ensemble members' predictions~\cite{single_model_uncert}.

\begin{figure}[tb]
\centering
    \includegraphics[width=\linewidth]{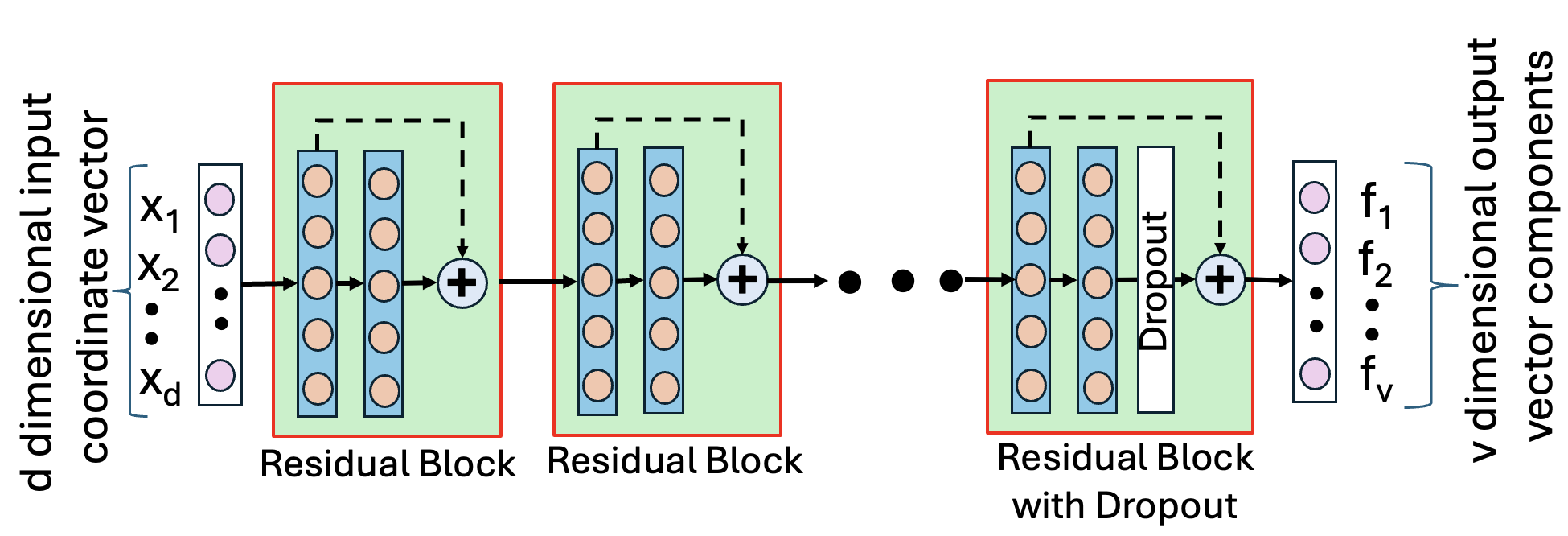}
\caption{The schematic of the MCDropout-enabled INR model which uses a residual block-based MLP architecture. A dropout layer is added at the last residual block to generate uncertainty estimates during inference time. The INR architecture for the Ensemble method is identical to this, except there is no dropout layer.}
\label{model_arch}
\end{figure}

\subsubsection{Ensemble Method}
In ensemble-based methods, several models work harmoniously to produce predictions that are of superior quality compared to the predictions from any individual model~\cite{ensemble.survey}. By doing this, ensemble methods improve the generalization error, and by estimating the variability among ensemble member predictions, the model uncertainty can be estimated robustly. Following this principle, Lakshminarayanan et al. propose Deep Ensembles~\cite{deepEnsembles} for DNNs. Conceptually, Deep Ensemble learning can be shown as an approximation of \textit{Bayesian averaging}~\cite{UQEnsemble}. In  Bayesian averaging, the final model prediction is formulated as: $prediction = \int \mathscr{P}_w(x)\pi(w | \mathcal{D})$. 
Here $\mathscr{P}_w(x)$ is the prediction associated with sample $x$ and $\pi(w | \mathcal{D})$ represents the posterior probability distribution with $\mathcal{D}$ being the training data. In reality, estimation of this integral is extremely difficult, and it is also found that to estimate this integral sufficiently accurately, exploration of all the modes of $\pi(w | \mathcal{D})$ is not required. This observation imply that non-weighted averaged predictions generated by a set of ensemble members can be considered an approximation of the above expression. Note that shuffling of the training data and a random parameter initialization during the training process introduces a good variety in each learned ensemble member to predict the uncertainty robustly, an approach followed in this work to generate Deep Ensembles. Subsequent research works on Deep Ensembles~\cite{gustafsson2020evaluating, beluch2018power, ovadia2019can, vyas2018out} show that ensemble methods often outperform other uncertainty estimation techniques and are more immune to changes in data distribution.

\subsubsection{MCDropout Method} \label{mcdTheory}
Dropout~\cite{gagh16, hisk12} is a regularization technique that prevents models from overfitting during training and is achieved by randomly masking a subset of the weights during training. However, Gal et al.~\cite{gagh16} discovered that adding dropout in a DNN with arbitrary depth and non-linear activations makes it theoretically equivalent to an approximate Bayesian inference in deep Gaussian processes. Then activating dropout at test time is equivalent to sampling from the Bayesian posterior distribution, $p(\textbf{W} \ | \ \textbf{X}, \textbf{Y})$, where $\textbf{W}$ denotes the model's weights, $\textbf{X}$ is the training data and $\textbf{Y}$ is the target output. Finally, the mean of these sampled predictions is considered as the expected model output, and by measuring the variance in these sampled predictions, model uncertainty can be estimated~\cite{gagh16}. Such probabilistic predictions are derived by collecting Monte Carlo (MC) samples from the dropout-enabled trained model, known as Monte Carlo Dropout (MCDropout) method, by performing multiple forward passes during inference.

\section{Uncertainty-Aware Neural Representation for Vector Field Data}
\subsection{Implicit Neural Representation}
Implicit neural representations with periodic activation functions have been identified as a promising solution for learning representations of coordinate-based data sets, where the mapping from any input coordinate in the data domain to the corresponding output quantity values is learned. Sitzmann et al.~\cite{siren} in their work depicted that a feed-forward neural network with sinusoidal activation function, termed as SIREN (sinusoidal representation network), can be used to build such INRs~\cite{siren}. Several variations of this SIREN have recently been used to solve many challenging problems in the scientific data visualization domain and obtain state-of-the-art results~\cite{levine_neural_compression, coordnet, stsrinr, neuralflowmap}. The success of these recent research efforts has prompted us to build our uncertainty-aware model using SIRENs as the base architecture. Besides analyzing the uncertainty estimates produced by the Deep Ensemble and MCDopout methods using a SIREN-based model, we study the accuracy such a network can achieve in representing intricate steady vector fields.

\subsection{Model Architecture}
Using an implicit neural network, we aim to learn the function that represents the mapping from the input data coordinate domain to the corresponding vector value space. To achieve this, we build our base model as a multilayer perceptron consisting of $d$ neurons as inputs, followed by $l$ hidden layers and an output layer containing $v$ neurons. To enrich the model's learning capability and train a deep network in a stable fashion,  we enhance the base SIREN architecture by incorporating residual blocks and skip connections~\cite{resnet} as was suggested in~\cite{levine_neural_compression}. Our model learns a function that takes a $d$ dimensional coordinate vector as input and predicts a $v$ dimensional vector output where $v$ is the number of components in the vector field data. Essentially, the network learns a function $\mathcal{F(\theta)}:\mathbb{R}^d \mapsto \mathbb{R}^v$, where $\theta$ represents the parameters of the neural network. When the size of the flow data is large, then using a single DNN to learn all the vector components jointly also allows us to produce a compact neural representation of the flow data set since the model parameters are shared across the vector components. Hence, the model size will be sufficiently smaller than the raw flow data set.

\subsection{Uncertainty Quantification Using MCDropout Method}
The model architecture used for the MCDropout method is shown in Fig.~\ref{model_arch}. We observe that a post-activation dropout layer is added at the last residual block to build a dropout-enabled model that can be used conveniently during inference to estimate prediction uncertainty. The addition of the dropout layer also helps in regularization during training. Theoretically, if one wants to simulate a fully Bayesian network, a dropout layer must be added at each residual block~\cite{segnet}. However, such a large number of dropout layers often acts as a strong regularizer and hinders the learning process, resulting in lower accuracy during test time~\cite{segnet}. We observe similar behavior when the dropout layer is added to every residual block. Hence, following Kendall et al.'s. suggestion~\cite{segnet}, we build a model by adding dropout at the deepest residual block, giving us robust uncertainty estimates and a high-quality vector field reconstruction. The impact of using different numbers of dropout layers has been studied and presented later in Table~\ref{drop_layers_study}.

As outlined in Section~\ref{mcdTheory}, inference involves generating a set of Monte Carlo samples through multiple forward passes of the trained model when dropout is enabled. In our approach, we produce $m$ instances of the vector field and subsequently calculate the average vector field, serving as the predicted vector field. The grid-pointwise standard deviation, computed using vector components estimated by $m$ forward passes, denotes the prediction uncertainty.  We separately compute standard deviation for each vector component and then add the standard deviation values to obtain the final prediction uncertainty.

\begin{table}[thb]
\centering
\caption{Description of data sets used in the experimentation.}
\label{datadesc_table}
\resizebox{0.8\linewidth}{!}{
\begin{tabular}{|c|c|c|}
\hline
\textbf{Data set} & \textbf{Dimensionality} & \textbf{Spatial Resolution} \\ \hline
Heated Cylinder (T=750) & 2D             & 150 $\times$ 450              \\ \hline
Heated Cylinder (T=1500) & 2D             & 150 $\times$ 450              \\ \hline
Fluid (T=750)      & 2D             & 512 $\times$ 512              \\ \hline
Hurricane Isabel (T=25) & 3D             & 250 $\times$ 250 $\times$ 50  \\ \hline
Tornado       & 3D             & 128 $\times$ 128 $\times$ 128 \\ \hline
Turbine          & 3D             & 151 $\times$ 71 $\times$ 56   \\ \hline
Tangaroa (T=150)       & 3D             & 300 $\times$ 180 $\times$ 120 \\ \hline
\end{tabular}
}
\end{table}
\begin{figure}[thb]
\centering
\begin{subfigure}[t]{0.3\linewidth}
    \centering
    \includegraphics[width=\linewidth]{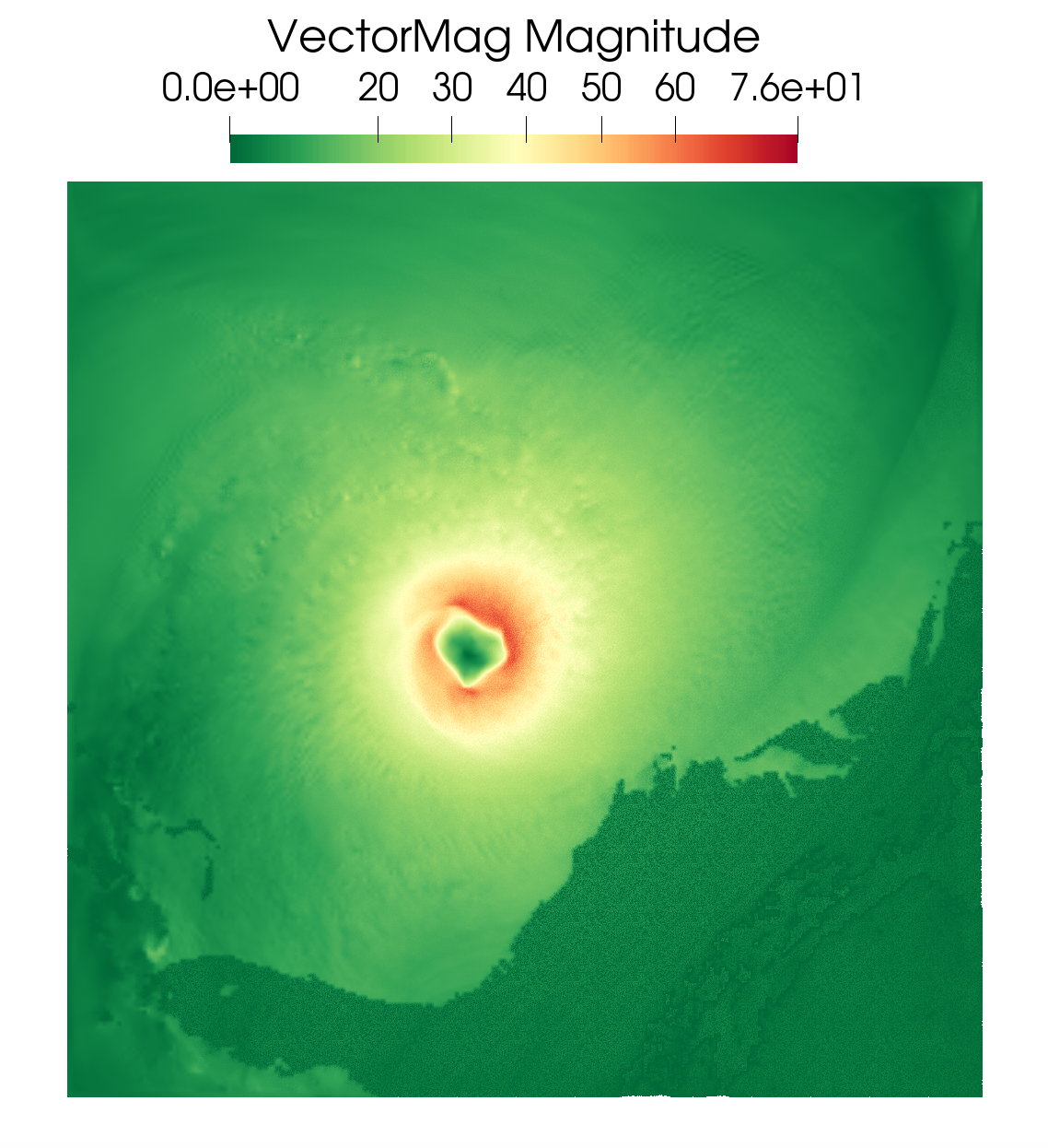}
    \caption{Ground truth of vector magnitude.}
    \label{isabel_GT}
\end{subfigure}
~
\begin{subfigure}[t]{0.3\linewidth}
    \centering
    \includegraphics[width=\linewidth]{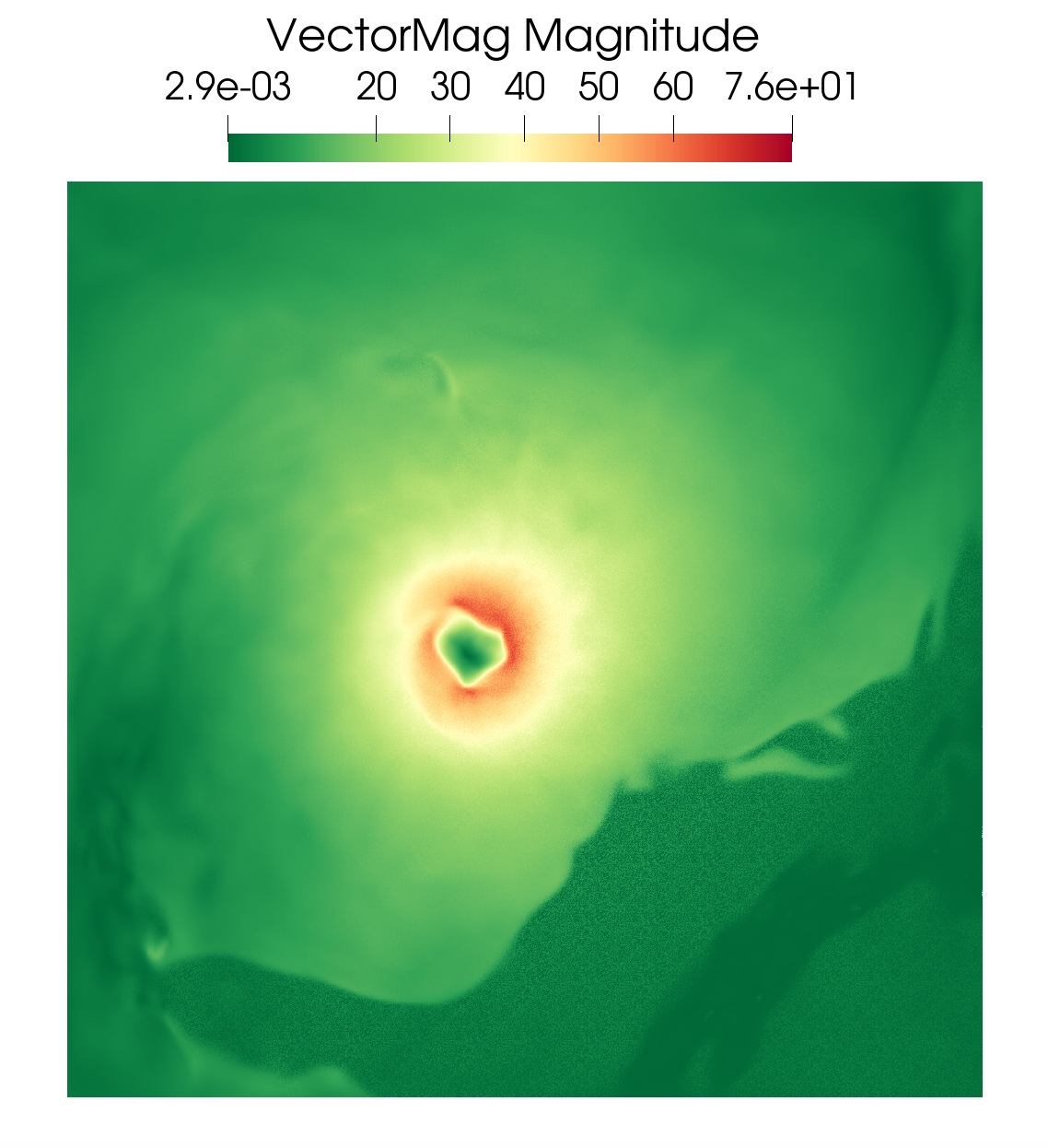}
    \caption{MCDropout reconstructed vector magnitude.}
    \label{isabel_MCD}
\end{subfigure}
~
\begin{subfigure}[t]{0.3\linewidth}
    \centering
    \includegraphics[width=\linewidth]{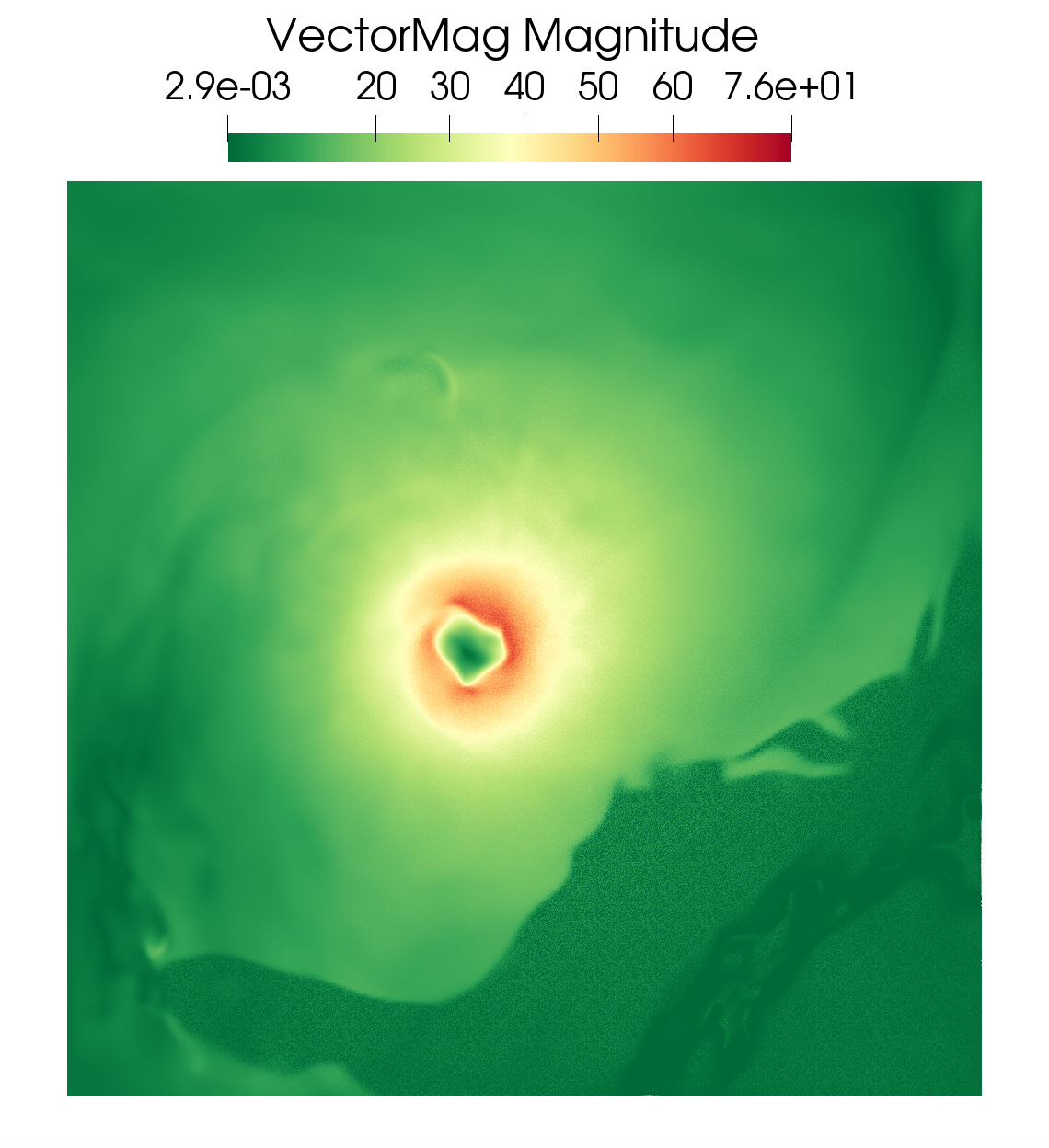}
    \caption{Ensemble reconstructed vector magnitude.}
    \label{isabel_ens}
\end{subfigure}
\caption{Volume visualization of the magnitude of reconstructed vector fields using MCDropout and Ensemble methods for Hurricane Isabel data set. The ground truth is shown in Fig.~\ref{isabel_GT}. We observe that both methods produce visually comparable results.}
\label{isabel_mag_vis}
\end{figure}

\subsection{Uncertainty Quantification Using Ensemble Method}
A schematic of our INR architecture is presented in Fig.~\ref{model_arch}. To build a Deep Ensemble~\cite{deepEnsembles} of INRs, we use this model architecture without the dropout layer. To produce an ensemble model comprising $m$ members, each member is trained separately, utilizing samples from the entire vector field data. During every ensemble creation, the training data is randomly shuffled for each member so that variability is introduced to each member. After the ensemble members' training is finished, each member generates a realization of the entire vector field by inferring vector values at the grid locations. Then, the averaged vector field represents the expected vector field, and the standard deviation computed at each grid point from the predicted vector values indicates the epistemic uncertainty of the ensemble model. We separately compute standard deviation for each vector component and then add the values to quantify the final uncertainty.

\textbf{Loss Function and Hyperparameters.}
For both MCDropout and Ensemble methods, we utilize $14$ residual blocks for 3D and $10$  for 2D vector field data sets. Each hidden layer comprises $120$ neurons for 3D and $100$ neurons for 2D data sets. The training uses conventional mean squared error loss ($\mathcal{L}_{mse}$). \rmark{In our work, we focus on using a consistent network architecture and hyperparameter combination for both uncertainty estimation methods to produce comparable results  across multiple data sets. Hence, by empirical experimentation, we identify a suitable learning rate and batch size combination that produce stable, consistent, and high-quality results across all data sets. We employ a batch size of $2048$ with the Adam optimizer~\cite{kiba14}, setting the learning rate at $5e-5$ and the two Adam optimizer coefficients $\beta_1$ and $\beta_2$ to their default values at $0.9$ and $0.999$, respectively. The learning rate is decayed by a factor of $0.1$ if the loss value does not decrease for $10$ consecutive epochs. During MCDropout method, a training dropout probability of $\eta=0.05$ and a testing dropout probability of $\eta=0.1$ is applied consistently, while no dropout is used for training ensemble models. Both MCDropout and Ensemble methods undergo training for $500$ epochs. Further discussion on the hyperparameter selection and impact of network architecture on reconstruction quality and uncertainty estimates are provided in Section~\ref{discussion_section}.}

\section{Uncertainty-Informed Flow Field Visualization} 
We conduct a thorough study of our models using six vector field data sets. The dimensionality and spatial resolution of these data sets are reported in Table~\ref{datadesc_table}. We use a GPU server with NVIDIA GeForce GTX $1080$Ti GPUs with $12$GB GPU memory for all the experimentation. All the models are implemented in PyTorch~\cite{pytorch2019}. The Heated Cylinder data set~\cite{heatedCylinder} and Fluid data set~\cite{fluid} are generated using Gerris flow solver~\cite{gerrisflowsolver}. Hurricane Isabel data was produced by the Weather Research and Forecast model, courtesy of NCAR and NSF. Turbine data set~\cite{stall_chen} and Tornado data~\cite{tornado} set are made available by Dr. Jen-Ping Chen and Dr. Roger Crawfis, respectively, at the Ohio State University. Tangaroa data set~\cite{tangaroa} is a simulation of an incompressible 3D flow around a CAD model of the research vessel Tangaroa.

\begin{figure}[thb]
\centering
\begin{subfigure}[t]{0.3\linewidth}
    \centering
    \includegraphics[width=\linewidth]{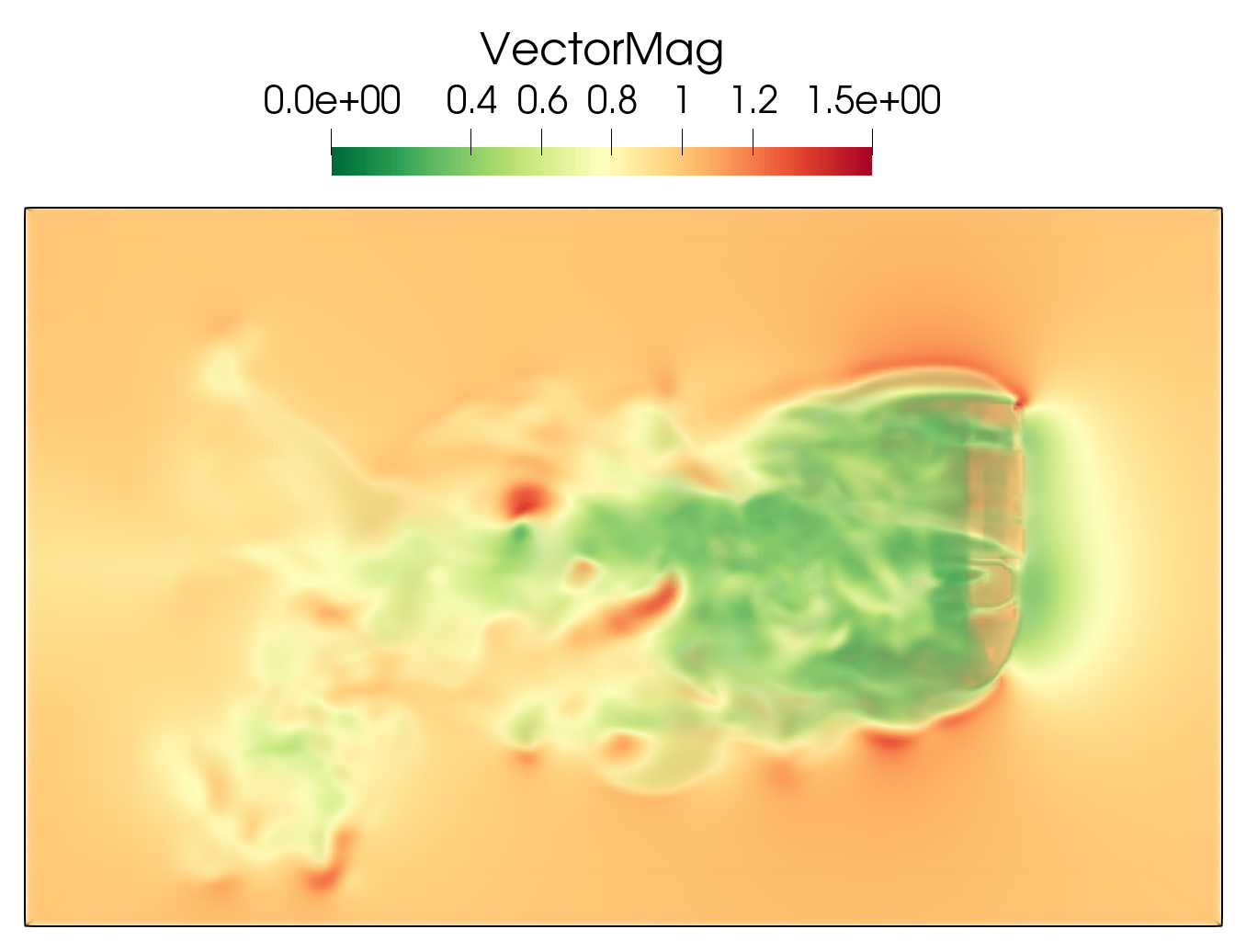}
    \caption{Ground truth of vector magnitude.}
    \label{tangaroa_GT}
\end{subfigure}
~
\begin{subfigure}[t]{0.3\linewidth}
    \centering
    \includegraphics[width=\linewidth]{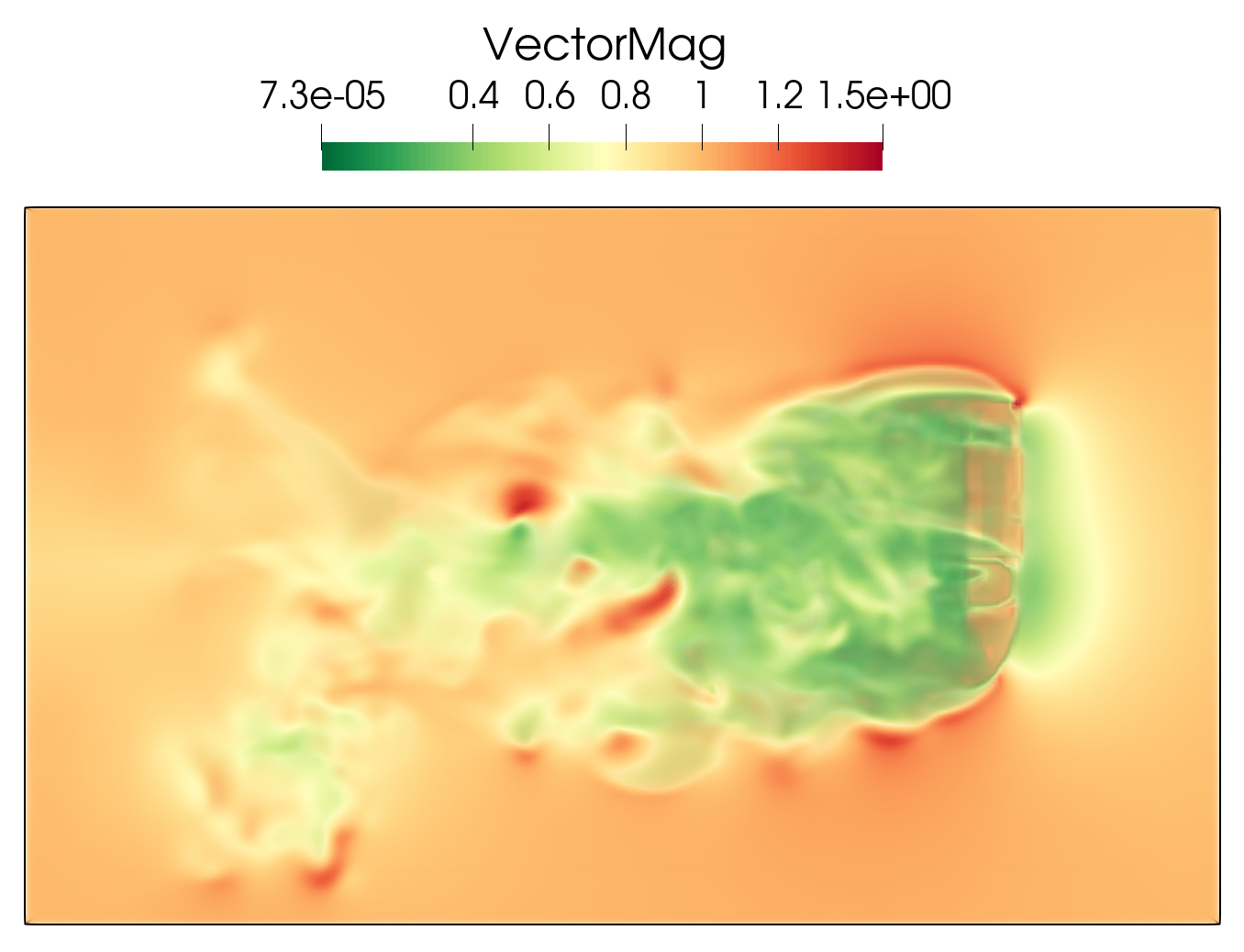}
    \caption{MCDropout reconstructed vector magnitude.}
    \label{tangaroa_MCD}
\end{subfigure}
~
\begin{subfigure}[t]{0.3\linewidth}
    \centering
    \includegraphics[width=\linewidth]{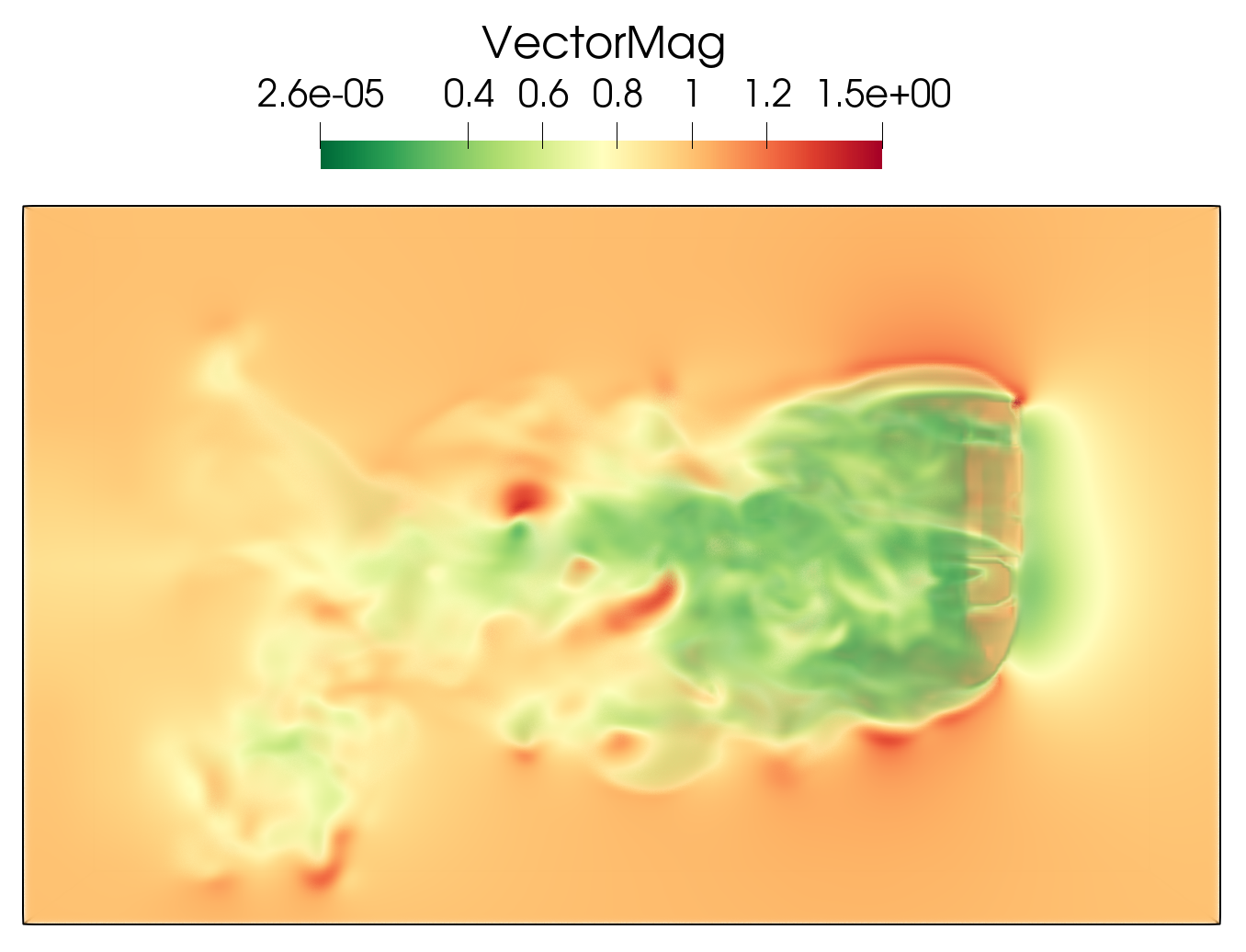}
    \caption{Ensemble reconstructed vector magnitude.}
    \label{tangaroa_ens}
\end{subfigure}
\caption{Volume visualization of the magnitude of reconstructed vector fields using MCDropout and Ensemble methods for Tangaroa data set. The ground truth is shown in Fig.~\ref{tangaroa_GT}. We observe that both methods produce visually comparable results.}
\label{tangaroa_mag_vis}
\end{figure}
\subsection{Visual Analysis of Reconstructed Field, Prediction Uncertainty, and Error}
We reconstruct the entire vector field to thoroughly assess the model's reconstruction quality, prediction uncertainty, and error. Through experimentation, we ascertain that utilizing $100$ Monte Carlo (MC) samples for the MCDropout method and $30$ ensemble members for the Ensemble method yields a robust estimation of vectors, error, and uncertainty. The determination of these numbers of MC samples and ensemble members is based on comparing the reconstruction quality across various quantities of MC samples and ensemble members. Detailed results can be found in Table~\ref{MC_samples_psnr} and Table~\ref{ensemble_samples_psnr}. Thus, unless specified otherwise, we consistently employ $100$ MC samples for MCDropout and $30$ ensemble members for the Ensemble method in all presented results.

To derive the final vector field, we calculate the average vectors at each grid point, where the averaging process entails computing the mean over $100$ vectors for the MCDropout method and over $30$ vectors for the Ensemble method. We compute the error at each grid point by contrasting the predicted value with the ground truth vectors to estimate fine-grained prediction error. \rmark{This error computation involves assessing each vector component individually and then adding the component-wise errors, i.e., the L1 norm, to obtain the total error at a grid point.} Similarly, we compute the component-wise uncertainty at each grid point (by computing the component-wise standard deviation) for both MCDropout and Ensemble methods, subsequently estimating the total uncertainty by summing the component-wise uncertainty values.

\begin{figure*}[thb]
\centering
\begin{subfigure}[t]{0.18\linewidth}
    \centering
    \includegraphics[width=\linewidth]{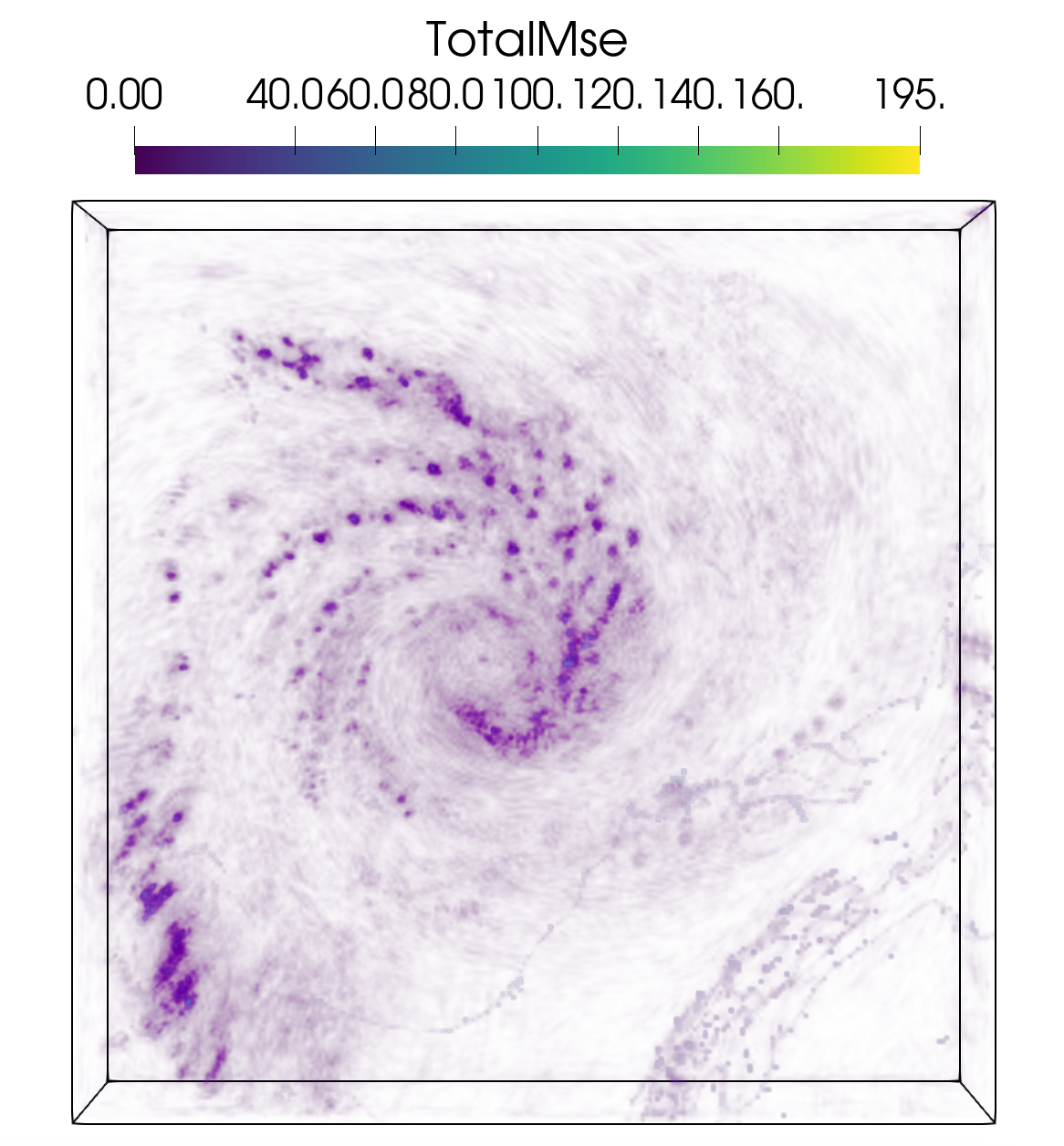}
    \caption{Spatial MSE visualization for MCDropout.}
    \label{isabel_MSE_MCD}
\end{subfigure}
~
\begin{subfigure}[t]{0.18\linewidth}
    \centering
    \includegraphics[width=\linewidth]{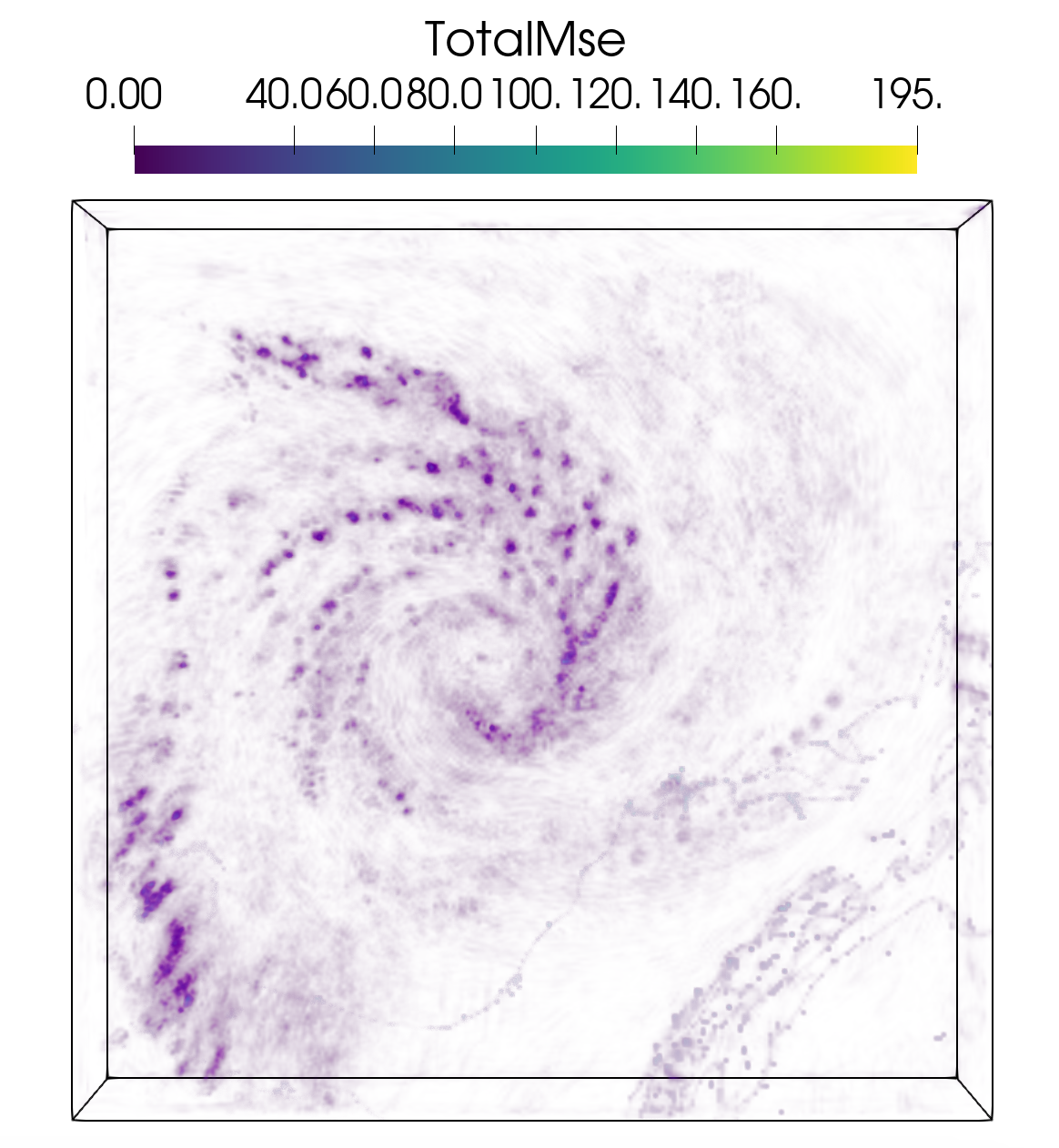}
    \caption{Spatial MSE visualization for Ensemble.}
    \label{isabel_MSE_ENS}
\end{subfigure}
~
\begin{subfigure}[t]{0.18\linewidth}
    \centering
    \includegraphics[width=\linewidth]{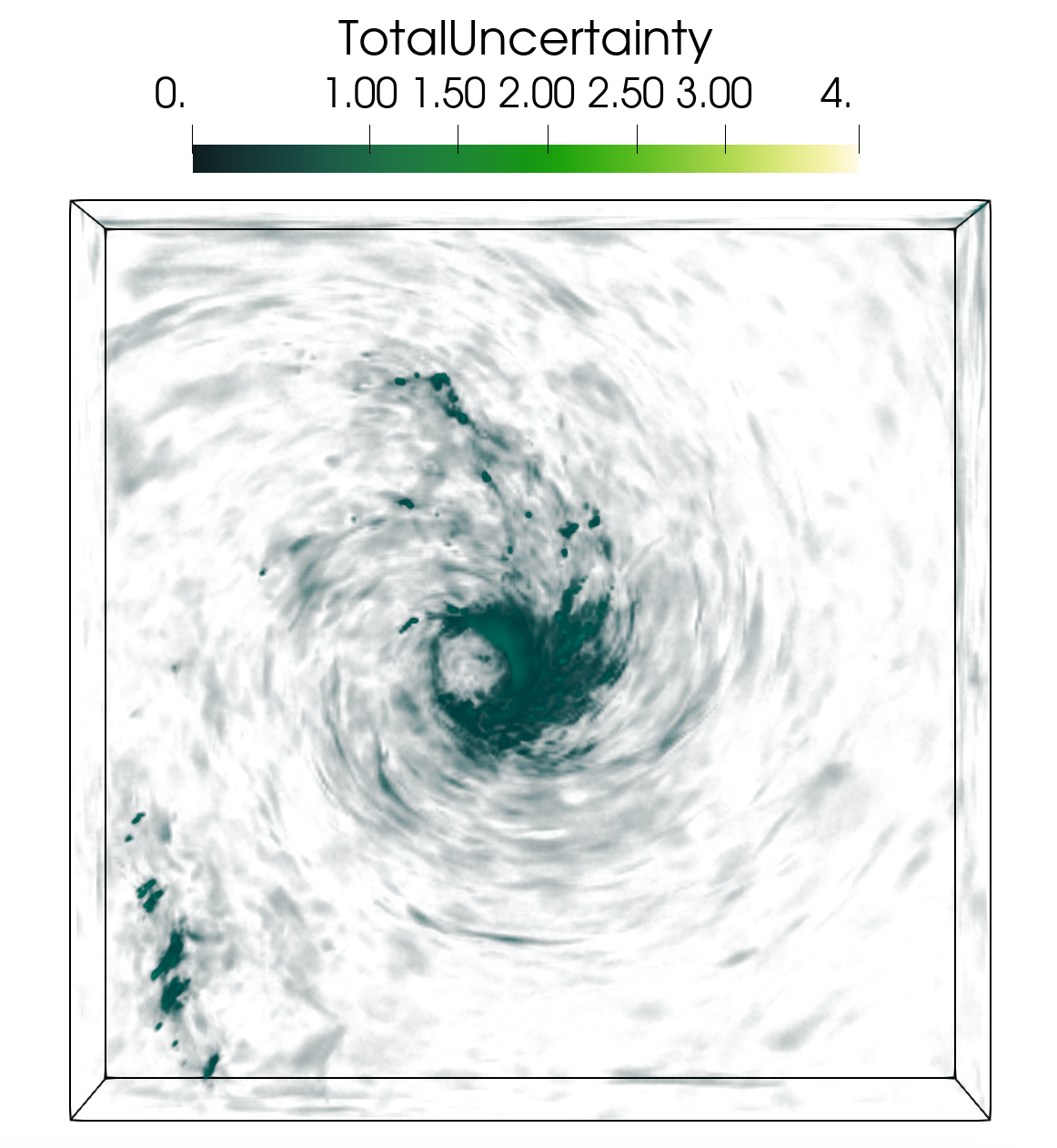}
    \caption{Spatial Uncertainty visualization for MCDropout.}
    \label{isabel_Uncrt_MCD}
\end{subfigure}
~
\begin{subfigure}[t]{0.18\linewidth}
    \centering
    \includegraphics[width=\linewidth]{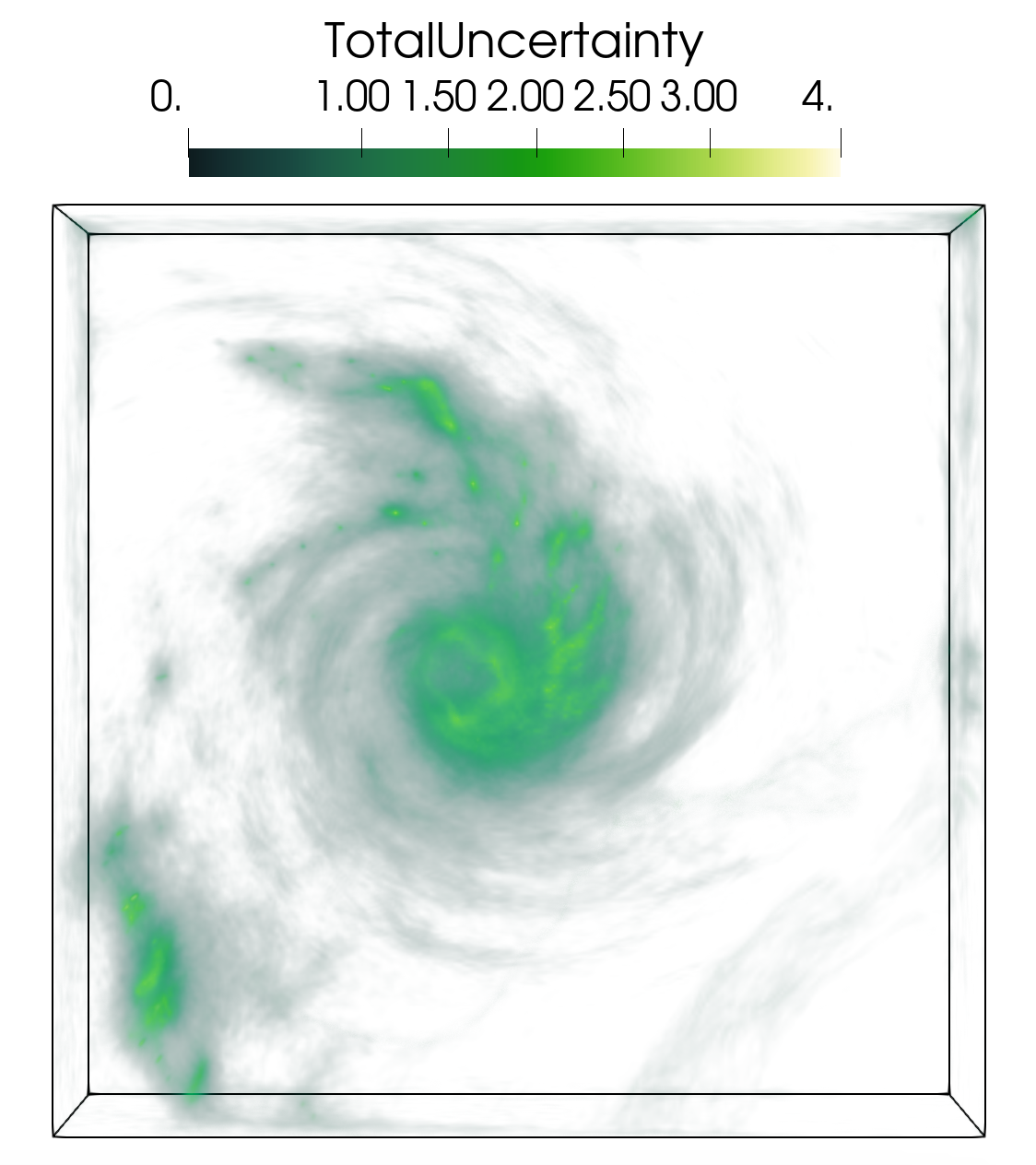}
    \caption{Spatial Uncertainty visualization for Ensemble.}
    \label{isabel_Uncrt_ENS}
\end{subfigure}
\caption{Visualization of uncertainty and error fields for Hurricane Isabel data set. The MSE and prediction uncertainty is estimated at each grid point between the predicted and ground truth vectors for MCDropout and Ensemble methods. Fig.~\ref{isabel_MSE_MCD} and Fig.~\ref{isabel_MSE_ENS} show the rendering of MSE fields, and Fig.~\ref{isabel_Uncrt_MCD} and Fig.~\ref{isabel_Uncrt_ENS} present the rendering of uncertainty fields. We observe that the locations with higher MSE correspond to similar spatial regions for both methods. In contrast, the vortex region is detected as a region with higher prediction uncertainty for both methods.}
\label{isabel_error_uncert_vis}
\end{figure*}

\begin{figure*}[thb]
\centering
\begin{subfigure}[t]{0.18\linewidth}
    \centering
    \includegraphics[width=\linewidth]{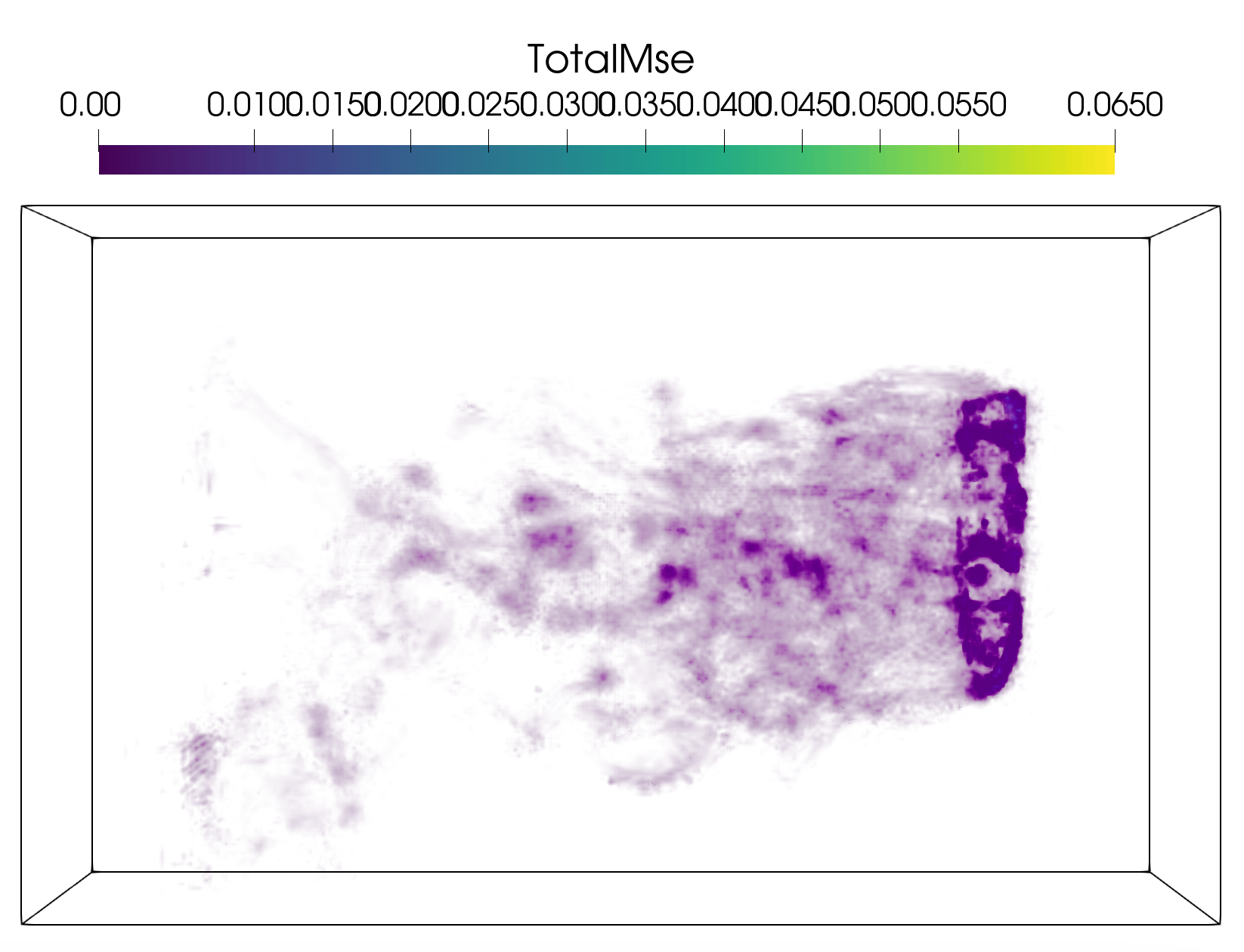}
    \caption{Spatial MSE visualization for MCDropout.}
    \label{tangaroa_MSE_MCD}
\end{subfigure}
~
\begin{subfigure}[t]{0.18\linewidth}
    \centering
    \includegraphics[width=\linewidth]{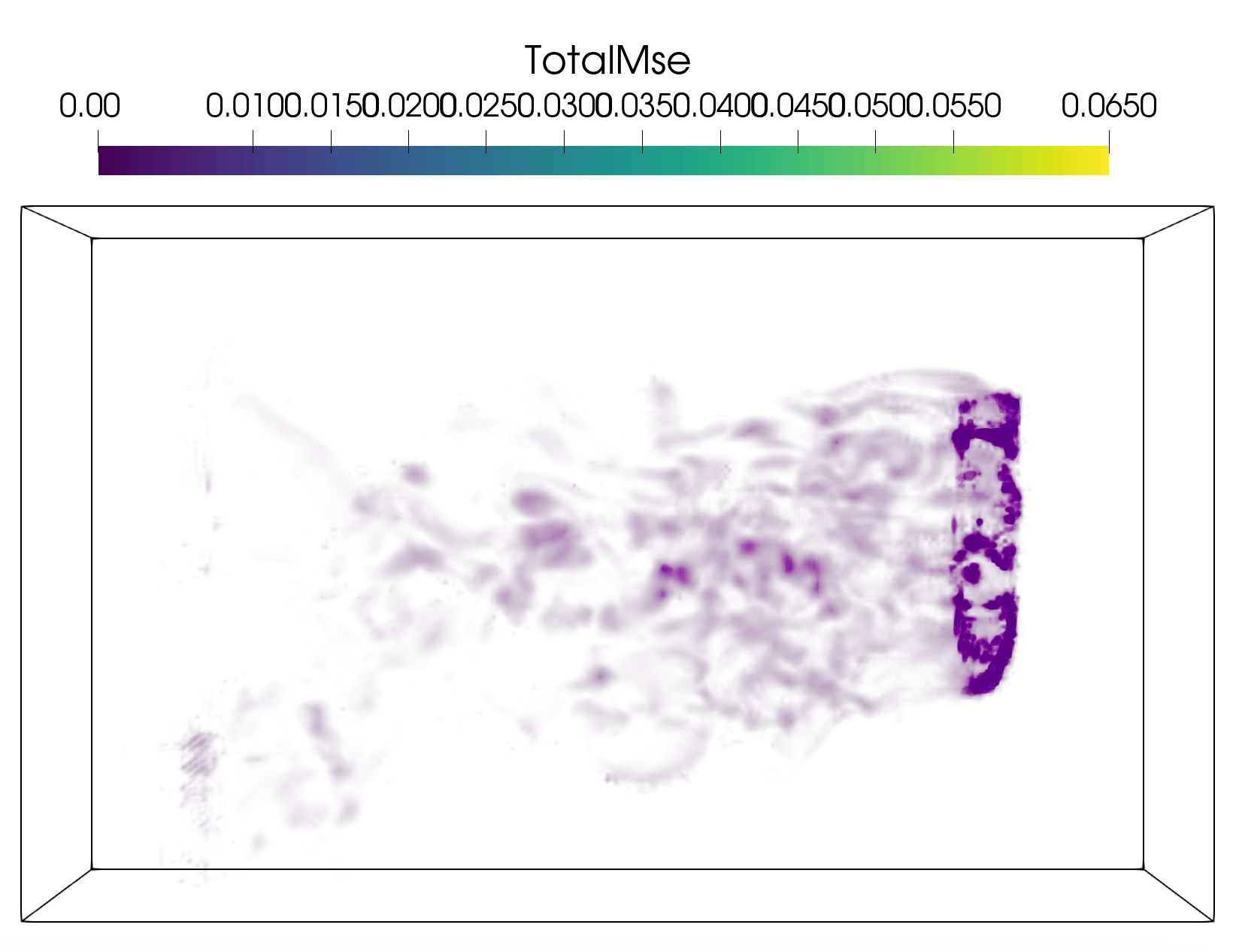}
    \caption{Spatial MSE visualization for Ensemble.}
    \label{tangaroa_MSE_ENS}
\end{subfigure}
~
\begin{subfigure}[t]{0.18\linewidth}
    \centering
    \includegraphics[width=\linewidth]{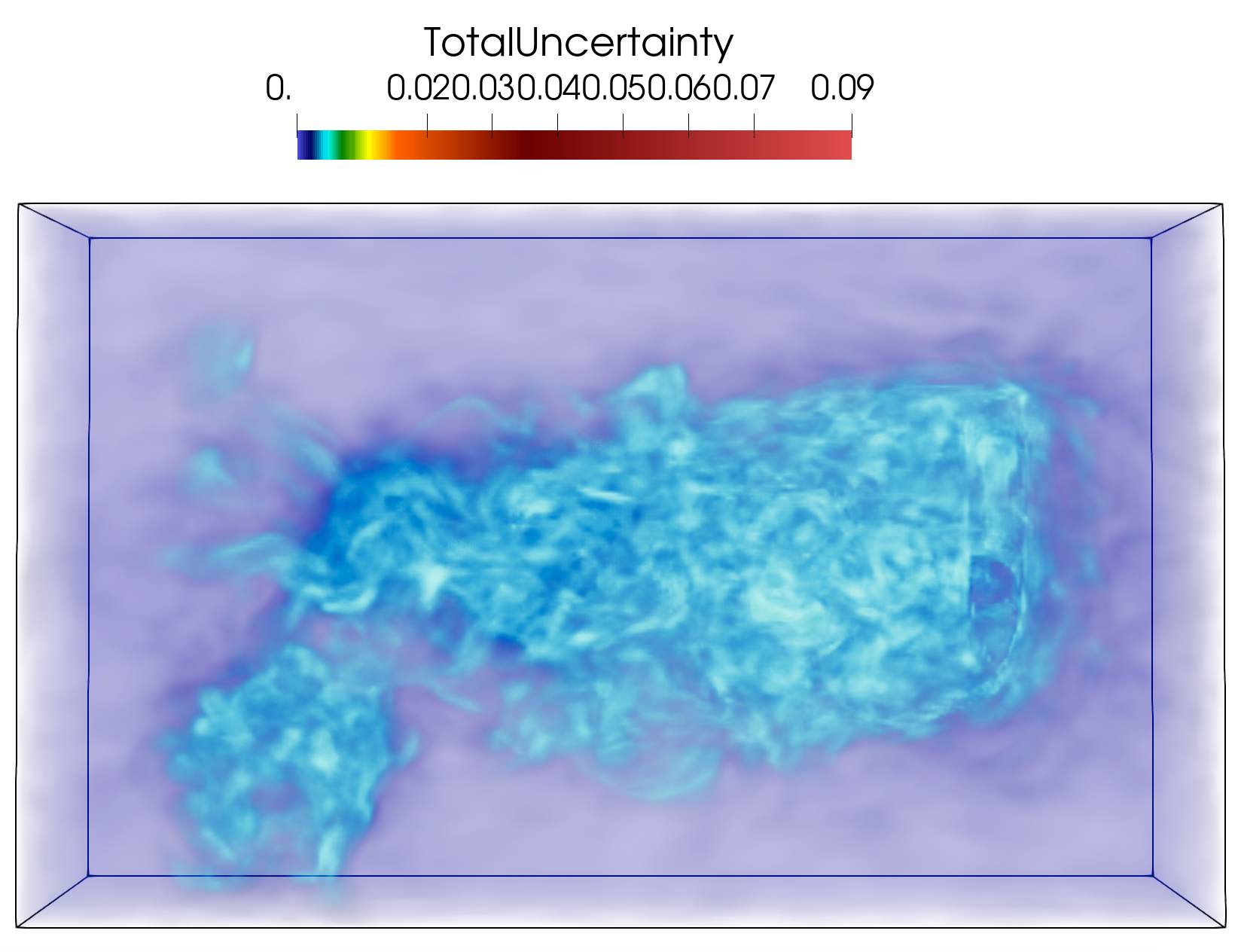}
    \caption{Spatial Uncertainty visualization for MCDropout.}
    \label{tangaroa_Uncert_MCD}
\end{subfigure}
~
\begin{subfigure}[t]{0.18\linewidth}
    \centering
    \includegraphics[width=\linewidth]{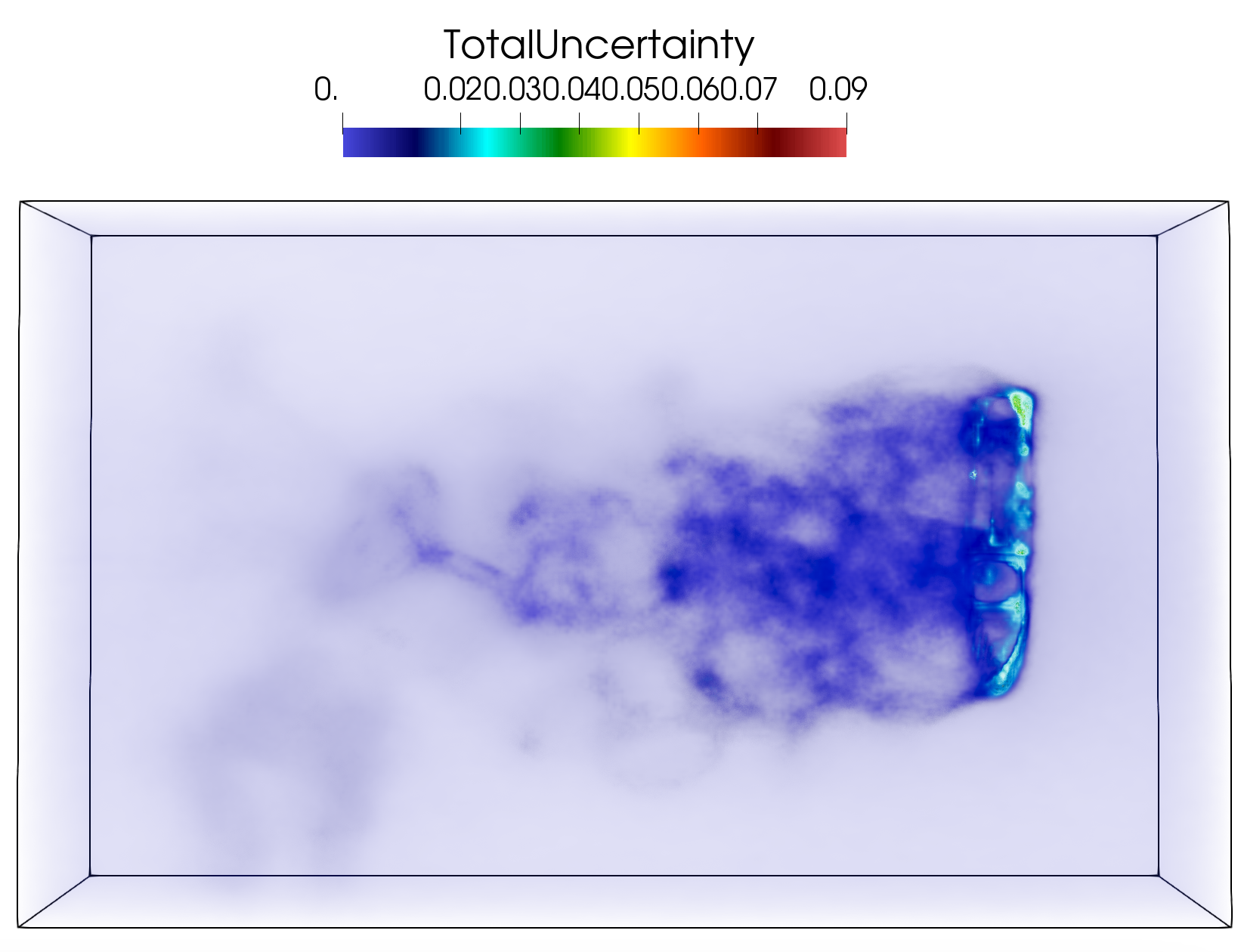}
    \caption{Spatial Uncertainty visualization for Ensemble.}
    \label{tangaroa_Uncert_ENS}
\end{subfigure}
\caption{Visualization of uncertainty and error fields for Tangaroa data set. The MSE and prediction uncertainty are estimated at each grid point between the predicted and ground truth vectors for MCDropout and Ensemble methods. Fig.~\ref{tangaroa_MSE_MCD} and Fig.~\ref{tangaroa_MSE_ENS} show the rendering of MSE fields, and Fig.~\ref{tangaroa_Uncert_MCD} and Fig.~\ref{tangaroa_Uncert_ENS} present the rendering of uncertainty fields. Results indicate that regions with higher MSE align spatially for both methods, while areas with higher prediction uncertainty are more widespread for MCDropout compared to the Ensemble method.}
\label{tangaroa_error_uncert_vis}
\end{figure*}

This section focuses on a qualitative comparison, presenting volume visualizations of the reconstructed vector magnitude fields, estimated prediction uncertainty, and error values for the Isabel and Tangaroa data sets. Fig.~\ref{isabel_mag_vis} and Fig.~\ref{tangaroa_mag_vis} display the volume rendering of ground truth, MCDropout, and Ensemble predicted vector magnitude fields for Isabel and Tangaroa data sets, respectively. Both methods accurately reconstruct the vector components for these data sets. Subsequently, Fig.~\ref{isabel_error_uncert_vis} and Fig.~\ref{tangaroa_error_uncert_vis} present volume renderings of predicted error and uncertainty fields for the Isabel and Tangaroa data sets. It is noted that error and uncertainty do not exhibit strong spatial correlation. While the error patterns for MCDropout and Ensemble methods appear visually similar, spatial uncertainty visualizations differ. In the Isabel data set, regions with higher prediction uncertainty correspond to vortex regions in the vector field for both MCDropout and Ensemble methods (see Fig.~\ref{isabel_Uncrt_MCD} and Fig.~\ref{isabel_Uncrt_ENS}). However, in the Tangaroa data set, the relatively higher uncertainty valued regions are confined to smaller spatial domain for the Ensemble method (Fig.~\ref{tangaroa_Uncert_ENS}). For MCDropout method, we see that the moderately uncertain regions are widespread (Fig.~\ref{tangaroa_Uncert_MCD}), compared to the Ensemble method.

\subsection{Visual Analysis of Flow Features Using Uncertainty-Aware Streamlines}

\begin{figure}[thb]
\centering
\begin{subfigure}[t]{0.38\linewidth}
    \centering
    \includegraphics[width=\linewidth]{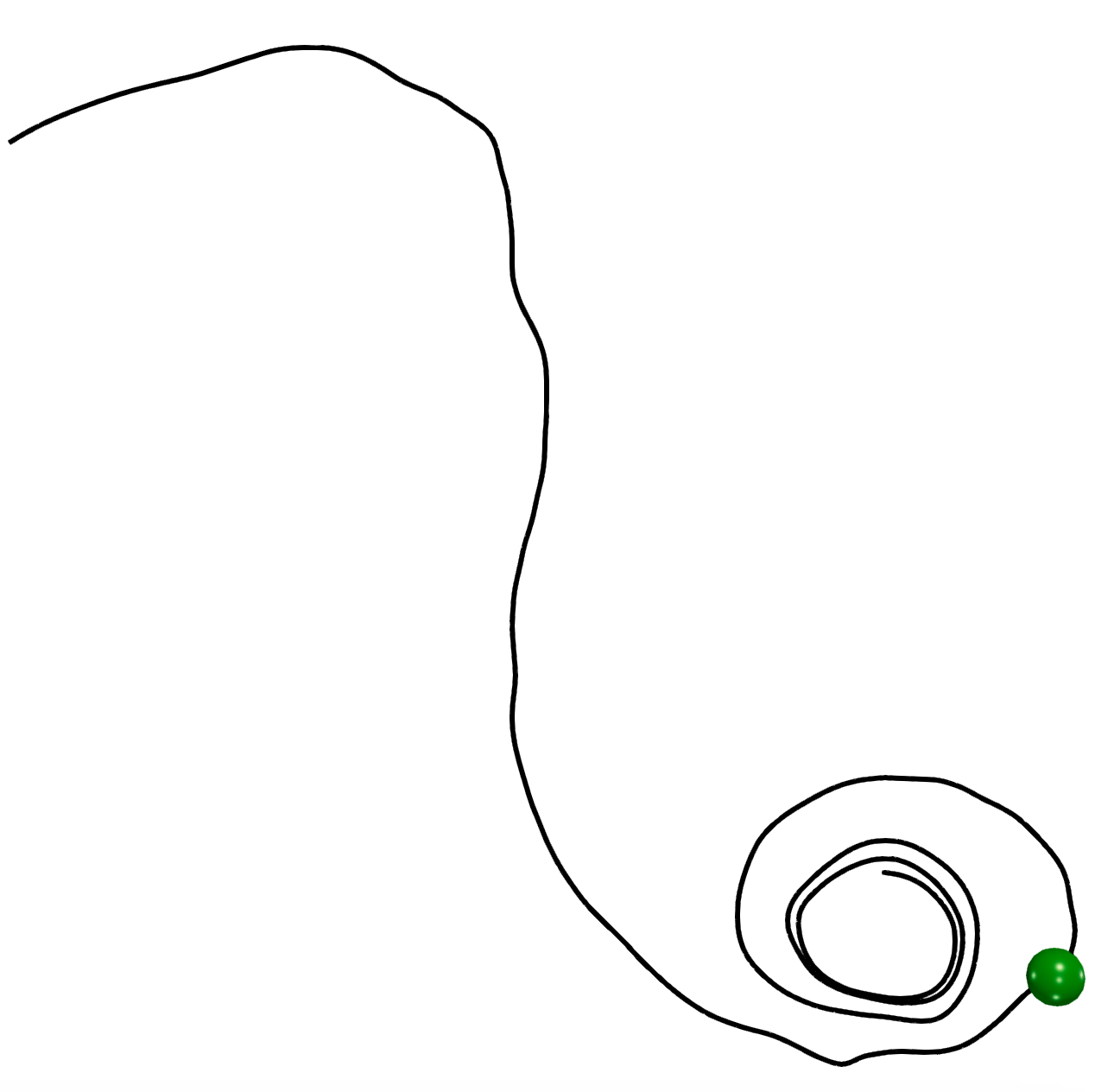}
    \caption{Ground truth streamline.}
    \label{streamline_demo_GT}
\end{subfigure}
~
\begin{subfigure}[t]{0.38\linewidth}
    \centering
    \includegraphics[width=\linewidth]{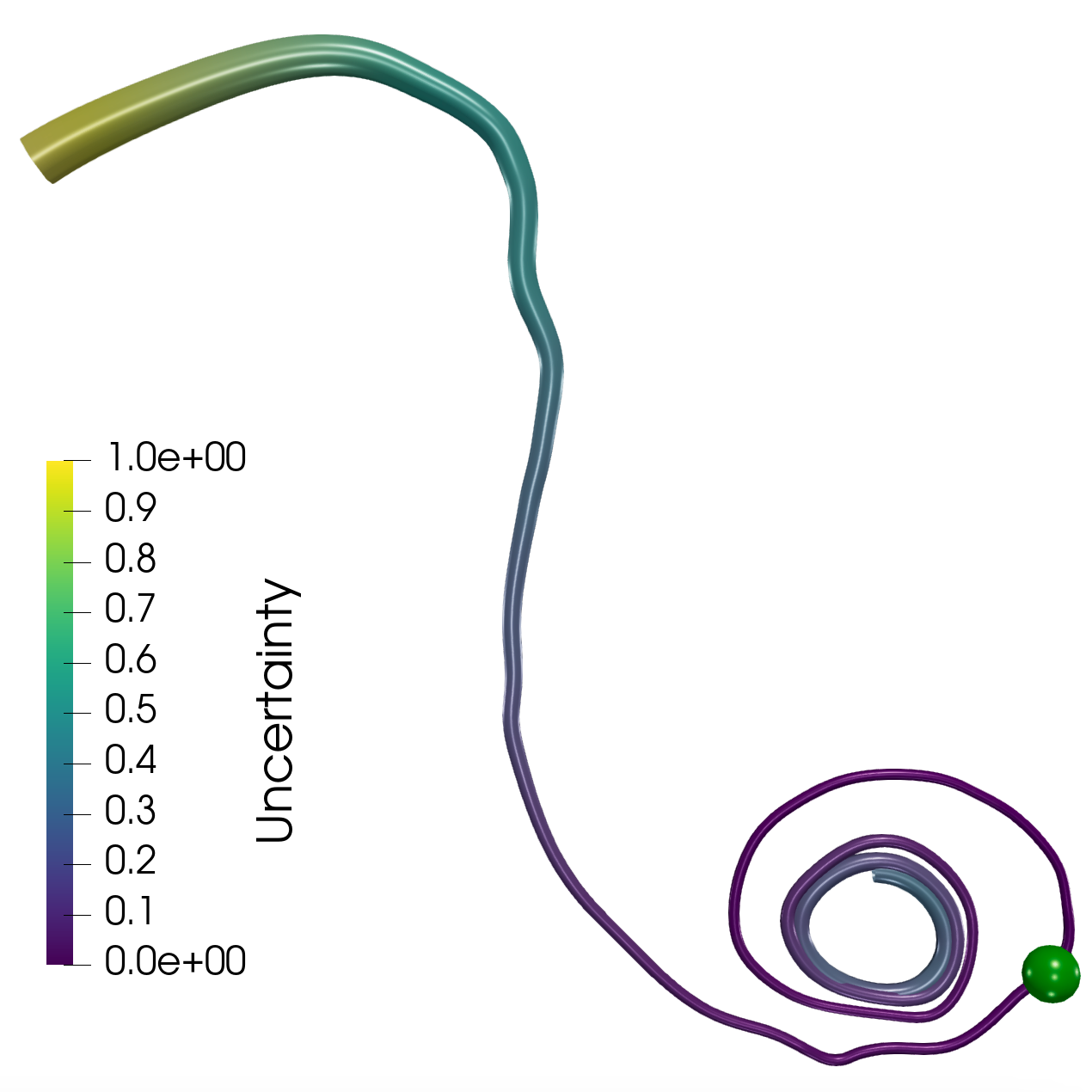}
    \caption{Uncertainty-aware visualization of streamline.}
    \label{streamline_demo_ens}
\end{subfigure}
\caption{Fig.~\ref{streamline_demo_GT} shows the ground truth. Fig.~\ref{streamline_demo_ens} shows the uncertainty-aware streamline from the Ensemble method, where the streamline is visualized as a stream tube, colored with uncertainty, and the diameter of the tube is varied using prediction uncertainty.}
\label{streamline_vis}
\end{figure}

\begin{figure}[thb]
\centering
\begin{subfigure}[t]{0.3\linewidth}
    \centering
    \includegraphics[width=\linewidth]{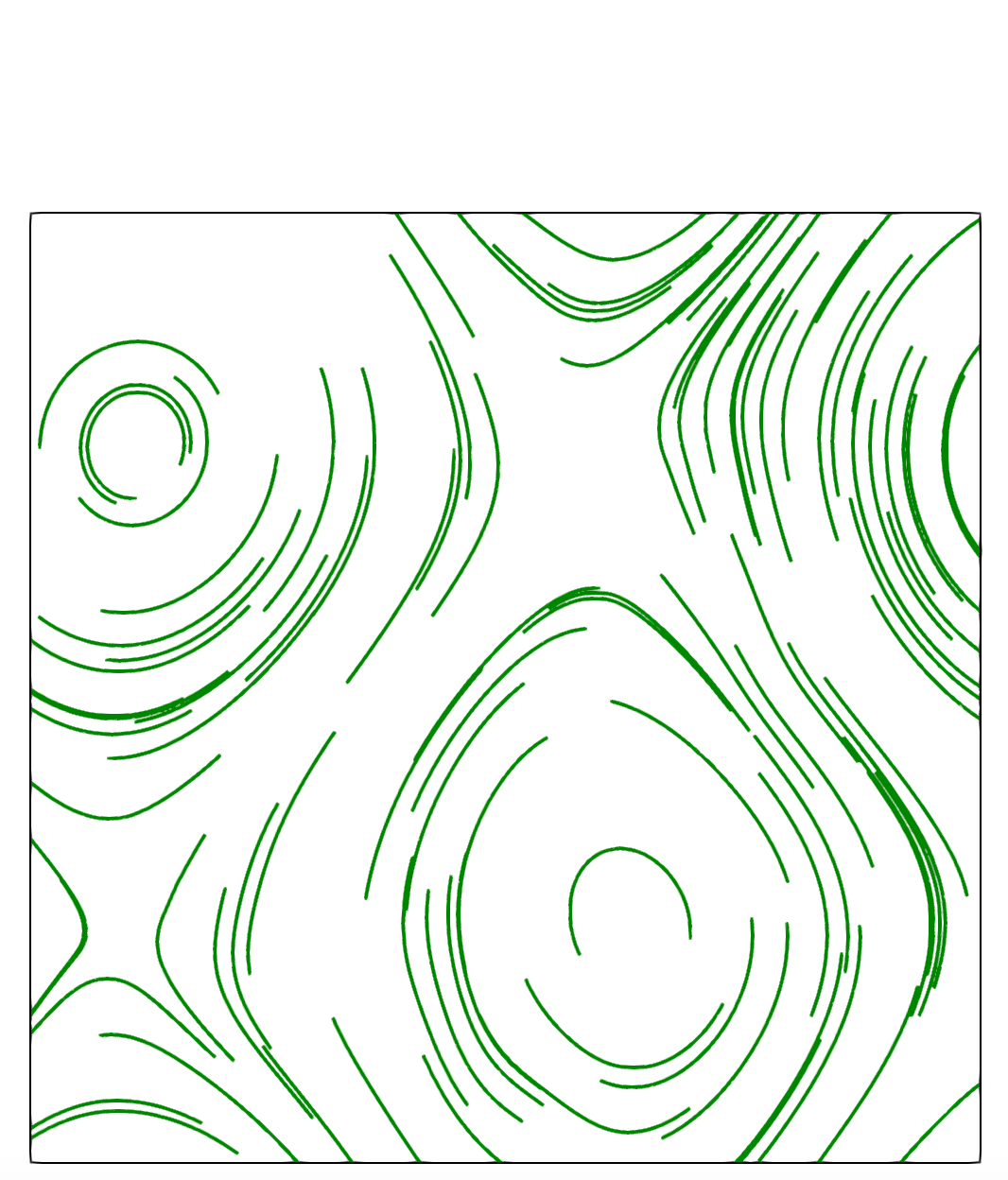}
    \caption{Ground truth streamlines.}
    \label{fluid_streamline_gt}
\end{subfigure}
~
\begin{subfigure}[t]{0.3\linewidth}
    \centering
    \includegraphics[width=\linewidth]{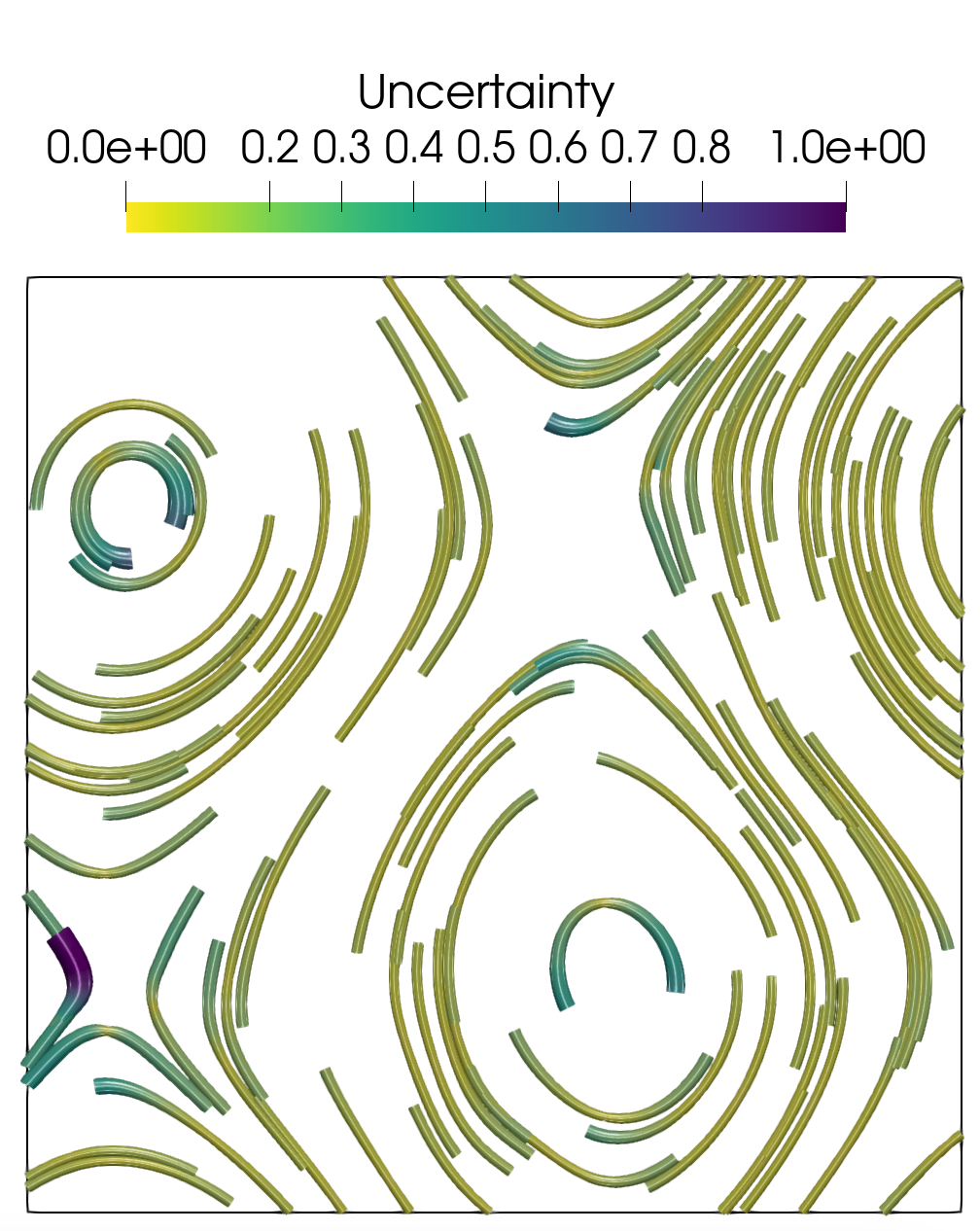}
    \caption{Uncertainty-aware streamline visualization generated by MCDropout method.}
    \label{fluid_streamline_mcd}
\end{subfigure}
~
\begin{subfigure}[t]{0.3\linewidth}
    \centering
    \includegraphics[width=\linewidth]{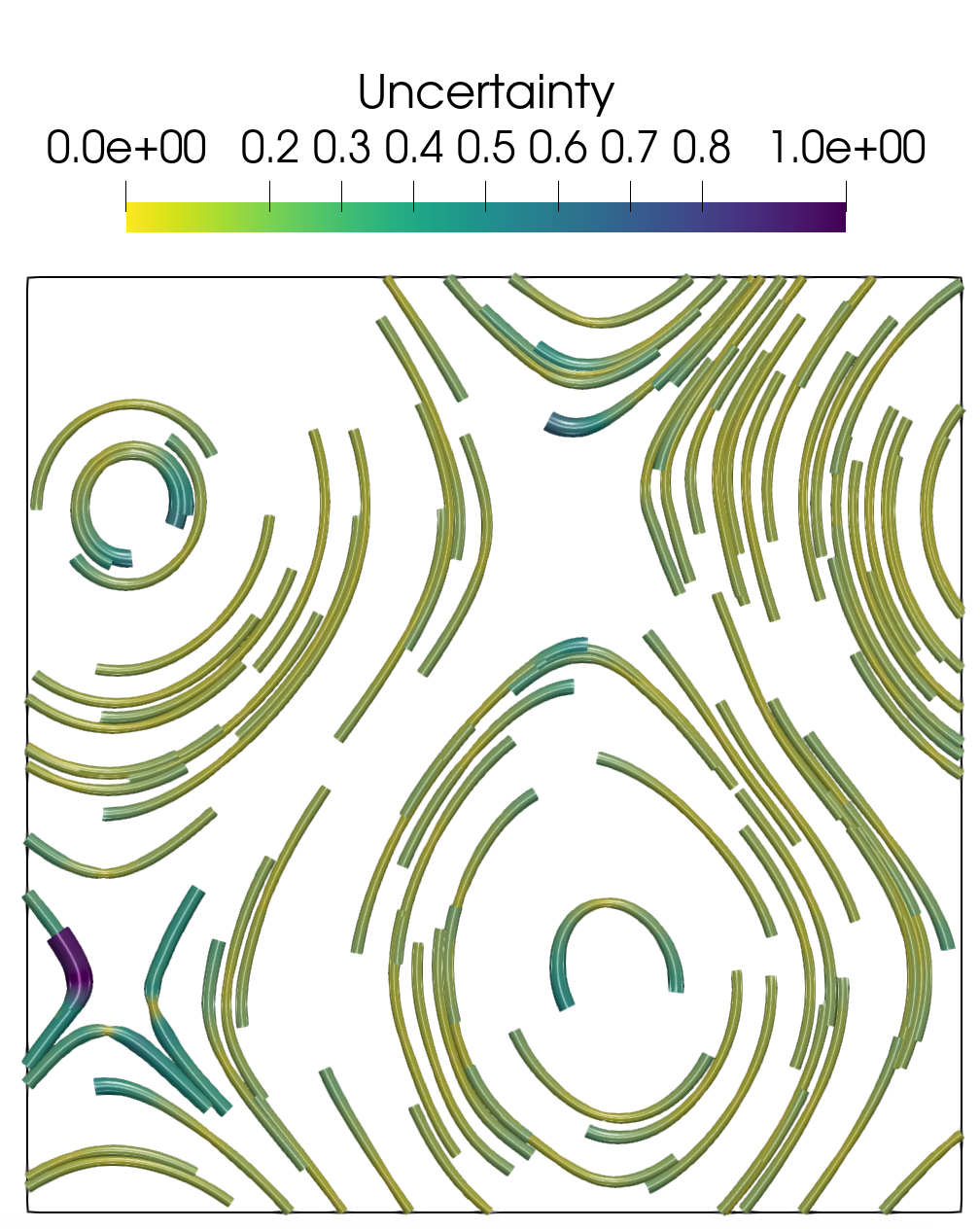}
    \caption{Uncertainty-aware streamline visualization generated by Ensemble method.}
    \label{fluid_streamline_ens}
\end{subfigure}
\caption{Uncertainty-aware streamline visualization for Fluid 2D data. The streamlines are generated using $100$ uniformly randomly generated seeds. Fig.~\ref{fluid_streamline_gt} shows the ground truth streamlines, Fig.~\ref{fluid_streamline_mcd} and Fig.~\ref{fluid_streamline_ens} show streamlines generated by MCDropout and Ensemble method, respectively. The streamlines are colored using uncertainty values and rendered using stream tube visualization, where the diameter of the tube is varied using the prediction uncertainty values.}
\label{fluid_streamline_vis}
\end{figure}

\begin{figure}[thb]
\centering
\begin{subfigure}[t]{0.31\linewidth}
    \centering
    \includegraphics[width=\linewidth]{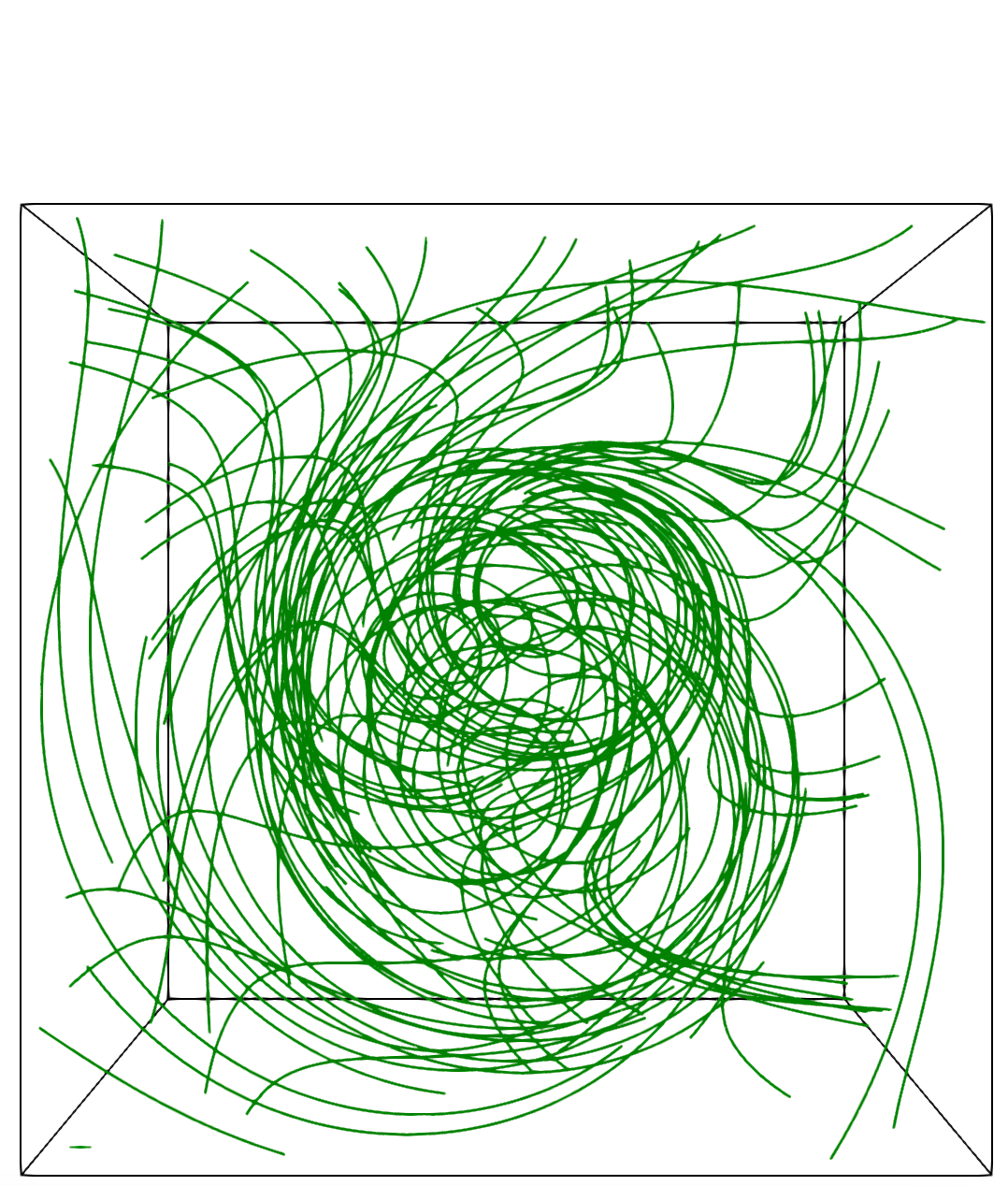}
    \caption{Ground truth streamlines.}
    \label{tornado_streamline_gt}
\end{subfigure}
~
\begin{subfigure}[t]{0.31\linewidth}
    \centering
    \includegraphics[width=\linewidth]{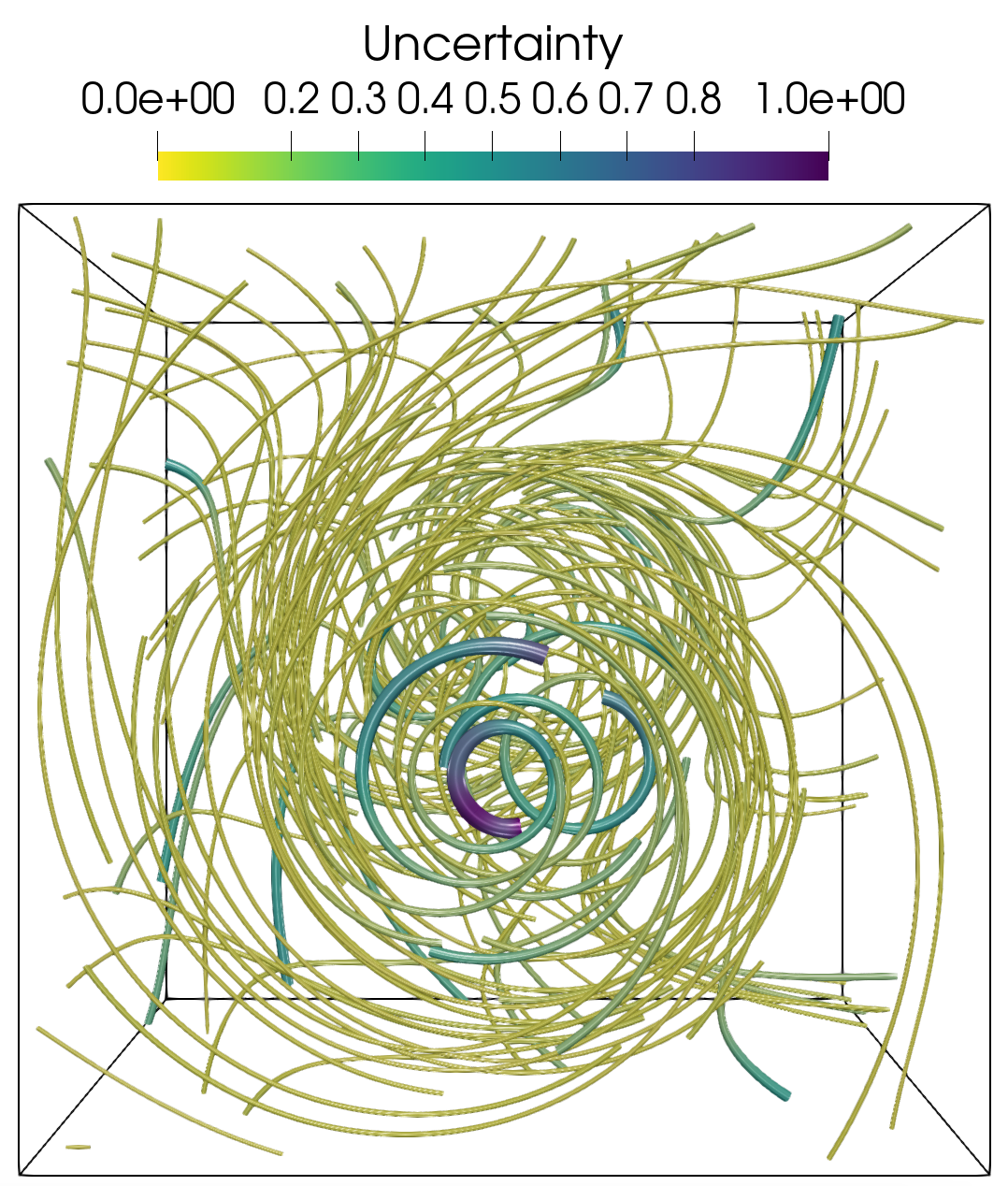}
    \caption{Uncertainty-aware streamline visualization generated by MCDropout method.}
    \label{tornado_streamline_mcd}
\end{subfigure}
~
\begin{subfigure}[t]{0.31\linewidth}
    \centering
    \includegraphics[width=\linewidth]{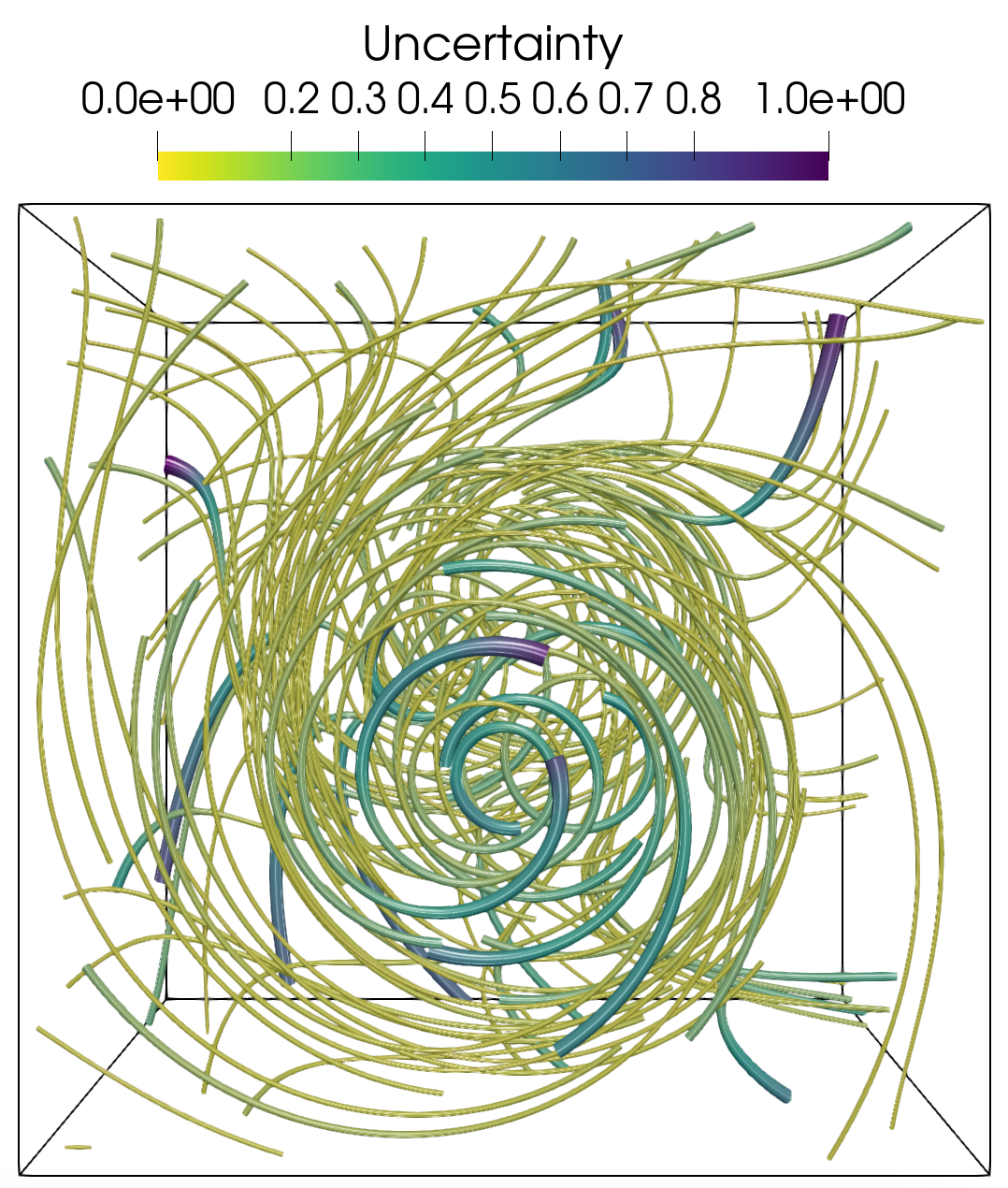}
    \caption{Uncertainty-aware streamline visualization generated by Ensemble method.}
    \label{tornado_streamline_ens}
\end{subfigure}
\caption{Uncertainty-aware streamline visualization for Tornado data. The streamlines are generated using $100$ uniformly randomly generated seeds. Fig.~\ref{tornado_streamline_gt} shows the ground truth streamlines, Fig.~\ref{tornado_streamline_mcd} and Fig.~\ref{tornado_streamline_ens} show streamlines generated by MCDropout and Ensemble method, respectively. The streamlines are colored using uncertainty values and rendered using stream tube visualization, where the diameter of the tube is varied using the prediction uncertainty values.}
\label{tornado_streamline_vis}
\end{figure}

\begin{figure}[thb]
\centering
\begin{subfigure}[t]{0.31\linewidth}
    \centering
    \includegraphics[width=\linewidth]{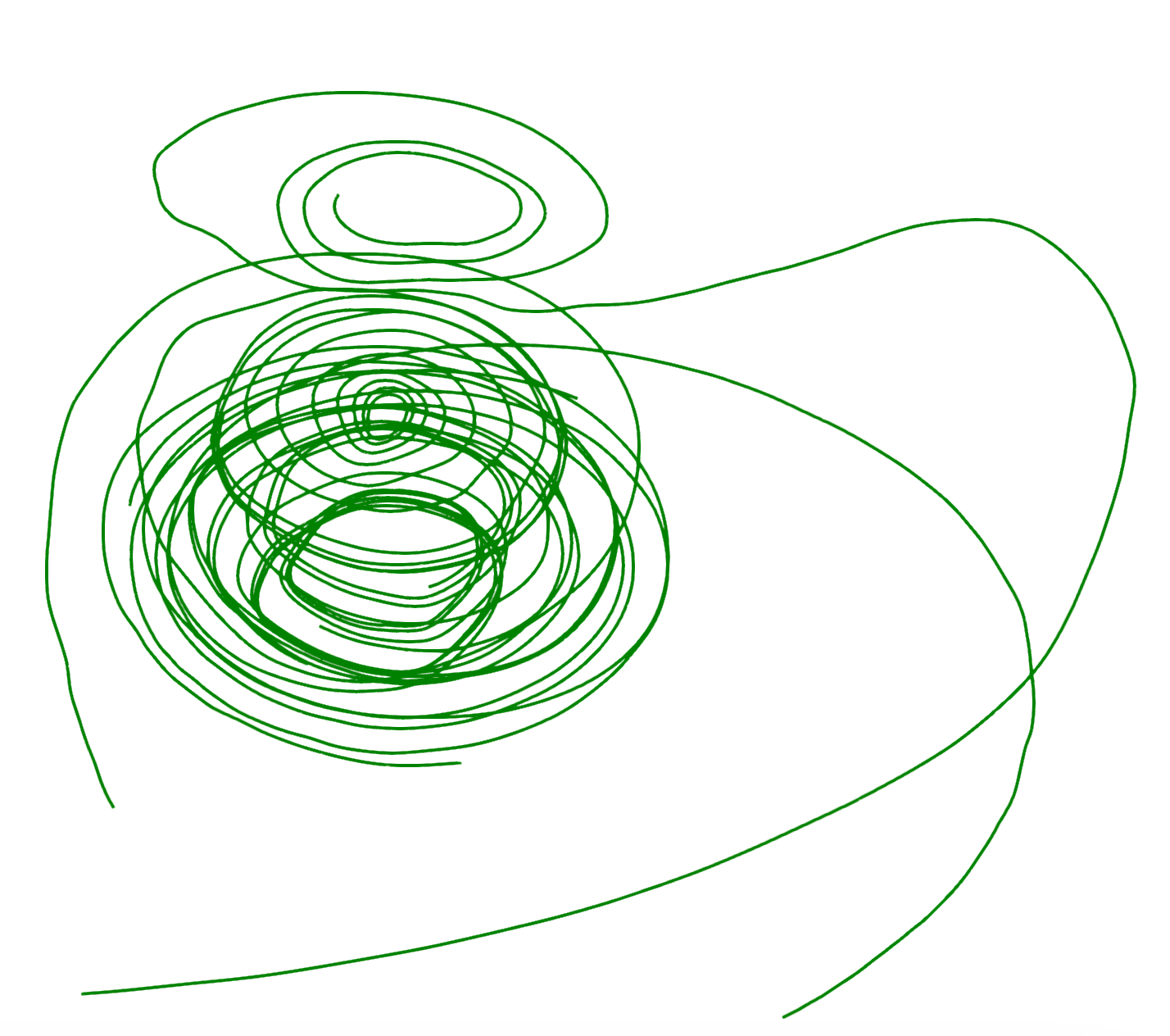}
    \caption{Ground truth streamlines.}
    \label{isabel_streamline_gt}
\end{subfigure}
~
\begin{subfigure}[t]{0.31\linewidth}
    \centering
    \includegraphics[width=\linewidth]{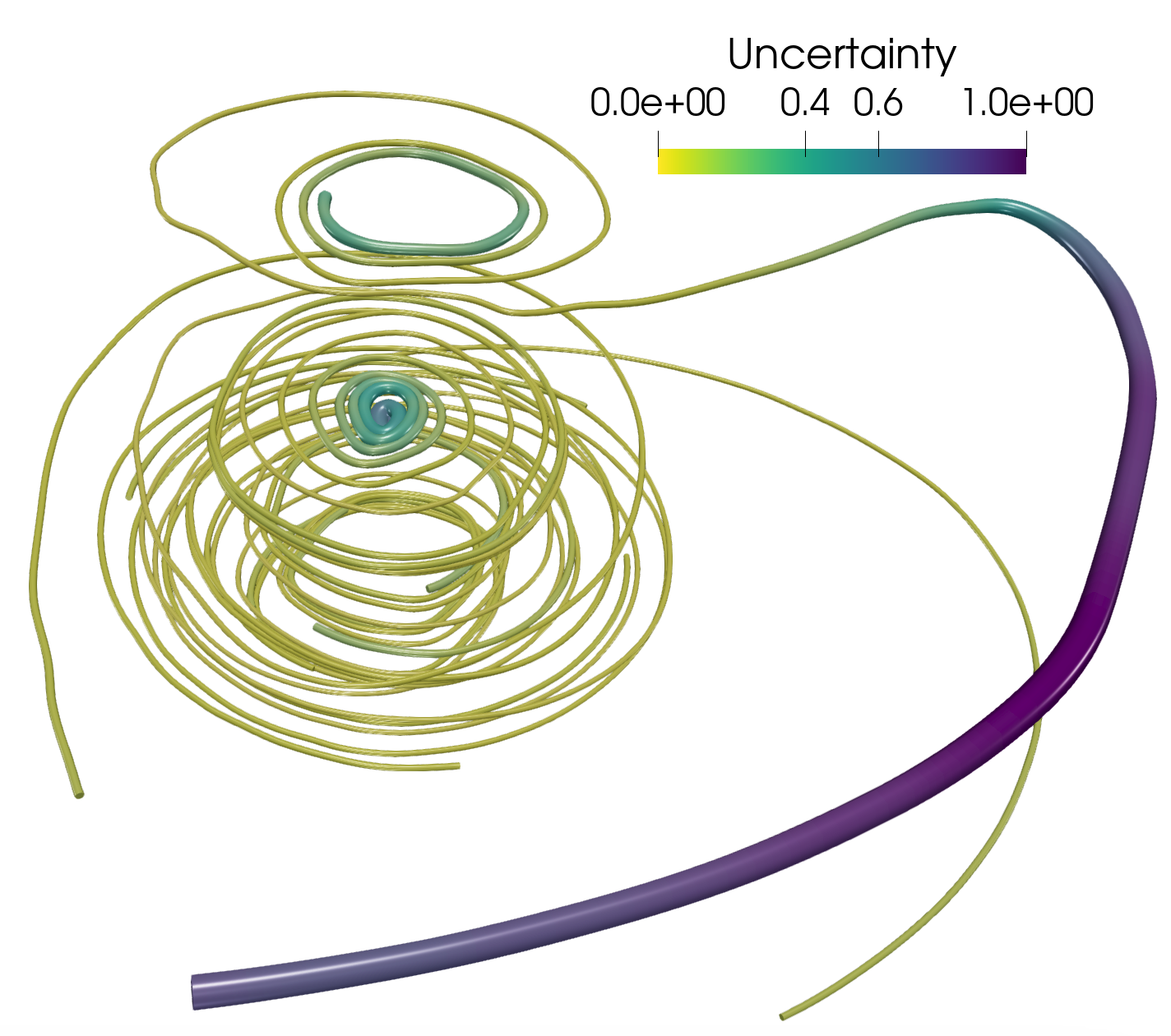}
    \caption{Uncertainty-aware streamline visualization generated by MCDropout method.}
    \label{isabel_streamline_mcd}
\end{subfigure}
~
\begin{subfigure}[t]{0.31\linewidth}
    \centering
    \includegraphics[width=\linewidth]{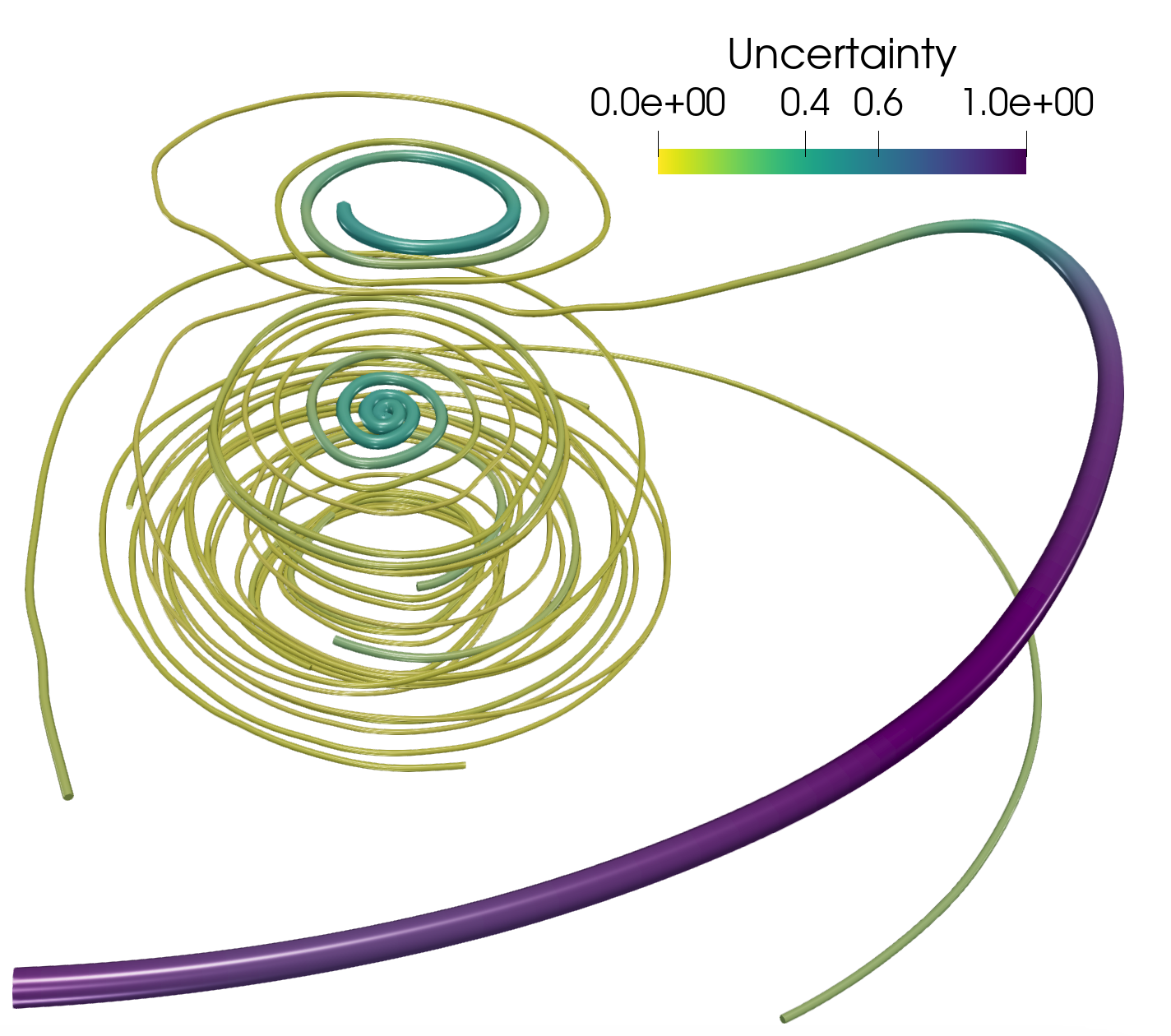}
    \caption{Uncertainty-aware streamline visualization generated by Ensemble method.}
    \label{isabel_streamline_ens}
\end{subfigure}
\caption{Uncertainty-aware streamline visualization for Hurricane Isabel data. The streamlines are generated using $8$ seed points placed around the vortex region in the data. Fig.~\ref{isabel_streamline_gt} shows the ground truth streamlines, Fig.~\ref{isabel_streamline_mcd} and Fig.~\ref{isabel_streamline_ens} show streamlines generated by MCDropout and Ensemble method respectively. The streamlines are colored using uncertainty values and rendered using stream tube visualization, where the diameter of the tube is varied using the prediction uncertainty values.}
\label{isabel_streamline_vis}
\end{figure}

\begin{figure}[thb]
\centering
\begin{subfigure}[t]{0.4\linewidth}
    \centering
    \includegraphics[width=\linewidth]{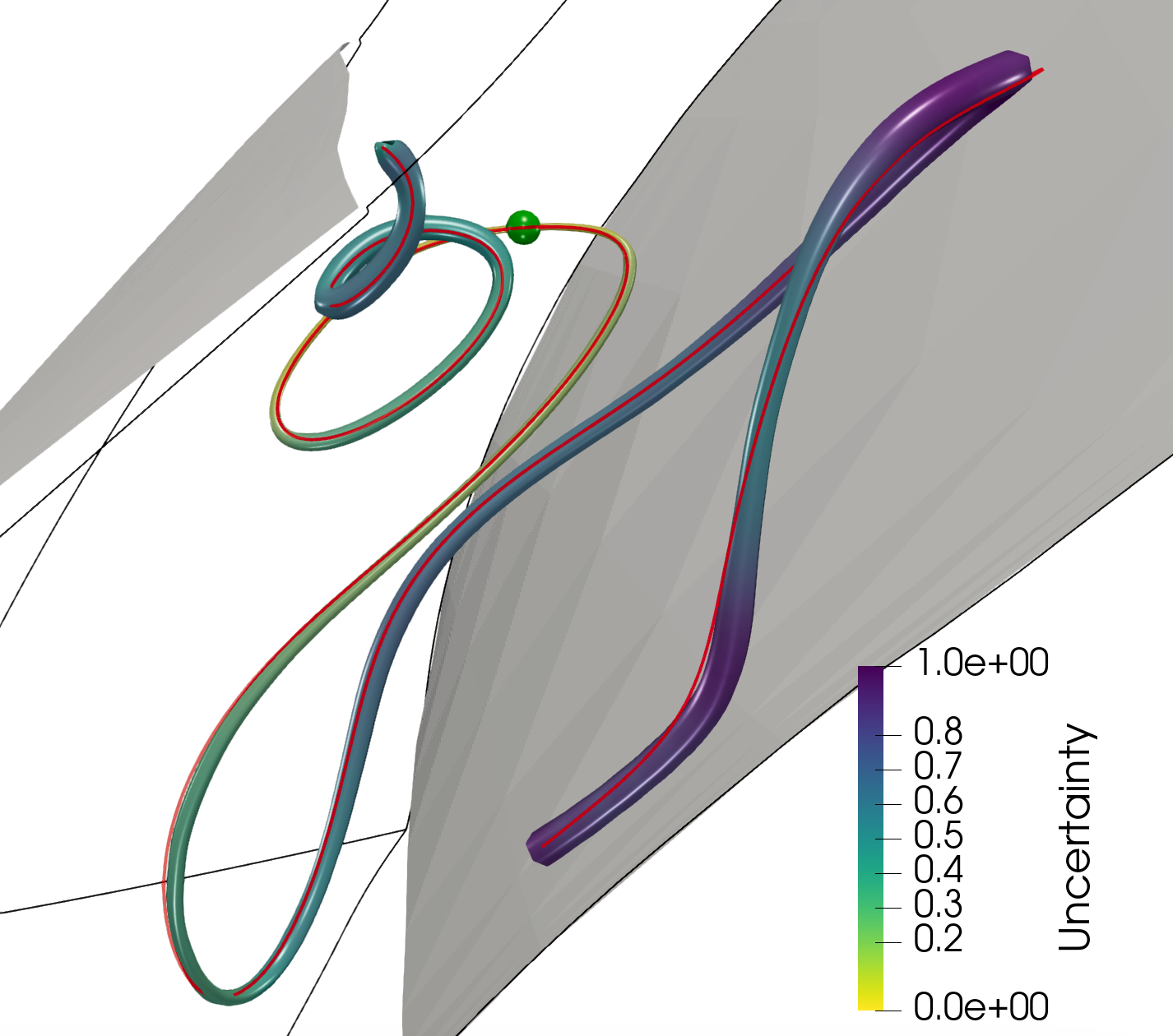}
    \caption{Uncertainty-aware streamline visualization generated by MCDropout method.}
    \label{turbine_streamline_mcd}
\end{subfigure}
~
\begin{subfigure}[t]{0.4\linewidth}
    \centering
    \includegraphics[width=\linewidth]{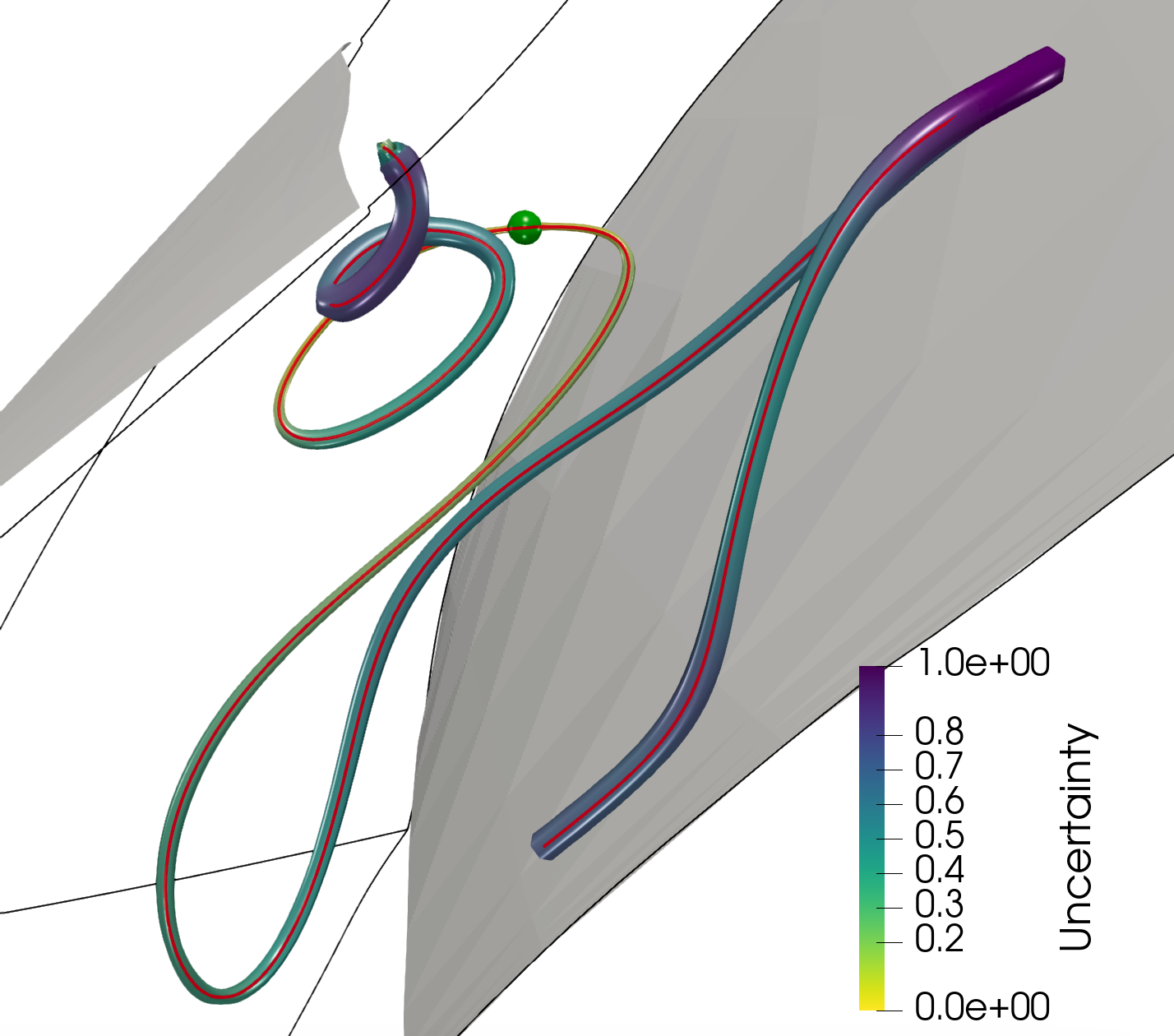}
    \caption{Uncertainty-aware streamline visualization generated by Ensemble method.}
    \label{turbine_streamline_ens}
\end{subfigure}
\caption{Uncertainty-aware streamline visualization for Turbine data. We show an uncertainty-informed streamline of a selected seed point. The ground truth (colored in red) is overlayed with the uncertain stream-tube as a reference, and the seed is shown as a green sphere. By comparing the streamline generated by the MCDropout method (Fig.~\ref{turbine_streamline_mcd}) and the Ensemble method (Fig.~\ref{turbine_streamline_ens}), it is observed that both the methods produce visually similar and comparable uncertainty.}
\label{turbine_streamline_vis}
\end{figure}

Streamlines serve as valuable tools for visualizing flow patterns, identifying recirculation regions, and pinpointing stagnation points, aiding in understanding fluid behavior across engineering and scientific domains. To enhance the effectiveness of flow feature visualization, we employ both MCDropout and Ensemble methods with our model's predictions. These methods not only generate streamlines but also allow us to quantify uncertainty in predicted vector values at each step of the integration process. To generate streamlines, we use RK-4 integration and trace the streamline in both \rmark{forward} and backward direction. Using the MCDropout approach, we generate $n$ realizations of the same streamline from a given seed point (with $n$=100 in our study) through MC sampling. \rmark{To estimate the expected streamline, we compute the mean streamline by averaging the coordinates at each integration step. However, due to variability among the MC streamline samples, some streamlines may go out of bounds. In such instances, we calculate the mean using only the existing streamlines at those integration steps. Our analysis method also supports the computation of the median streamline as an alternative to the mean streamline to mitigate the impact of outliers. The uncertainty at each step of the streamline is determined by the standard deviation computed from the $n$ samples following similar strategy as used to compute the mean streamline.} Similarly, employing the Ensemble method involves generating $n$ realizations of the streamline (where $n=30$ due to our use of 30 ensemble members), considering the mean streamline as the final result, and computing pointwise standard deviation as the uncertainty estimate. 
 
Utilizing models sensitive to uncertainty helps measure and communicate uncertainty associated with streamlines, aiding users in making informed decisions during flow feature analysis~\cite{Ferstl_streamline_variability}. This uncertainty information builds user confidence in the model's predictions and highlights segments where predictions may lack confidence. We visualize uncertainty in streamlines by rendering the mean (averaged) streamline as a stream-tube, with the stream-tube diameter scaled by uncertainty estimates and colored by uncertainty values. Fig.\ref{streamline_vis} shows this stream-tube visualization, where Fig.\ref{streamline_demo_GT} displays the ground truth streamline without uncertainty, and Fig.~\ref{streamline_demo_ens} shows the uncertainty-aware visualization. Higher uncertainty is observed at the streamline's ends, with the seed location marked by a green sphere. Next, we qualitatively study the uncertainty characteristics estimated by MCDropout and Ensemble methods through several case studies.

\textbf{Fluid (2D) Data Set.} We use $100$ uniformly randomly generated seed points to generate streamlines for the Fluid data set using MCDropout and Ensemble methods. The results are shown in Fig.~\ref{fluid_streamline_vis}. The ground truth streamlines are shown in Fig.~\ref{fluid_streamline_gt}, and streamlines from MCDropout and Ensemble methods are provided in Fig.~\ref{fluid_streamline_mcd} and Fig.~\ref{fluid_streamline_ens} respectively. The streamlines are colored using prediction uncertainty, and the thickness of the stream tubes also varies according to the prediction uncertainty. It is observed that both MCDropout and Ensemble methods can generate accurate streamlines when compared against ground truth streamlines (Fig.~\ref{fluid_streamline_gt}). Furthermore, by comparing Fig.~\ref{fluid_streamline_mcd} and Fig.~\ref{fluid_streamline_ens}, we also learn that both MCDropout and Ensemble methods estimate comparable uncertainty estimates across the entire data set.

\textbf{Tornado Data Set.} Next, we conduct a similar study using the Tornado data set to study the generated uncertainty patterns. In this study, streamlines on $100$ uniformly randomly generated seed points are shown, where the ground truth streamlines are shown in Fig.~\ref{tornado_streamline_gt}. \rmark{We find that both MCDropout and Ensemble methods produce visually accurate streamlines. We also notice that while the uncertainty patterns are generally comparable, several streamlines demonstrate different uncertainty patterns between MCDropout and Ensemble methods as seen from Fig.~\ref{tornado_streamline_mcd} and Fig.~\ref{tornado_streamline_ens} respectively.} 

\textbf{Hurricane Isabel Data Set.} To study the uncertainty-aware streamlines in the Hurricane Isabel data set, we generate eight streamlines by seeding around the vortex core region in the data set. The ground truth streamlines are shown in Fig.~\ref{isabel_streamline_gt}. Fig.~\ref{isabel_streamline_mcd} and Fig.~\ref{isabel_streamline_ens} depict the uncertainty-aware streamlines generated by MCDropout and Ensemble methods respectively. Again, we observe a comparable uncertainty pattern for both uncertainty estimation methods.

\textbf{Turbine Data Set.} Finally, uncertainty-aware streamline visualization for the Turbine data set is presented in Fig.~\ref{turbine_streamline_vis}. We show a visualization of a selected streamline, generated by the MCDropout method, in Fig.~\ref{turbine_streamline_mcd}. Visualization of the same generated by the Ensemble method is given in Fig.~\ref{turbine_streamline_ens}. The rotor blades in the Turbine data set are shown as a context using gray surface rendering, and the ground truth streamline (in red color) is overlaid for visual comparison. Similar to the previous case studies, we observe comparable uncertainty estimates from the MCDropout and Ensemble methods.

\subsection{Visual Analysis of Flow Features Using Uncertainty-Aware Critical Points}
\begin{figure}[thb]
\centering
\begin{subfigure}[t]{0.3\linewidth}
    \centering
    \includegraphics[width=\linewidth]{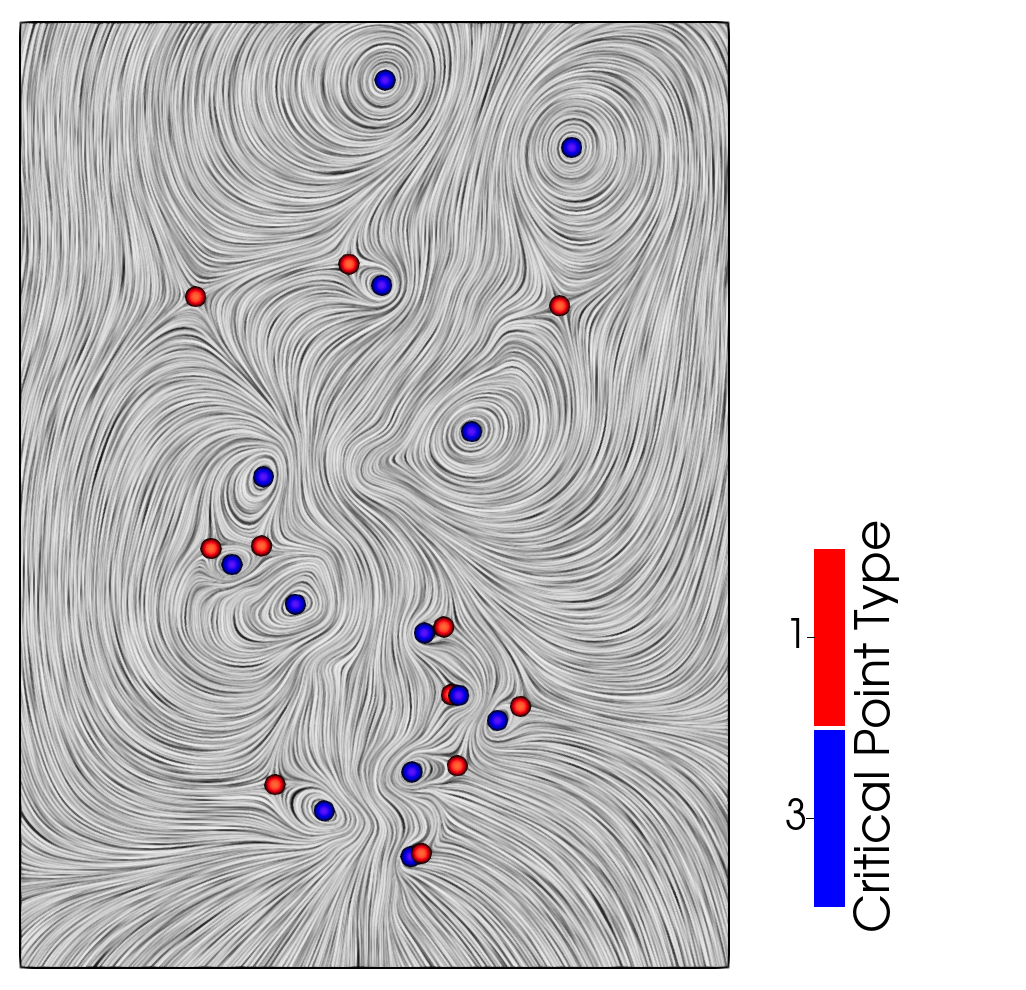}
    \caption{Ground truth critical points with the vector field visualized as LIC as context.}
    \label{boussinesq_CP_GT}
\end{subfigure}
~
\begin{subfigure}[t]{0.3\linewidth}
    \centering
    \includegraphics[width=\linewidth]{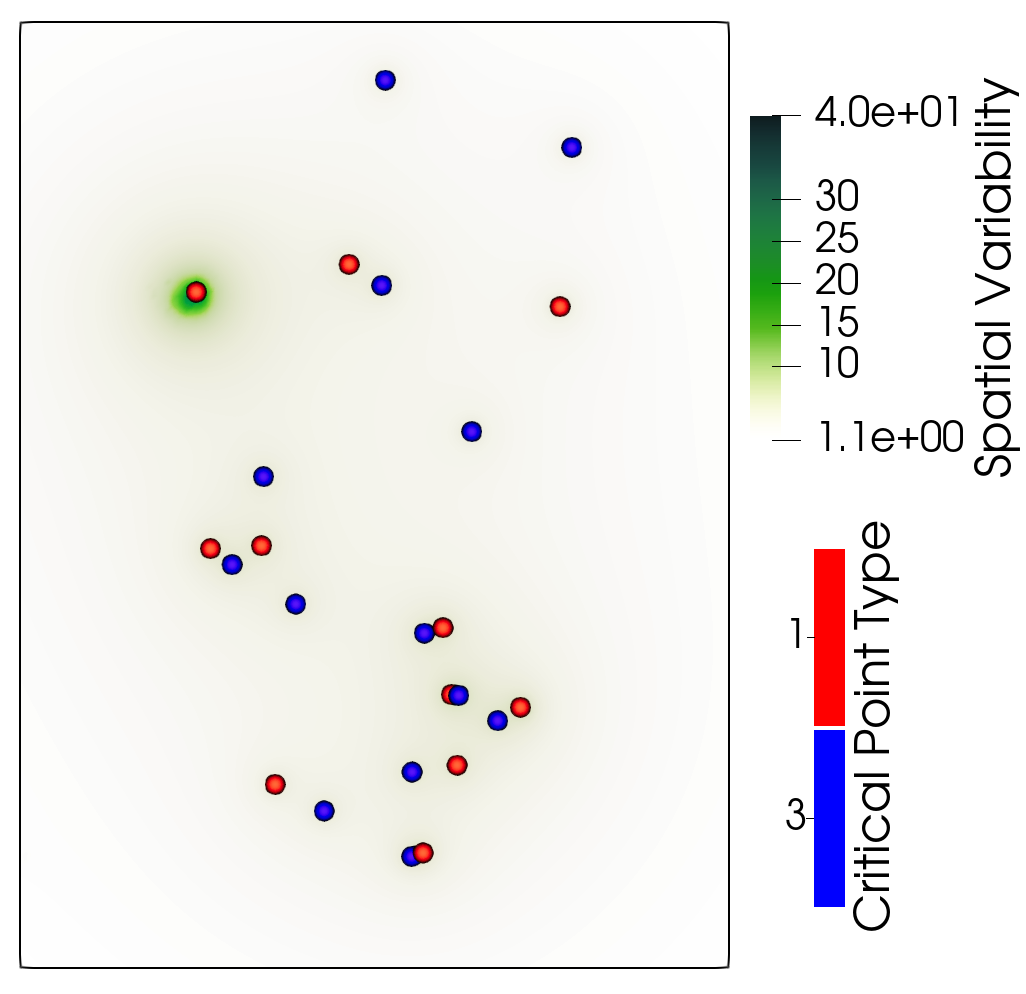}
    \caption{Critical points detected by the MCDropout method. The positional variability is shown with the green shade in the background, indicating prediction uncertainty.}
    \label{boussinesq_CP_MCD}
\end{subfigure}
~
\begin{subfigure}[t]{0.3\linewidth}
    \centering
    \includegraphics[width=\linewidth]{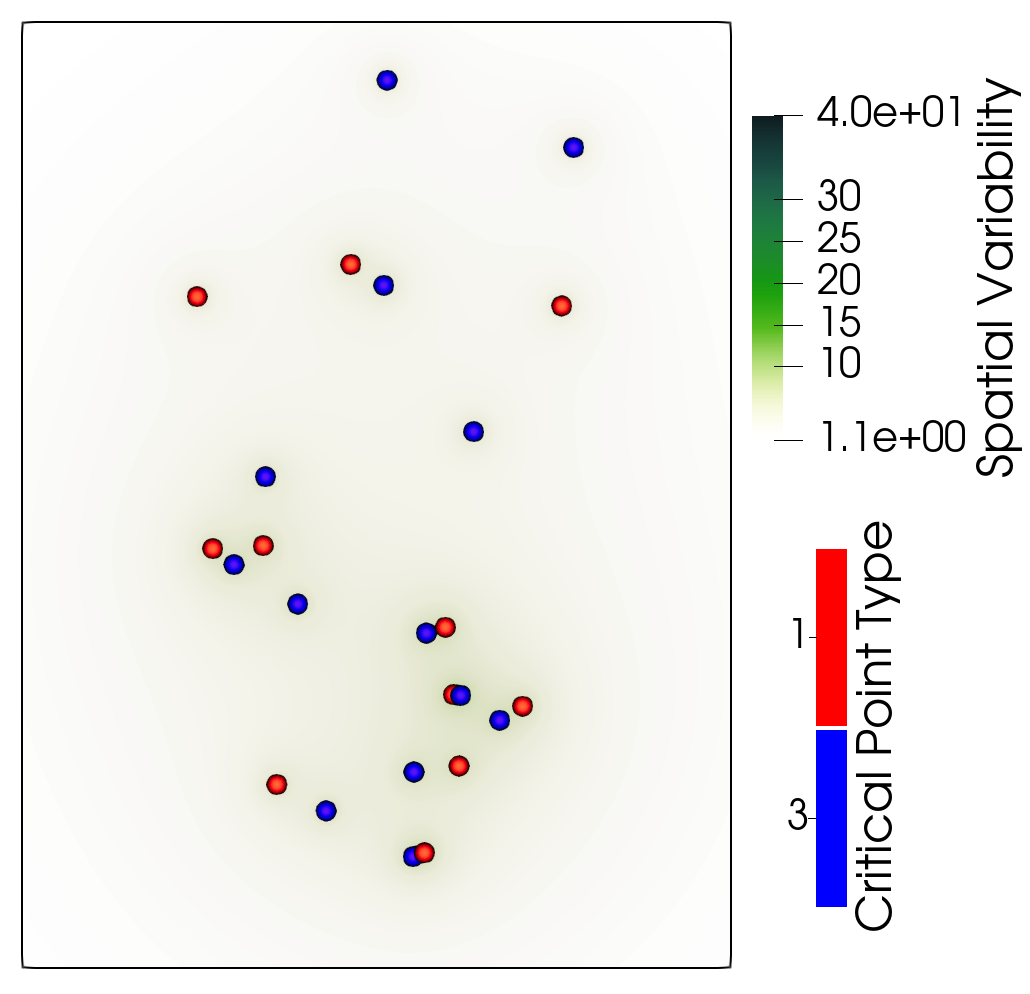}
    \caption{Critical points detected by the Ensemble method. The positional variability is shown with the green shade in the background, indicating prediction uncertainty.}
    \label{boussinesq_CP_ENS}
\end{subfigure}
\caption{Result of critical point analysis for Heated Cylinder data set (T=750). Critical points detected by MCDropout and Ensemble methods are shown in Fig.~\ref{boussinesq_CP_MCD} and Fig.~\ref{boussinesq_CP_ENS}, respectively. The green highlighted regions in the background show the variability in the predicted critical point's locations, with brighter green indicating higher variability. The ground truth is shown in Fig.~\ref{boussinesq_CP_GT}. We observe that both methods detect all the critical points, and the MCDropout method produces higher prediction variability than the Ensemble method.}
\label{boussinesq_crit_vis}
\end{figure}

\begin{figure}[thb]
\centering
\begin{subfigure}[t]{0.3\linewidth}
    \centering
    \includegraphics[width=\linewidth, height=1.2in]{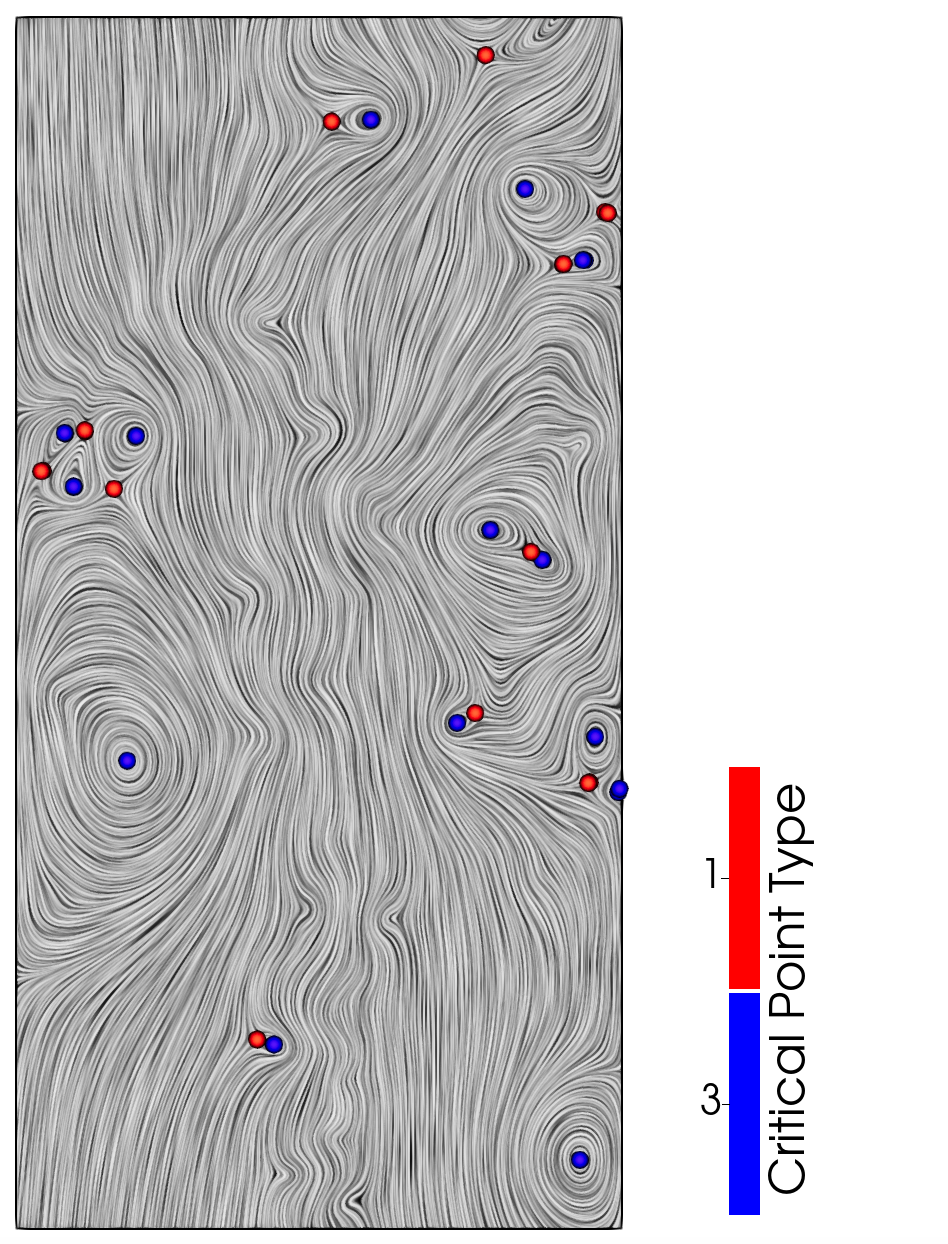}
    \caption{Ground truth critical points with the vector field visualized as LIC as context.}
    \label{boussinesq1_CP_GT}
\end{subfigure}
~
\begin{subfigure}[t]{0.3\linewidth}
    \centering
    \includegraphics[width=\linewidth, height=1.2in]{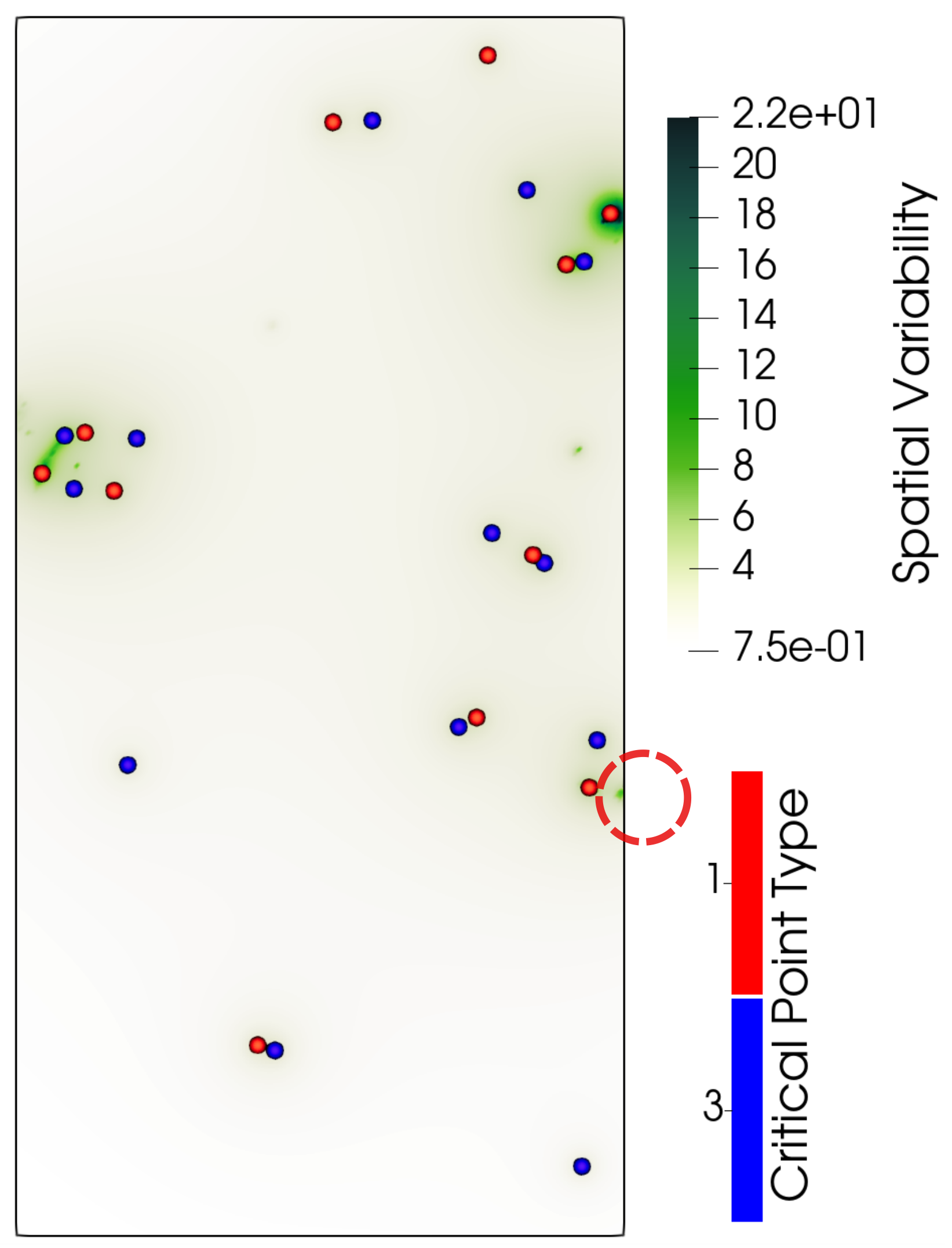}
    \caption{Critical points detected by the MCDropout method. The positional variability is shown with the green shade in the background, indicating prediction uncertainty.}
    \label{boussinesq1_CP_MCD}
\end{subfigure}
~
\begin{subfigure}[t]{0.3\linewidth}
    \centering
    \includegraphics[width=\linewidth, height=1.2in]{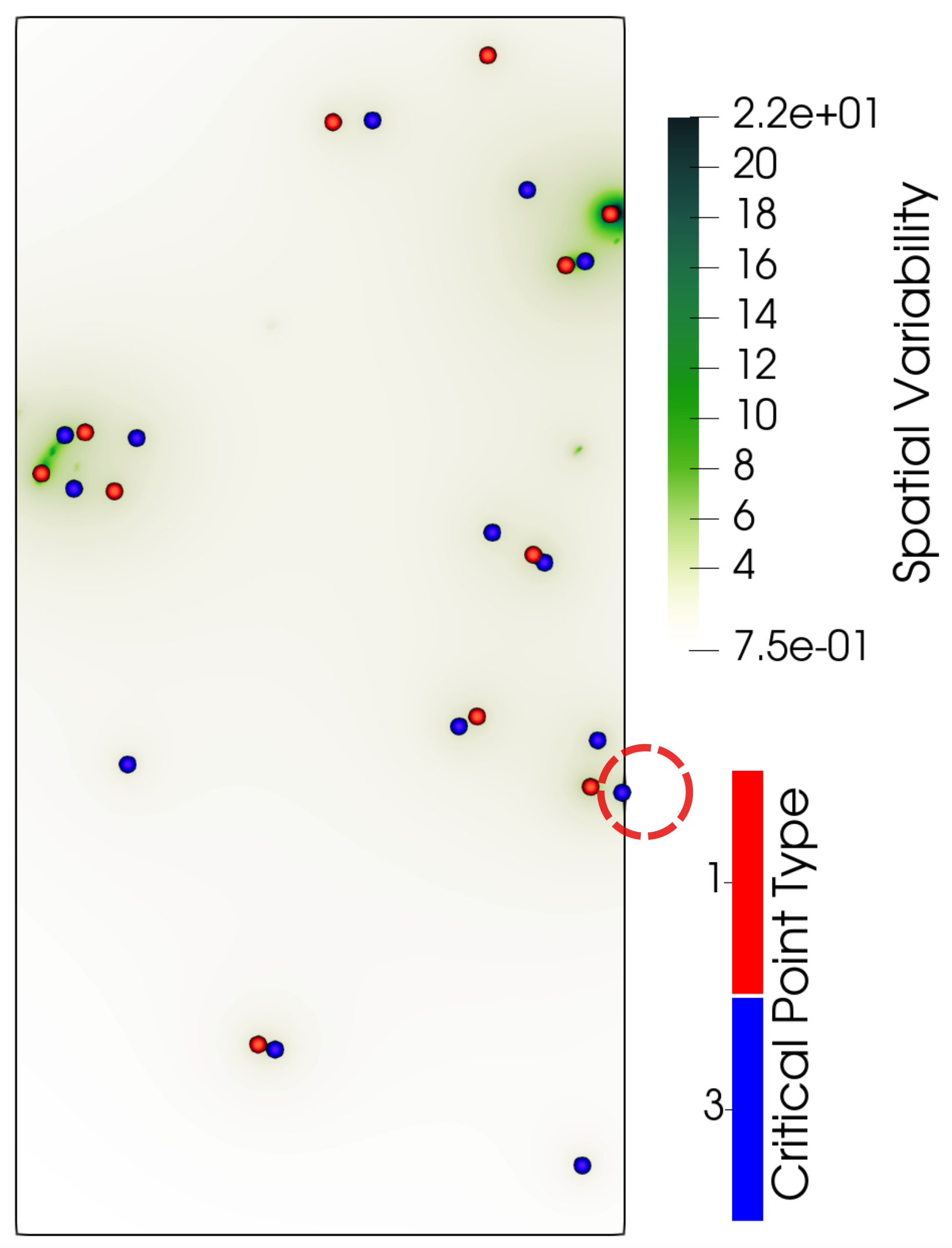}
    \caption{Critical points detected by the Ensemble method. The positional variability is shown with the green shade in the background, indicating prediction uncertainty.}
    \label{boussinesq1_CP_ENS}
\end{subfigure}
\caption{\rmark{Result of critical point analysis for Heated Cylinder data set (T=1500). Critical points detected by MCDropout and Ensemble methods are shown in Fig.~\ref{boussinesq1_CP_MCD} and Fig.~\ref{boussinesq1_CP_ENS}, respectively. The green highlighted regions in the background show the variability in the predicted critical point's locations, with brighter green indicating higher prediction variability. The ground truth critical points are shown in Fig.~\ref{boussinesq1_CP_GT}. We further observe that the MCDropout method fails to detect a critical point at the boundary, as shown by the red dotted circle in Fig.~\ref{boussinesq1_CP_MCD}, which is successfully detected by the Ensemble method.}}
\label{boussinesq1_crit_vis}
\end{figure}

Critical points play an important role in studying vector field characteristics as it helps identify points where the flow behavior undergoes significant changes. These points, including saddles, sources, centers, and sinks, provide valuable insights into the dynamics of the vector field. By observing  critical points, users determine flow patterns, locate stagnation points, and comprehend the overall flow behavior. Comprehensive understanding of robustness of critical points can lead to the accurate flow features study~\cite{crit_point_robustness, gunther2014mandatory, crit_point_ensemble} and flow data compression~\cite{vector_compression_crit_point}.

\rmark{In this study, we evaluate the accuracy of uncertainty-aware neural models in predicting critical point locations and quantify the spatial variability of these predictions using the MCDropout and Ensemble methods. Initially, we compute the mean vector field using both techniques to ensure the robustness of the estimated critical points. From this mean-field, we identify the critical point locations. To assess the spatial variability of each critical point's location derived from the mean field, we analyze data from individual vector field realizations: $100$ fields for the MCDropout and $30$ for the Ensemble method. For each method, first, we identify critical points in each vector field realization. Then, we create a new scalar field where the scalar value at each grid point reflects the cumulative contribution of all critical points from all the field realizations. Here, we do not distinguish between critical points detected from different MC samples (or ensemble members). Each critical point deposits its contribution to all the grid points. The contribution of a critical point to a grid point is calculated as the inverse Euclidean distance between the critical point and the grid point. Hence, the contribution of a critical point to the nearby grid points will be higher and gradually fall off for far away grid points. Formally, for a grid point $P$, we compute its scalar value as follows: $v_P = \sum_{i=1}^{N}\frac{1}{D(P, C_i)}$, where $v_P$ denotes the value at grid point $P$ and $D(P, C_i)$ is the Euclidean distance between point $P$ and critical point $C_i$ with $N$ being the total number of critical points detected across all realizations.

Therefore, a grid point with a higher value will indicate a relatively large number of critical points near that grid point generated from all the realizations. Now, consider a spatial region in this newly constructed scalar field around a given mean critical point extracted from the mean field. Suppose we observe relatively high scalar values for the grid points in this region. In that case, it will indicate that when extracted from each realization, this critical point's location is scattered within that region, resulting in higher scalar values for most grid points. Conversely, if the detected critical points across all realizations are almost identical with minimal variability, they will only contribute heavily to a few nearby grid points. Thus, by examining the spatial spread of relatively high-valued regions around each mean critical point in this scalar field, we can determine whether the detected critical points across all the field realizations are concentrated (i.e., high certainty) or spatially scattered (i.e., low certainty).}


In Fig.~\ref{boussinesq_crit_vis} and Fig.~\ref{boussinesq1_crit_vis}, we show the results of the critical point analysis using two time steps (T=750 and T=1500) of the Heated Cylinder data set. We show the ground truth critical points with the vector field rendered using surface LIC visualization in Fig.~\ref{boussinesq_CP_GT} and Fig.~\ref{boussinesq1_CP_GT} for time step 750 and 1500, respectively. The saddles and centers are shown using red and blue dots. The results obtained from the MCDropout and Ensemble methods are shown in Fig.~\ref{boussinesq_CP_MCD} and Fig.~\ref{boussinesq_CP_ENS} for time step 750. We observe that both methods correctly predict all the critical points from the reconstructed vector fields. The green shade in the background highlights the spatial variability that we computed as a scalar field, as discussed above. We observe that the MCDropout method produces high prediction variability for a critical point indicated by the bright green spot around it. However, the Ensemble method makes highly confident predictions for all the critical points.

\rmark{Similarly, the visualization of critical points for time step $1500$ for the Heated Cylinder data is depicted in Fig.~\ref{boussinesq1_crit_vis}. Here, we observe that the MCDropout method fails to detect a critical point at the boundary of the data set (see red dotted region in Fig.~\ref{boussinesq1_CP_MCD}). This shows a potential limitation of computing the final predicted flow field as the average over multiple instances of predicted flow fields. Investigating individual predicted sampled fields and visualizing the median flow field can help recover such missing critical points. In contrast, the Ensemble method successfully retains all the critical points (see Fig.~\ref{boussinesq1_CP_ENS}) when compared against the ground truth (Fig.~\ref{boussinesq1_CP_GT}).} Both methods show similar spatial variability, as indicated by the bright green spots around several critical points. A quantitative comparison on the accuracy of the predicted critical points is provided later in Section~\ref{crit_point_accuracy_study}.

\section{Quantitative Evaluation and Parameter Study}
\begin{table}[thb]
\centering
\caption{Comparison of reconstruction quality (PSNR), error (RMSE), and model storage overhead between MCDropout and Ensemble methods.}
\label{model_size_psnr}
\resizebox{\linewidth}{!}{
\begin{tabular}{|c|ccc|ccc|}
\hline
\multirow{2}{*}{\textbf{Data Set}} &
  \multicolumn{3}{c|}{\textbf{\begin{tabular}[c]{@{}c@{}}Deep Ensemble\\  (30 Members)\end{tabular}}} &
  \multicolumn{3}{c|}{\textbf{\begin{tabular}[c]{@{}c@{}}MCDropout \\ (100 MC samples)\end{tabular}}} \\ \cline{2-7} 
 &
  \multicolumn{1}{c|}{\textbf{\begin{tabular}[c]{@{}c@{}}Model Size\\  (KB)\end{tabular}}} &
  \multicolumn{1}{c|}{\textbf{\begin{tabular}[c]{@{}c@{}}PSNR\\  (dB) $\uparrow$\end{tabular}}} &
  \textbf{RMSE $\downarrow$} &
  \multicolumn{1}{c|}{\textbf{\begin{tabular}[c]{@{}c@{}}Model Size\\  (KB)\end{tabular}}} &
  \multicolumn{1}{c|}{\textbf{\begin{tabular}[c]{@{}c@{}}PSNR \\ (dB) $\uparrow$\end{tabular}}} &
  \textbf{RMSE $\downarrow$} \\ \hline
Heated Cylinder (T=750)  & \multicolumn{1}{c|}{24364} & \multicolumn{1}{c|}{\textbf{57.799}} & \textbf{0.00209} & \multicolumn{1}{c|}{812}  & \multicolumn{1}{c|}{55.32}  & 0.00278 \\ \hline
Heated Cylinder (T=1500) & \multicolumn{1}{c|}{24364} & \multicolumn{1}{c|}{\textbf{58.417}} & \textbf{0.00251} & \multicolumn{1}{c|}{812}  & \multicolumn{1}{c|}{54.913} & 0.00375 \\ \hline
Fluid                    & \multicolumn{1}{c|}{24364} & \multicolumn{1}{c|}{\textbf{76.897}} & \textbf{0.00019} & \multicolumn{1}{c|}{812}  & \multicolumn{1}{c|}{73.953} & 0.00027 \\ \hline
Hurricane Isabel         & \multicolumn{1}{c|}{48612} & \multicolumn{1}{c|}{\textbf{53.602}} & \textbf{0.51843} & \multicolumn{1}{c|}{1620} & \multicolumn{1}{c|}{52.176} & 0.61607 \\ \hline
Tornado                  & \multicolumn{1}{c|}{48608} & \multicolumn{1}{c|}{\textbf{72.832}} & \textbf{0.00493} & \multicolumn{1}{c|}{1620} & \multicolumn{1}{c|}{69.5}   & 0.00721 \\ \hline
Turbine                  & \multicolumn{1}{c|}{48604} & \multicolumn{1}{c|}{\textbf{53.361}} & \textbf{0.00811} & \multicolumn{1}{c|}{1620} & \multicolumn{1}{c|}{51.545} & 0.01053 \\ \hline
Tangaroa                 & \multicolumn{1}{c|}{48572} & \multicolumn{1}{c|}{\textbf{68.665}} & \textbf{0.00152} & \multicolumn{1}{c|}{1620} & \multicolumn{1}{c|}{65.008} & 0.00232 \\ \hline
\end{tabular}
}
\end{table}

\begin{table}[thb]
\centering
\caption{PSNR (dB) values when the varying number of MC samples are used for computation in the MCDropout method. PSNR increases slowly as the number of MC samples increases. Here, we use $100$ MC samples to produce stable reconstruction and uncertainty estimates.}
\label{MC_samples_psnr}
\resizebox{0.8\linewidth}{!}{
\begin{tabular}{|c|ccc|}
\hline
\multirow{2}{*}{\textbf{Data Set}} & \multicolumn{3}{c|}{\textbf{PSNR ($\uparrow$) for diff. \#MCSamples}}                          \\ \cline{2-4} 
                                   & \multicolumn{1}{c|}{50}     & \multicolumn{1}{c|}{100}    & 150    \\ \hline
Heated Cylinder (T=750)  & \multicolumn{1}{c|}{55.182} & \multicolumn{1}{c|}{55.32}  & \textbf{55.37}  \\ \hline
Heated Cylinder (T=1500) & \multicolumn{1}{c|}{54.781} & \multicolumn{1}{c|}{54.913} & \textbf{54.955} \\ \hline
Fluid                              & \multicolumn{1}{c|}{73.29}  & \multicolumn{1}{c|}{73.953} & \textbf{74.204} \\ \hline
Hurricane Isabel         & \multicolumn{1}{c|}{52.175} & \multicolumn{1}{c|}{\textbf{52.176}} & \textbf{52.176} \\ \hline
Tornado                            & \multicolumn{1}{c|}{69.497} & \multicolumn{1}{c|}{69.5}   & \textbf{69.501} \\ \hline
Turbine                            & \multicolumn{1}{c|}{51.519} & \multicolumn{1}{c|}{51.545} & \textbf{51.553} \\ \hline
Tangaroa                           & \multicolumn{1}{c|}{65.003} & \multicolumn{1}{c|}{65.008} & \textbf{65.01}  \\ \hline
\end{tabular}
}
\end{table}

\begin{table}[thb]
\centering
\caption{PSNR (dB) values when varying numbers of ensemble members are used for the Ensemble method. We observe that PSNR increase is almost insignificant beyond $15$ members upto $30$.}
\label{ensemble_samples_psnr}
\resizebox{0.85\linewidth}{!}{
\begin{tabular}{|c|cccc|}
\hline
\multirow{2}{*}{\textbf{Data Set}} & \multicolumn{4}{c|}{\textbf{PSNR ($\uparrow$) for diff. \#Ensemble members}}                                                 \\ \cline{2-5} 
                                   & \multicolumn{1}{c|}{15}     & \multicolumn{1}{c|}{20}     & \multicolumn{1}{c|}{25}     & 30     \\ \hline
Heated Cylinder (T=750)  & \multicolumn{1}{c|}{57.203} & \multicolumn{1}{c|}{57.347} & \multicolumn{1}{c|}{57.595} & \textbf{57.799} \\ \hline
Heated Cylinder (T=1500) & \multicolumn{1}{c|}{58.193} & \multicolumn{1}{c|}{58.312} & \multicolumn{1}{c|}{58.305} & \textbf{58.417} \\ \hline
Fluid                              & \multicolumn{1}{c|}{76.795} & \multicolumn{1}{c|}{76.85}  & \multicolumn{1}{c|}{76.734} & \textbf{76.897} \\ \hline
Hurricane Isabel         & \multicolumn{1}{c|}{\textbf{53.626}} & \multicolumn{1}{c|}{53.589} & \multicolumn{1}{c|}{53.57}  & 53.602 \\ \hline
Tornado                            & \multicolumn{1}{c|}{72.735} & \multicolumn{1}{c|}{72.81}  & \multicolumn{1}{c|}{\textbf{72.899}} & 72.832 \\ \hline
Turbine                            & \multicolumn{1}{c|}{\textbf{53.366}} & \multicolumn{1}{c|}{53.433} & \multicolumn{1}{c|}{53.301} & 53.361 \\ \hline
Tangaroa                           & \multicolumn{1}{c|}{68.443} & \multicolumn{1}{c|}{68.536} & \multicolumn{1}{c|}{68.556} & \textbf{68.665} \\ \hline
\end{tabular}
}
\end{table}

\begin{table}[tbh]
\caption{Streamline error analysis for MCDropout and Ensemble method using Chamfer and Hausdorff distance metrics. We generate streamlines from $100$ uniformly randomly selected seed points for each data set and report their average Chamfer and Hausdorff distances when compared against ground truth streamlines. We observe that, on average, the Ensemble method produces more accurate streamlines.}
\label{streamline_error}
\resizebox{\linewidth}{!}{
\begin{tabular}{|c|cc|cc|}
\hline
\multirow{2}{*}{\textbf{Data Set}}                                  & \multicolumn{2}{c|}{\textbf{Deep Ensemble}} & \multicolumn{2}{c|}{\textbf{MCDropout}} \\ \cline{2-5} 
 &
  \multicolumn{1}{c|}{\textbf{\begin{tabular}[c]{@{}c@{}}Avg. Chamfer \\ Distance $\downarrow$ \end{tabular}}} &
  \textbf{\begin{tabular}[c]{@{}c@{}}Avg. Hausdorff \\ Distance $\downarrow$\end{tabular}} &
  \multicolumn{1}{c|}{\textbf{\begin{tabular}[c]{@{}c@{}}Avg. Chamfer \\ Distance $\downarrow$\end{tabular}}} &
  \textbf{\begin{tabular}[c]{@{}c@{}}Avg. Hausdorff \\ Distance $\downarrow$\end{tabular}} \\ \hline
\begin{tabular}[c]{@{}c@{}}Heated Cylinder\\  (T=750)\end{tabular}  & \multicolumn{1}{c|}{\textbf{0.00104}}    & \textbf{0.00095}   & \multicolumn{1}{c|}{0.00776}  & 0.004   \\ \hline
\begin{tabular}[c]{@{}c@{}}Heated Cylinder \\ (T=1500)\end{tabular} & \multicolumn{1}{c|}{\textbf{0.00062} }   & \textbf{0.00062}   & \multicolumn{1}{c|}{0.0037}   & 0.00189 \\ \hline
Fluid                                                               & \multicolumn{1}{c|}{\textbf{0.00453}}    & \textbf{0.00469}   & \multicolumn{1}{c|}{0.01261}  & 0.01314 \\ \hline
Hurricane Isabel                                                    & \multicolumn{1}{c|}{\textbf{1.816}}      & \textbf{1.95277}   & \multicolumn{1}{c|}{2.54561}  & 2.86682 \\ \hline
Tornado                                                             & \multicolumn{1}{c|}{\textbf{0.00261}}    & \textbf{0.00296}   & \multicolumn{1}{c|}{0.0096}   & 0.00938 \\ \hline
Turbine                                                             & \multicolumn{1}{c|}{\textbf{0.00167}}    & \textbf{0.00127}   & \multicolumn{1}{c|}{0.00215}  & 0.00342 \\ \hline
Tangaroa                                                            & \multicolumn{1}{c|}{\textbf{0.00024}}    & \textbf{0.00036}   & \multicolumn{1}{c|}{0.0004}   & 0.00063 \\ \hline
\end{tabular}
}
\end{table}

\begin{table}[thb]
\centering
\caption{Average RMSE in predicted critical point locations are reported for MCDropout and Ensemble methods. We observe that the Ensemble method produces more accurate critical points.}
\label{crit_error}
\resizebox{0.9\linewidth}{!}{
\begin{tabular}{|c|c|c|}
\hline
\multirow{2}{*}{\textbf{Data Set}} &
  \multirow{2}{*}{\textbf{\begin{tabular}[c]{@{}c@{}}Avg. RMSE ($\downarrow$) for \\ Deep Ensemble\end{tabular}}} &
  \multirow{2}{*}{\textbf{\begin{tabular}[c]{@{}c@{}}Avg. RMSE ($\downarrow$) for \\ MCDropout\end{tabular}}} \\
                         &        &        \\ \hline
Heated Cylinder (T=750)  & \textbf{0.01866} & 0.02896 \\ \hline
Heated Cylinder (T=1500) & \textbf{0.03234} & 0.03481 \\ \hline
Fluid                    & \textbf{0.22243} & 0.24127 \\ \hline
\end{tabular}
}
\end{table}

\begin{table}[thb]
\centering
\caption{We report how the reconstruction quality (PSNR) changes when different numbers of Dropout layers are used for the MCDropout method. Three configurations are studied: Dropout used (1) only at the last residual block; (2) at the last half of the residual blocks; (3) at all the residual blocks. We observe that PSNR decreases with an increasing number of Dropout layers as it acts as a strong regularizer. Hence, Dropout only at the last residual block is used.}
\label{drop_layers_study}
\resizebox{0.9\linewidth}{!}{
\begin{tabular}{|c|c|c|c|}
\hline
\multirow{2}{*}{\textbf{Data Set}} & \multicolumn{3}{c|}{\textbf{PSNR ($\uparrow$) with different Dropout Layer Location}}  \\ \cline{2-4} 
 & \multicolumn{1}{c|}{\textbf{Last Res. block}} & \multicolumn{1}{c|}{\textbf{Last half Res. blocks}} & \textbf{All Res. blocks} \\ \hline
Hurricane Isabel          & \multicolumn{1}{c|}{\textbf{52.176}} & \multicolumn{1}{c|}{51.686} & 49.595 \\ \hline
Turbine                   & \multicolumn{1}{c|}{\textbf{51.545}} & \multicolumn{1}{c|}{50.16}  & 39.274 \\ \hline
\end{tabular}
}
\end{table}

\begin{table}[thb]
\centering
\caption{PSNR (dB) of 100 MC samples with different Dropout probabilities during inference.}
\label{varied_dropout}
\resizebox{0.9\linewidth}{!}{
\begin{tabular}{|c|cccccc|}
\hline
\multirow{2}{*}{\textbf{Data Set}} &
  \multicolumn{6}{c|}{\textbf{\begin{tabular}[c]{@{}c@{}}PSNR ($\uparrow$) for different test-time Dropout probabilities\\  (100 MC Samples)\end{tabular}}} \\ \cline{2-7} 
 &
  \multicolumn{1}{c|}{\textbf{0.05}} &
  \multicolumn{1}{c|}{\textbf{0.1}} &
  \multicolumn{1}{c|}{\textbf{0.2}} &
  \multicolumn{1}{c|}{\textbf{0.3}} &
  \multicolumn{1}{c|}{\textbf{0.4}} &
  \textbf{0.5} \\ \hline
Heated Cylinder (T=750) &
  \multicolumn{1}{c|}{\textbf{55.391}} &
  \multicolumn{1}{c|}{55.32} &
  \multicolumn{1}{c|}{55.171} &
  \multicolumn{1}{c|}{54.935} &
  \multicolumn{1}{c|}{54.678} &
  54.369 \\ \hline
Heated Cylinder (T=1500) &
  \multicolumn{1}{c|}{\textbf{54.983}} &
  \multicolumn{1}{c|}{54.913} &
  \multicolumn{1}{c|}{54.758} &
  \multicolumn{1}{c|}{54.564} &
  \multicolumn{1}{c|}{54.306} &
  53.984 \\ \hline
Fluid &
  \multicolumn{1}{c|}{\textbf{74.352}} &
  \multicolumn{1}{c|}{73.953} &
  \multicolumn{1}{c|}{73.139} &
  \multicolumn{1}{c|}{72.282} &
  \multicolumn{1}{c|}{71.328} &
  70.283 \\ \hline
Hurricane Isabel &
  \multicolumn{1}{c|}{\textbf{52.176}} &
  \multicolumn{1}{c|}{\textbf{52.176}} &
  \multicolumn{1}{c|}{52.175} &
  \multicolumn{1}{c|}{52.173} &
  \multicolumn{1}{c|}{52.172} &
  52.169 \\ \hline
Tornado &
  \multicolumn{1}{c|}{\textbf{69.501}} &
  \multicolumn{1}{c|}{69.5} &
  \multicolumn{1}{c|}{69.497} &
  \multicolumn{1}{c|}{69.493} &
  \multicolumn{1}{c|}{69.487} &
  69.479 \\ \hline
Turbine &
  \multicolumn{1}{c|}{\textbf{51.556}} &
  \multicolumn{1}{c|}{51.545} &
  \multicolumn{1}{c|}{51.514} &
  \multicolumn{1}{c|}{51.472} &
  \multicolumn{1}{c|}{51.424} &
  51.347 \\ \hline
Tangaroa &
  \multicolumn{1}{c|}{\textbf{65.011}} &
  \multicolumn{1}{c|}{65.008} &
  \multicolumn{1}{c|}{65.002} &
  \multicolumn{1}{c|}{64.994} &
  \multicolumn{1}{c|}{64.983} &
  64.968 \\ \hline
\end{tabular}
}
\end{table}

\begin{table}[thb]
\centering
\caption{Training time (Hours) for MCDropout and Ensemble (30 ensemble members) for 500 epochs with a batch size of 2048 and inference time (seconds) for generating final vector field using MCDropout (100 MC samples) and Ensemble (30 ensemble members) methods.}
\label{train_inference_time}
\resizebox{\linewidth}{!}{
\begin{tabular}{|c|cc|cc|}
\hline
\multirow{2}{*}{\textbf{Data set}} &
  \multicolumn{2}{c|}{\textbf{\begin{tabular}[c]{@{}c@{}}Training time\\ (Hours)\end{tabular}}} &
  \multicolumn{2}{c|}{\textbf{\begin{tabular}[c]{@{}c@{}}Inference time\\ (Seconds)\end{tabular}}} \\ \cline{2-5} 
 &
  \multicolumn{1}{c|}{\textbf{MCDropout}} &
  \textbf{\begin{tabular}[c]{@{}c@{}}Ensemble\\ (30 members)\end{tabular}} &
  \multicolumn{1}{c|}{\textbf{\begin{tabular}[c]{@{}c@{}}MCDropout\\ (100 MC samples)\end{tabular}}} &
  \textbf{\begin{tabular}[c]{@{}c@{}}Ensemble\\ (30 members)\end{tabular}} \\ \hline
Heated Cylinder (T=750)  & \multicolumn{1}{c|}{0.27}  & 7.12   & \multicolumn{1}{c|}{2.84}   & 1.83   \\ \hline
Heated Cylinder (T=1500) & \multicolumn{1}{c|}{0.27}  & 7.12   & \multicolumn{1}{c|}{2.85}   & 1.83   \\ \hline
Fluid                    & \multicolumn{1}{c|}{0.64}  & 18.06  & \multicolumn{1}{c|}{12.33}  & 3.99   \\ \hline
Hurricane Isabel         & \multicolumn{1}{c|}{7.03}  & 218.06 & \multicolumn{1}{c|}{182.14} & 54.48  \\ \hline
Tornado                  & \multicolumn{1}{c|}{4.73}  & 151.8  & \multicolumn{1}{c|}{122.13} & 37.99  \\ \hline
Turbine                  & \multicolumn{1}{c|}{1.6}   & 59.46  & \multicolumn{1}{c|}{32.39}  & 11.51  \\ \hline
Tangaroa                 & \multicolumn{1}{c|}{14.86} & 462.68 & \multicolumn{1}{c|}{389.89} & 114.13 \\ \hline
\end{tabular}
}
\end{table}
\textbf{Reconstruction Quality and Prediction Error.}
Table~\ref{model_size_psnr} shows a comparison between the Ensemble and MCDropout methods concerning model size, PSNR, and RMSE. The Ensemble method outperforms MCDropout in terms of both PSNR and RMSE. However, achieving this superior reconstruction quality with the Ensemble method requires nearly $30$ times more storage space and training time compared to MCDropout, as it involves training $30$ ensemble members. It is worth noting that the PSNR of the Ensemble method is computed using predictions from $30$ ensemble members, while for MCDropout, it is calculated based on $100$ Monte Carlo (MC) samples. The rationale behind the chosen number of samples is elaborated on subsequently.

\textbf{Impact of Different Number of MC Samples on PSNR Value for MCDropout Method.}
Table~\ref{MC_samples_psnr} demonstrates the PSNR value of the vector field obtained by averaging different numbers of MC samples. There is a slight gain of PSNR as we increase the number of MC samples up to 150. After that, the gains in PSNR are not as significant to justify considering more number of samples. Hence, to find a good trade-off between computation time and prediction quality, we use $100$ MC samples for all experiments involving the MCDropout method.

\textbf{Impact of Different Number of Ensemble Members on PSNR Value for Ensemble Method.}
The effect of different numbers of ensemble members on PNSR is depicted in Table~\ref{ensemble_samples_psnr}. The number of ensemble members required to produce robust prediction is much lower than in the MCDropout method. It can be observed that PSNR gains are insignificant when increasing the number of ensemble members beyond $15$ up to $30$. Hence, at around $15$ ensemble members, the PSNR becomes primarily saturated. However, for consistency, we use $30$ ensemble members for the experiments in the work.

\textbf{Streamline Error Analysis for MCDropout and Ensemble Methods.}
We examine the precision of streamlines generated by both MCDropout and Ensemble methods, utilizing averaged mean streamlines for comparison. To conduct a thorough quantitative assessment, we generate mean streamlines from 100 randomly uniformly distributed seed points for each method. Subsequently, we gauge the streamline error by measuring Chamfer distance~\cite{champher} and Hausdorff distance between the predicted averaged streamlines and the ground truth counterparts. It's noteworthy that for the MCDropout method, we aggregate results over 100 Monte Carlo (MC) samples, while for the Ensemble method, we utilize streamlines predicted by 30 ensemble members. The average Chamfer distance and Hausdorff distance values computed over the 100 streamlines for each data set are presented in Table~\ref{streamline_error}. It is evident that the Ensemble method consistently yields more precise streamlines compared to the MCDropout method across all data sets, as evidenced by the lower Chamfer and Hausdorff distance values.

\textbf{Accuracy Analysis of The Predicted Critical Points for MCDropout and Ensemble Methods.} 
\label{crit_point_accuracy_study}
To quantitatively assess the precision of the identified critical point locations, we calculate the average root mean squared error (RMSE) between the predicted and the corresponding ground truth critical points for both MCDropout and Ensemble methods. The outcomes are documented in Table~\ref{crit_error} for the Heated Cylinder and Fluid data sets. It is apparent that, on average, the Ensemble method outperforms the MCDropout method in accurately determining critical points using the reconstructed vector field.

\textbf{Impact of Varying Number of Dropout Layers on MCDropout Model Performance.}
\begin{figure}[thb]
\centering
\begin{subfigure}[t]{0.29\linewidth}
    \centering
    \includegraphics[width=\linewidth]{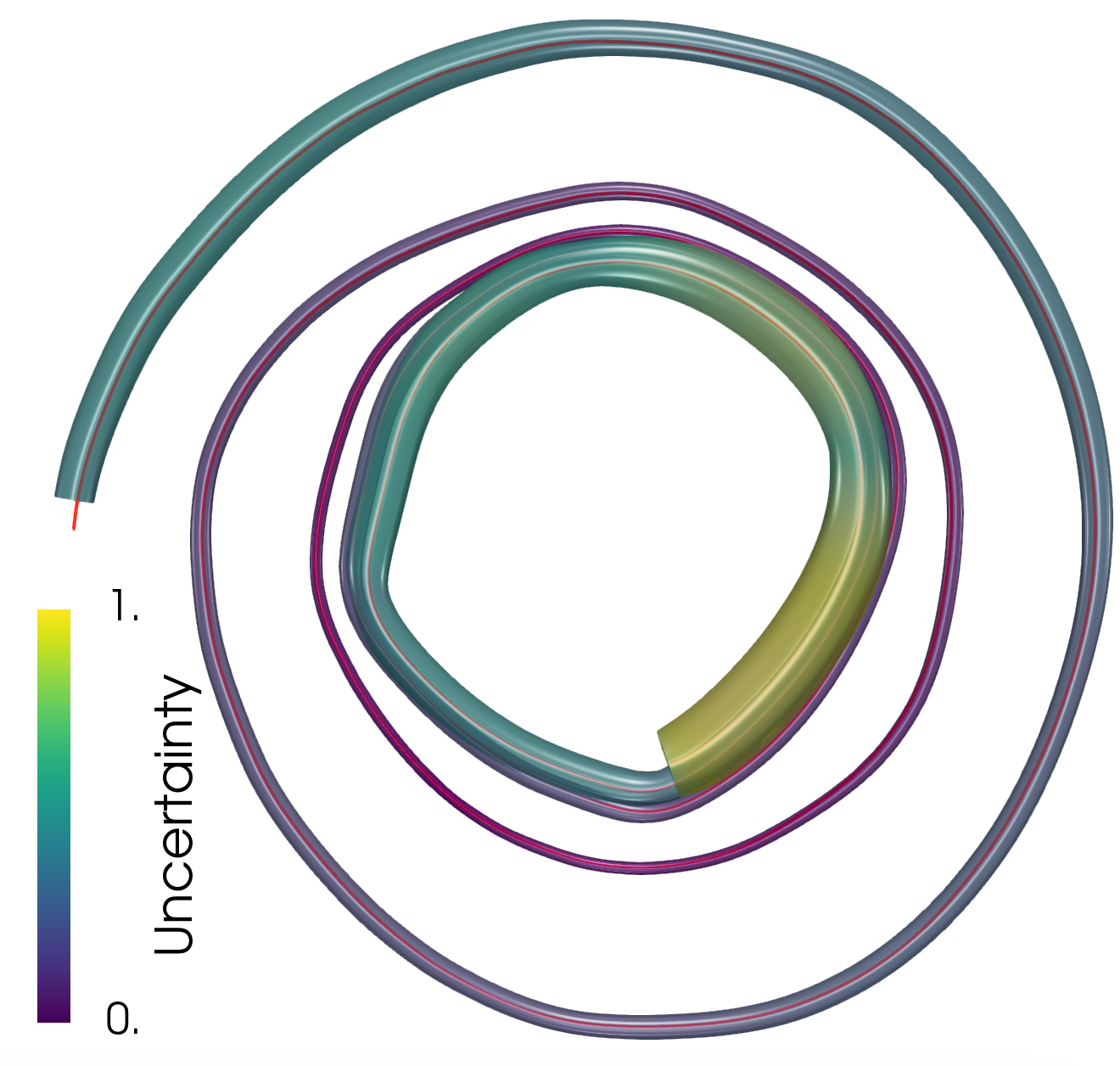}
    \caption{Dropout applied at the last RS block.}
    \label{last_drop_isabel}
\end{subfigure}
~
\begin{subfigure}[t]{0.29\linewidth}
    \centering
    \includegraphics[width=\linewidth]{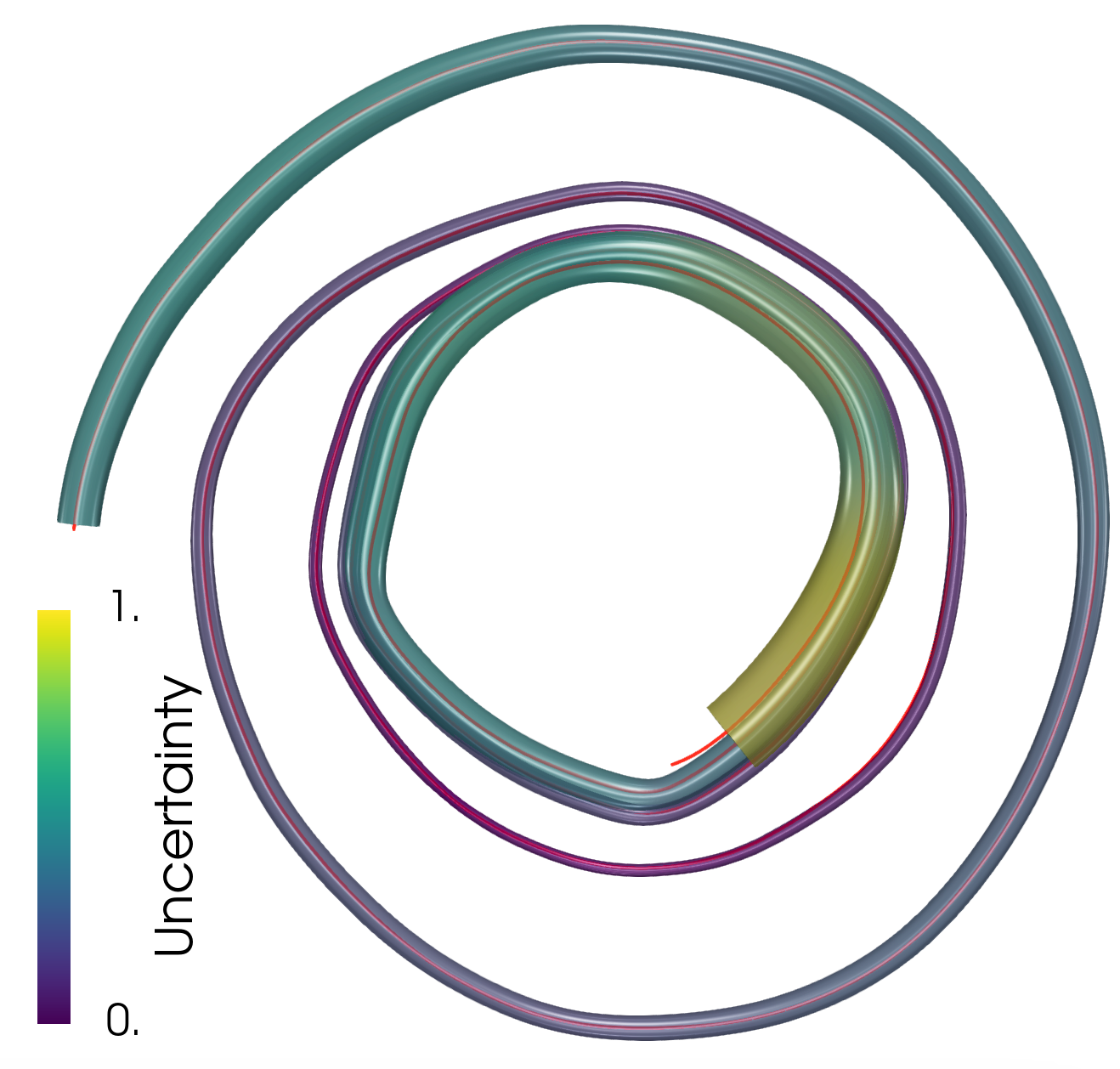}
    \caption{Dropout applied at the last half of the RS blocks.}
    \label{half_drop_isabel}
\end{subfigure}
~
\begin{subfigure}[t]{0.29\linewidth}
    \centering
    \includegraphics[width=\linewidth]{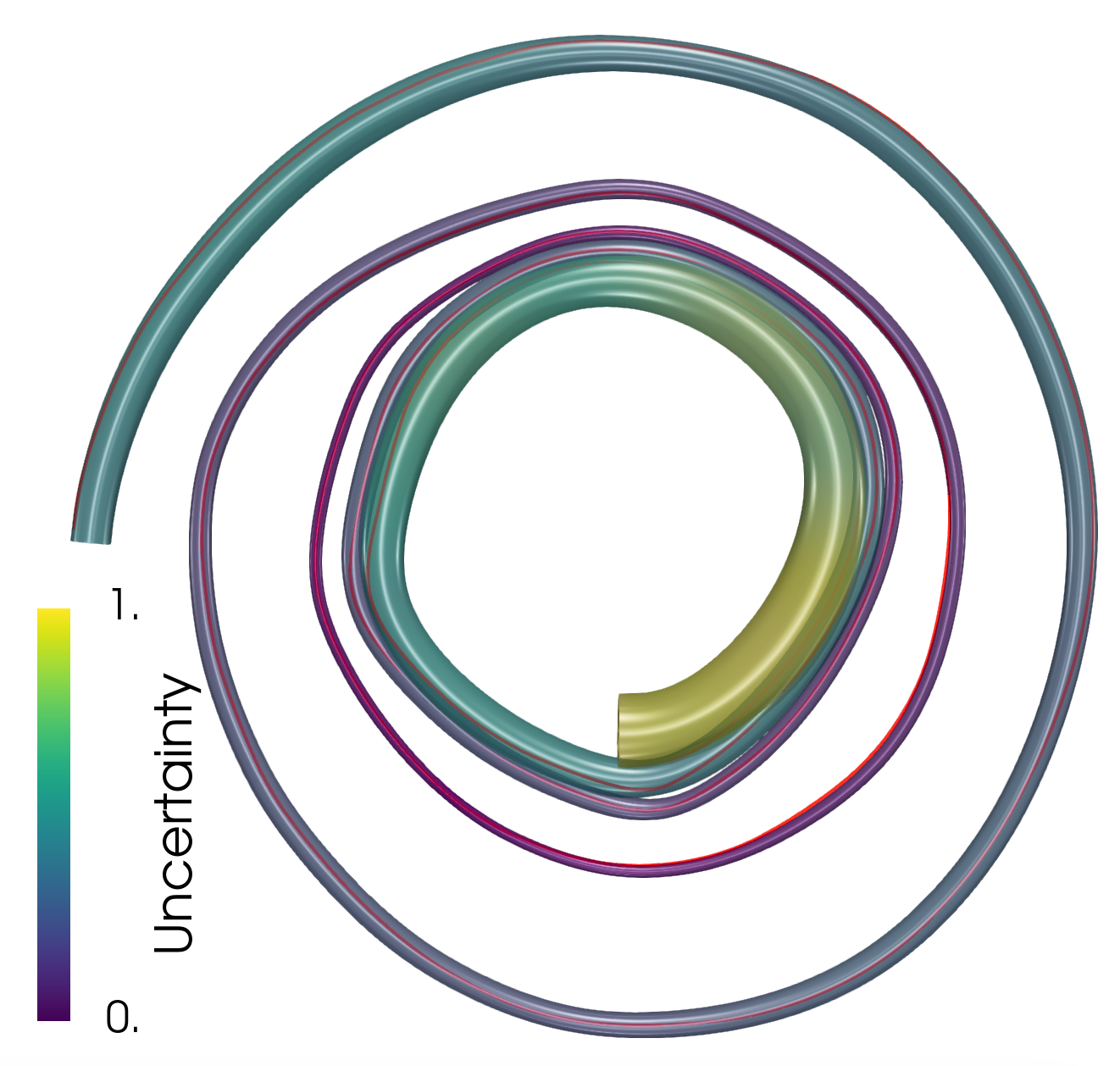}
    \caption{Dropout applied at all the RS blocks.}
    \label{all_drop_isabel}
\end{subfigure}
\caption{Comparing uncertainty patterns when dropout is added at the last (Fig.~\ref{last_drop_isabel}), last half (Fig.~\ref{half_drop_isabel}), and all residual blocks (Fig.~\ref{all_drop_isabel}). The uncertainty pattern remains identical for these configurations. Ground truth streamline is overlaid in red as reference.}
\label{diff_dropout_layer_vis}
\end{figure}
In~\cite{gagh16}, Gal et al. demonstrate that incorporating a dropout layer after each hidden layer in a DNN renders the model theoretically equivalent to conducting inference in a fully Bayesian neural network. However, Kendall et al.~\cite{segnet} argue that adding such dropout layers after every layer could hamper the model's learnability, as numerous dropout layers act as potent regularizers. This aligns with our observations of decreased reconstruction quality when applying dropout at each residual block in the MCDropout model. Kendall et al.~\cite{segnet} further propose to use fewer dropout layers, approximating a partial Bayesian network, to achieve robust uncertainty estimation. They suggest that a dropout layer after the final layer is sufficient to produce high prediction accuracy and robust uncertainty estimates. Consequently, we use dropout layer at the last residual block to ensure high-quality predictions and robust uncertainty estimates. In Fig.~\ref{diff_dropout_layer_vis}, we show results of uncertainty estimates for a streamline selected from Isabel data set. We present uncertainty visualization using model with dropout added at the (1) last residual block, (2) last half of the blocks, and (3) all residual blocks. We observe that all of these configurations produce similar uncertainty estimates. To assess the detailed impact of varying the number of dropout layers, we conduct experiments on the Isabel and Turbine data sets. The results, detailed in Table~\ref{drop_layers_study}, indicate a decrease in PSNR with increasing number of dropout layers.

\textbf{Impact of Varying Dropout Probabilities on MCDropout Model Performance.}
Table~\ref{varied_dropout} depicts a study examining the quality of vector field reconstruction (measured by PSNR) averaged over $100$ MC samples across various test-time dropout probabilities. The MCDropout model is trained with a dropout rate of $0.05$. Notably, it is observed that PSNR remains  largely consistent up to a dropout probability of $0.2$, consistent with findings by Gal et al.~\cite{gagh16}. They posit that once a model has reached convergence, minimal variations in dropout probability do not significantly impact the predicted quality. However, as dropout is increased to as high as $0.5$, the  prediction quality gradually diminishes, as evidenced by lower PSNR values.

\begin{table}[thb]
\centering
\caption{\rmark{Evaluating reconstruction quality under different hyperparameter combinations for MCDropout method.}}
\label{mcd_hyp_study}
\resizebox{\linewidth}{!}{
\begin{tabular}{|c|ccc|ccc|ccc|}
\hline
\multirow{3}{*}{\textbf{\begin{tabular}[c]{@{}c@{}}Learning\\ Rate\end{tabular}}} &
  \multicolumn{3}{c|}{Turbine PSNR (100 MC samples)} &
  \multicolumn{3}{c|}{Tornado PSNR (100 MC samples)} &
  \multicolumn{3}{c|}{Isabel PSNR (100 MC samples)} \\ \cline{2-10} 
 &
  \multicolumn{3}{c|}{\textbf{Batch Size}} &
  \multicolumn{3}{c|}{\textbf{Batch Size}} &
  \multicolumn{3}{c|}{\textbf{Batch Size}} \\ \cline{2-10} 
 &
  \multicolumn{1}{c|}{\textbf{1024}} &
  \multicolumn{1}{c|}{\textbf{2048}} &
  \textbf{4096} &
  \multicolumn{1}{c|}{\textbf{1024}} &
  \multicolumn{1}{c|}{\textbf{2048}} &
  \textbf{4096} &
  \multicolumn{1}{c|}{\textbf{1024}} &
  \multicolumn{1}{c|}{\textbf{2048}} &
  \textbf{4096} \\ \hline
\textbf{0.0001} &
  \multicolumn{1}{c|}{20.791} &
  \multicolumn{1}{c|}{51.854} &
  50.9 &
  \multicolumn{1}{c|}{21.094} &
  \multicolumn{1}{c|}{21.103} &
  67.757 &
  \multicolumn{1}{c|}{27.1} &
  \multicolumn{1}{c|}{27.364} &
  27.604 \\ \hline
\textbf{0.0005} &
  \multicolumn{1}{c|}{19.498} &
  \multicolumn{1}{c|}{19.957} &
  20.074 &
  \multicolumn{1}{c|}{19.076} &
  \multicolumn{1}{c|}{19.716} &
  19.414 &
  \multicolumn{1}{c|}{25.598} &
  \multicolumn{1}{c|}{25.726} &
  25.881 \\ \hline
\textbf{1e−5} &
  \multicolumn{1}{c|}{46.338} &
  \multicolumn{1}{c|}{41.776} &
  43.662 &
  \multicolumn{1}{c|}{67.893} &
  \multicolumn{1}{c|}{67.135} &
  65.879 &
  \multicolumn{1}{c|}{52.286} &
  \multicolumn{1}{c|}{51.973} &
  51.446 \\ \hline
\textbf{5e−5} &
  \multicolumn{1}{c|}{51.934} &
  \multicolumn{1}{c|}{51.545} &
  50.43 &
  \multicolumn{1}{c|}{71.036} &
  \multicolumn{1}{c|}{69.5} &
  69.703 &
  \multicolumn{1}{c|}{30.888} &
  \multicolumn{1}{c|}{52.176} &
  52.195 \\ \hline
\end{tabular}
}
\end{table}
\rmark{
\textbf{Consistent Hyperparameter Selection.} In this work, we seek to identify a hyperparameter combination that ensures stable training of implicit neural models using both MCDropout and Ensemble methods across multiple flow data sets. Since both methods share the same base network architecture, we perform the hyperparameter selection experiment using the MCDropout method as it is a single-model based approach and requires less number of model training compared to the expensive ensemble method. After determining a stable hyperparameter combination, we apply these same parameters to train the ensemble models, aiming for stable, consistent, and comparable results. Table~\ref{mcd_hyp_study} presents the reconstruction quality, measured by PSNR, for 12 different hyperparameter combinations with varying learning rates and batch sizes on the Turbine, Isabel, and Tornado data sets. The combination of a learning rate of $5e-5$ and a batch size of $2048$ consistently produces high-quality, stable PSNR values across all three data sets. Consequently, we adopt this hyperparameter combination for subsequent training of both MCDropout and Ensemble models. Our results show that this combination yields consistent and comparable uncertainty-aware neural representations across all data sets.}

\begin{table}[thb]
\centering
\caption{\rmark{Reconstruction quality with varying network depth.}}
\label{arch_study}
\resizebox{0.9\linewidth}{!}{
\begin{tabular}{|c|cccc|cccc|}
\hline
 &
  \multicolumn{4}{c|}{\textbf{\begin{tabular}[c]{@{}c@{}}MCDropout\\ (PSNR ($\uparrow$) for 100 MC samples)\end{tabular}}} &
  \multicolumn{4}{c|}{\textbf{\begin{tabular}[c]{@{}c@{}}Ensemble\\ (PSNR ($\uparrow$) for 15 ensemble members)\end{tabular}}} \\ \hline
\begin{tabular}[c]{@{}c@{}}\#Res\\ Blocks\end{tabular} &
  \multicolumn{1}{c|}{10} &
  \multicolumn{1}{c|}{12} &
  \multicolumn{1}{c|}{14} &
  16 &
  \multicolumn{1}{c|}{10} &
  \multicolumn{1}{c|}{12} &
  \multicolumn{1}{c|}{14} &
  16 \\ \hline
Tornado &
  \multicolumn{1}{c|}{67.21} &
  \multicolumn{1}{c|}{68.79} &
  \multicolumn{1}{c|}{69.50} &
  \textbf{72.21} &
  \multicolumn{1}{c|}{71.58} &
  \multicolumn{1}{c|}{\textbf{73.05}} &
  \multicolumn{1}{c|}{72.70} &
  37.08 \\ \hline
Turbine &
  \multicolumn{1}{c|}{50.89} &
  \multicolumn{1}{c|}{50.68} &
  \multicolumn{1}{c|}{\textbf{51.55}} &
  49.97 &
  \multicolumn{1}{c|}{53.84} &
  \multicolumn{1}{c|}{54.96} &
  \multicolumn{1}{c|}{53.80} &
  \textbf{55.43} \\ \hline
Isabel &
  \multicolumn{1}{c|}{51.66} &
  \multicolumn{1}{c|}{51.74} &
  \multicolumn{1}{c|}{\textbf{52.18}} &
  28.39 &
  \multicolumn{1}{c|}{52.87} &
  \multicolumn{1}{c|}{52.68} &
  \multicolumn{1}{c|}{\textbf{53.69}} &
  35.24 \\ \hline
\end{tabular}
}
\end{table}

\begin{figure}[tb]
\centering
    \includegraphics[width=\linewidth]{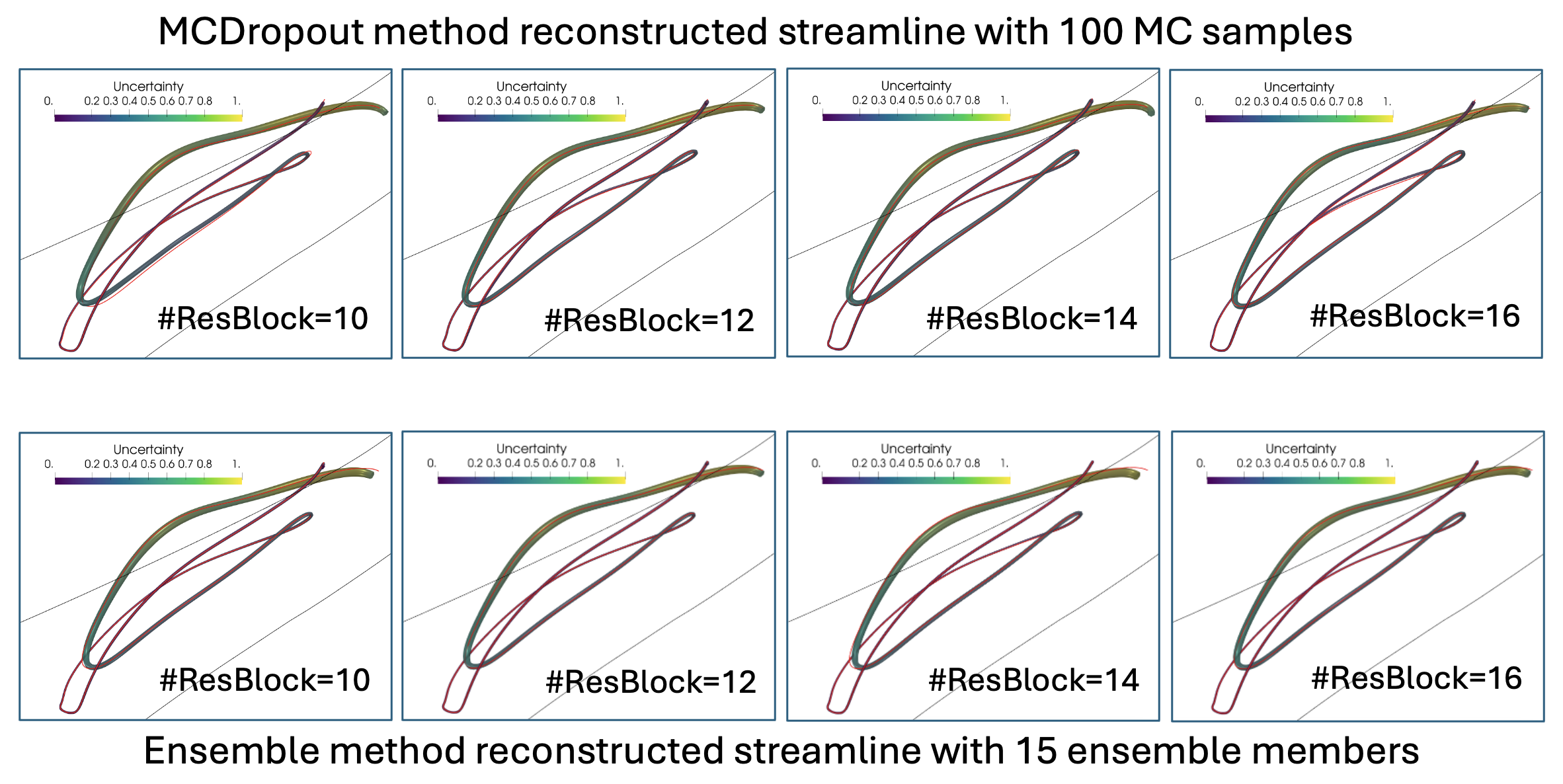}
\caption{\rmark{Visualization of a representative streamline from Turbine data set under varying numbers of residual blocks for MCDropout and Ensemble methods. It is observed that the estimated prediction uncertainty is robust and is not influenced when the number of residual blocks is varied.}}
\label{uncert_streamline_arch_study}
\end{figure}

\begin{figure}[thb]
\centering
\begin{subfigure}[t]{0.46\linewidth}
    \centering
    \includegraphics[width=\linewidth]{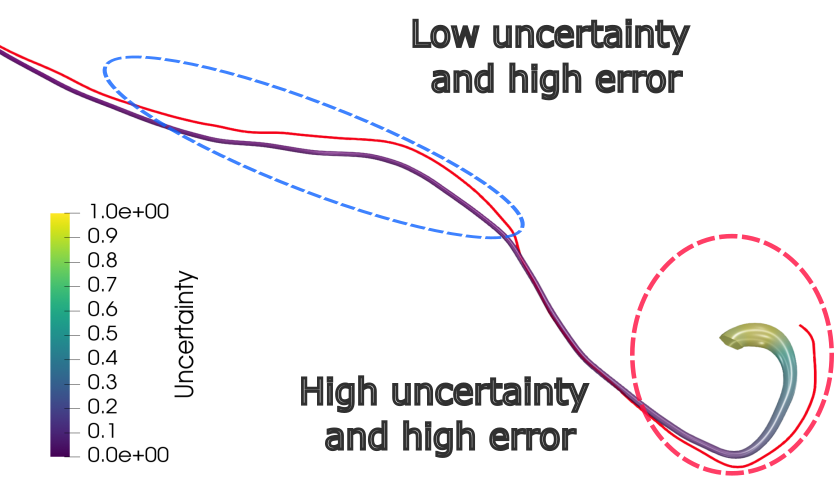}
    \caption{Ground truth (red) and predicted streamline from MCDropout method.}
    \label{uncert_error_isabel_MCD}
\end{subfigure}
~
\begin{subfigure}[t]{0.46\linewidth}
    \centering
    \includegraphics[width=\linewidth]{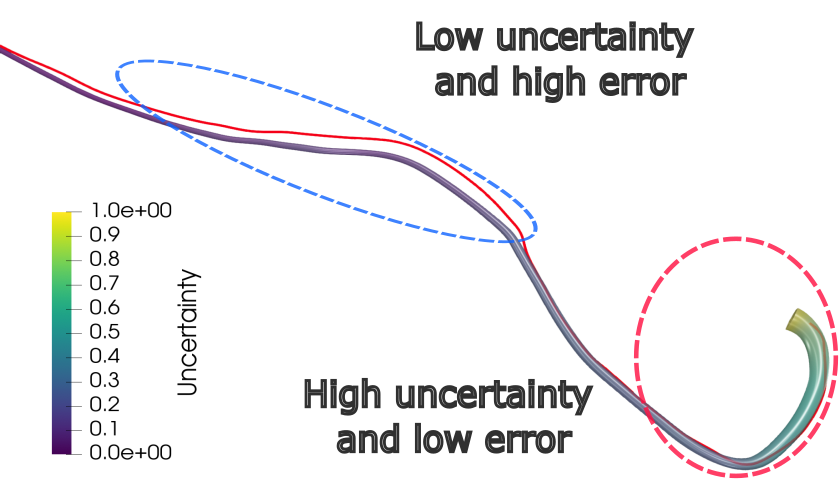}
    \caption{Ground truth (red) and predicted streamline from Ensemble method.}
    \label{uncert_error_isabel_ENS}
\end{subfigure}
\caption{The importance of error and uncertainty information for vector field feature analysis using streamlines generated by a DNN model.}
\label{uncert_error}
\end{figure}

\rmark{
\textbf{Impact of Varying Model Architecture on Accuracy and Uncertainty Estimates.}
We vary the network depth by adjusting the number of residual blocks, while keeping the number of neurons per layer fixed. The learning rate and batch size are maintained at $5e−5$ and $2048$, respectively, to ensure comparability of the results. From Table~\ref{arch_study}, we observe that PSNR values exhibit minimal variation across different numbers of residual blocks. However, an exception occurs with $16$ residual blocks: the PSNR drops for the Tornado data set using the Ensemble method and for the Isabel data set using the MCDropout method. This drop is due to unstable training, which would require further hyperparameter tuning. Additionally, we examine the impact of varying the number of residual blocks on the estimated prediction uncertainty. We find that prediction uncertainty remain unaffected by changes in the model architecture. In Fig.~\ref{uncert_streamline_arch_study}, we present a representative streamline from the Turbine data under different numbers of residual blocks for both  methods. It is evident that the prediction uncertainty is robust and remains consistent regardless of the number of residual blocks. However, for a much shallower network, we anticipate that the quality of reconstruction and uncertainty estimates will gradually degrade as the network will have less learning ability.}

\textbf{Training and Inference Time for MCDropout and Ensemble Methods.}
Table~\ref{train_inference_time} shows training and inference times for the two methods. The Ensemble approach has faster inference time since it uses predictions from only $30$ members, whereas MCDropout generates $100$ MC samples, resulting in longer inference times. However, Ensemble training is significantly more time and resource consuming because it requires training multiple models ($30$ in this study), unlike MCDropout, which needs to train a single model. This significant difference in training time often makes MCDropout preferable for timely results with robust uncertainty estimates.

\section{Discussion}
\label{discussion_section}

\textbf{Uncertainty-agnostic vs. Uncertainty-aware DNNs.} We advocate the use of DNNs equipped to quantify uncertainty when analyzing vector fields. We find  that uncertainty-informed neural networks offer valuable insights into the reliability of their predictions. By effectively conveying such uncertainty to experts, they can make well-informed decisions about the data features. This integration of uncertainty is crucial in fostering trust in the use of DNN-predicted results for scientific research. Particularly when dealing with large vector fields, scientists often have to rely on the model's outputs due to the impracticality of handling full-resolution data. In such cases, where ground truth data may be unavailable, estimating errors becomes challenging. However, uncertainty-aware DNNs can still provide reliable uncertainty estimates, offering experts greater confidence in interpreting predicted results. Next, we discuss the domain expert feedback to highlight the implications of our proposed techniques.

\rmark{
\textbf{Domain Expert Feedback.}
We interviewed a computational fluid dynamics scientist to collect expert feedback. The expert immediately liked the idea of using uncertainty-aware neural models and agreed that visualizing prediction uncertainty on predicted flow features is critical and can provide meaningful insights about the prediction's quality and help build trust in using deep learning techniques for scientific research. The expert found our uncertainty-informed streamline visualization intuitive and easy to comprehend. The expert was also intrigued to see that these neural models can be both under-confident and over-confident, as highlighted in Fig.~\ref{uncert_error}, and agreed that access to such results is critical for verifying and validating hypotheses from flow data. Next, the expert suggested that developing such uncertainty-aware models for time-varying flow data would be very useful. Finally, the expert observed that complex and turbulent regions tend to incur higher prediction uncertainty, which was expected since predicting intricate flow features is a more complicated task than predicting less turbulent flow. This is because of the highly nonlinear and multiscale nature of turbulent flows. However, the expert further noted that large-scale ocean and upper atmospheric flows tend to be non-turbulent, and hence the proposed method can serve as an effective uncertainty-aware compression method for such large-scale flow data.}

\rmark{
\textbf{Mean vs. Median Streamline.} Formally, both MCDropout and Ensemble methods consider the expected model output to be the mean and so we visualize the mean streamlines. However, to mitigate outlier sensitivity, our method also allows visualization of the median streamline, which is less affected by outliers. Experiments show that mean and median streamlines are nearly identical, with minor differences. In Fig.~\ref{mean_median_streamline}, results from the Isabel data show that for both methods, mean (green) and median (blue) streamlines generally overlap, though the zoomed inset reveals slight differences for the Ensemble method.
\begin{figure}[thb]
\centering
\begin{subfigure}[t]{0.47\linewidth}
    \centering
    \includegraphics[width=\linewidth]{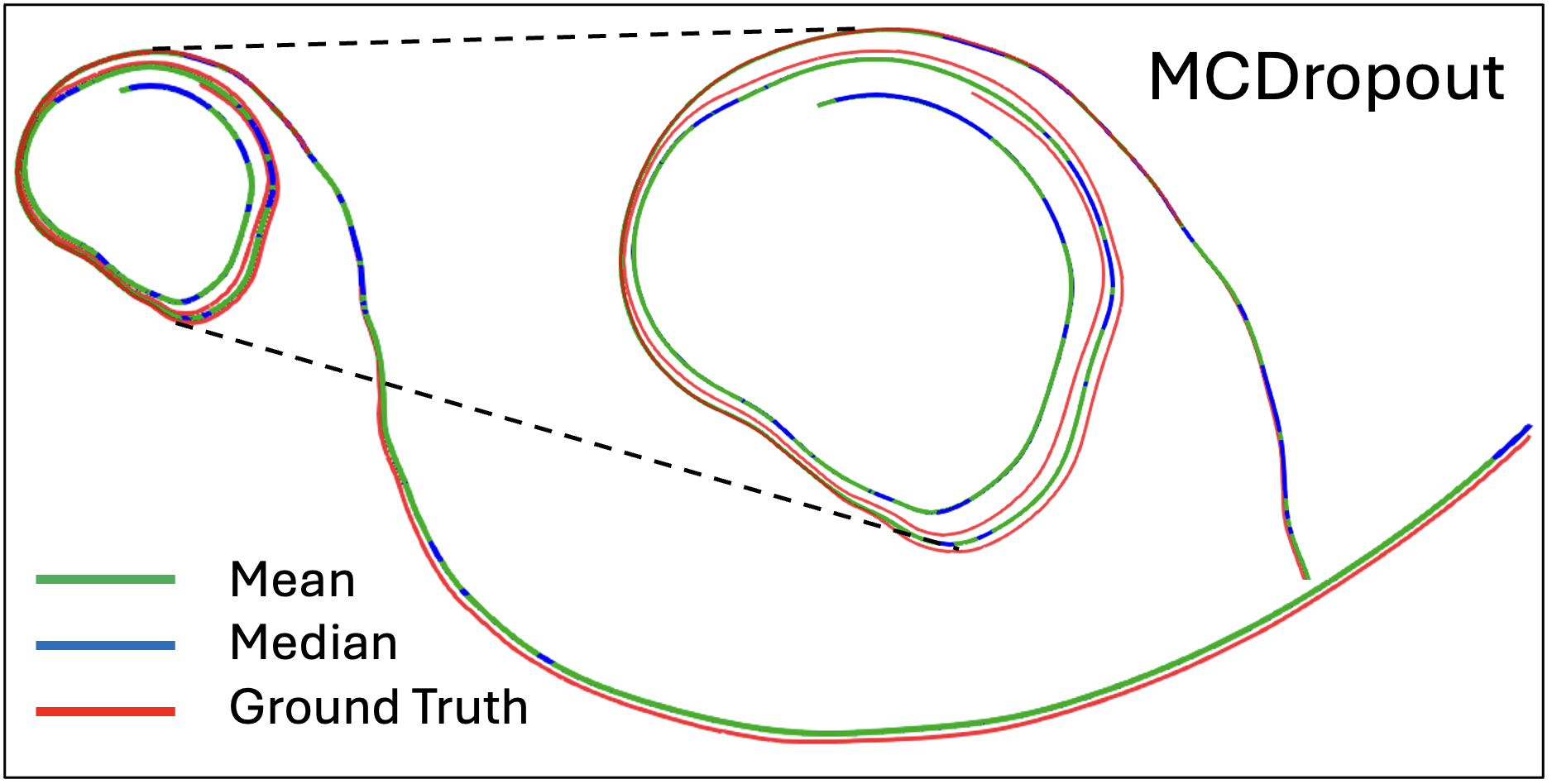}
    \caption{Mean vs. Median streamline for MCDropout.}
    \label{mean_median_MCD_Comb}
\end{subfigure}
~
\begin{subfigure}[t]{0.47\linewidth}
    \centering
    \includegraphics[width=\linewidth]{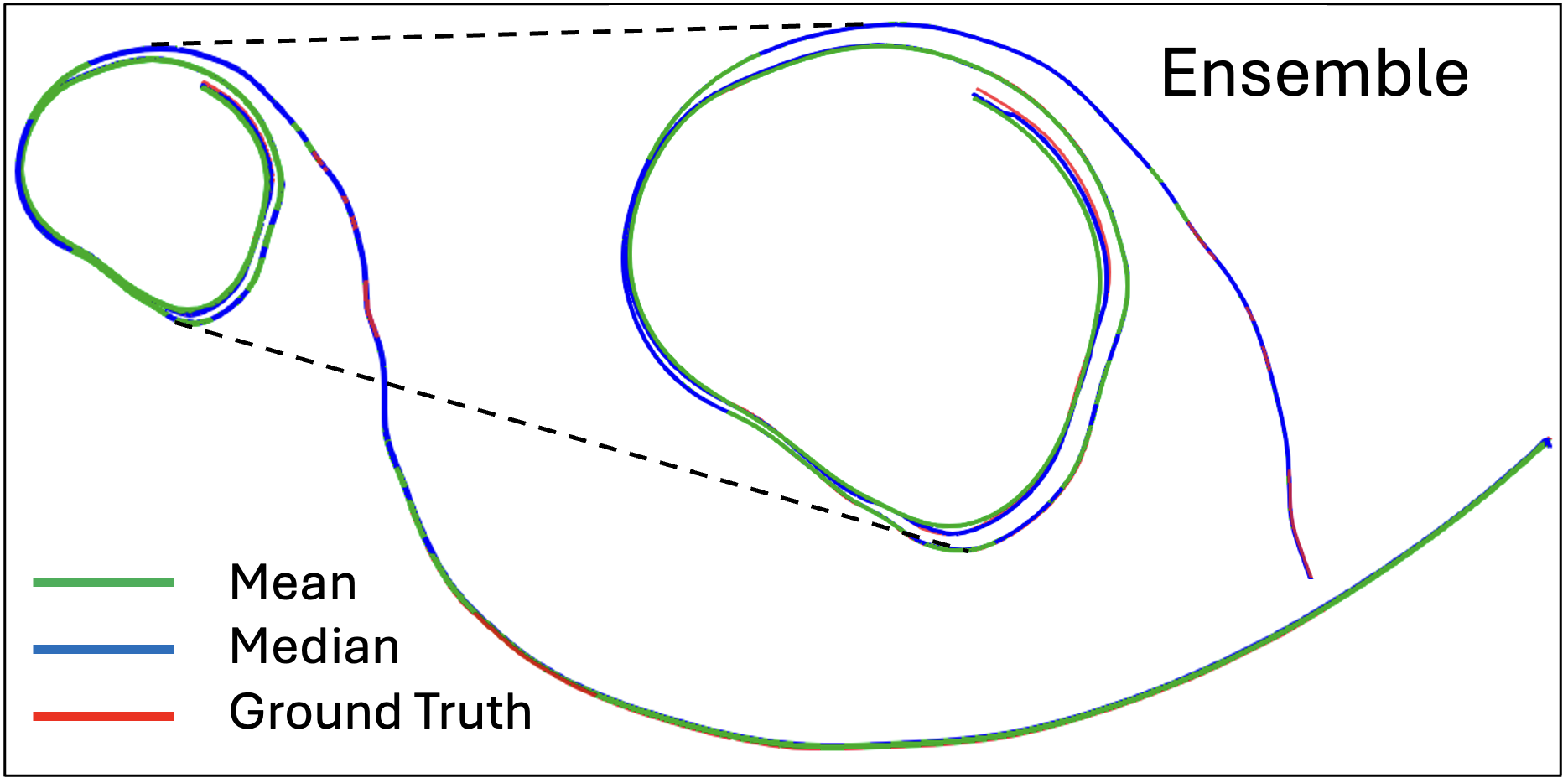}
    \caption{Mean vs. Median streamline for Ensemble.}
    \label{mean_median_ENS_Comb}
\end{subfigure}
\caption{\rmark{We show representative mean and median streamlines for MCDropout and Ensemble method. We observe that mean and median streamlines are almost identical while the zoomed view on the right shows minor differences between them for the Ensemble method.}}
\label{mean_median_streamline}
\end{figure}
}

\textbf{Error vs. Uncertainty.} Our experiments also reveal instances where the models may exhibit high confidence (low uncertainty) in predicting a segment of a streamline, yet this segment can have high errors. This poses a potential limitation for uncertainty-aware methods, as users might trust predictions guided by high model confidence, only to find the results partially erroneous. This observation is illustrated in Fig.\ref{uncert_error_isabel_MCD} and Fig.\ref{uncert_error_isabel_ENS}, where we present a streamline from the Isabel data set. Both MCDropout and Ensemble methods predict a segment of the streamline (marked by blue dotted region) with low uncertainty, but when compared against the ground truth streamline (depicted in red), we discover high prediction errors. Conversely, we identify another segment of the streamline (highlighted by the red dotted region), where the MCDropout method exhibits high uncertainty, and the segment also shows high error. However, for the Ensemble method, this segment demonstrates low error despite the method yielding high uncertainty. These findings suggest that while uncertainty can aid in communicating the confidence of predicted results, it may also lead to situations where the model is confident but the result is erroneous. Therefore,  additional methods of model explainability will be necessary for such cases.

\textbf{MCDropout vs. Ensemble Method.} This work investigates two approaches for estimating uncertainty using an INR to analyze steady-state vector fields. We use Deep Ensemble, a widely accepted standard~\cite{huzw23, gpt4} despite its extensive training time, and explore single-model-based MCDropout method to mitigate computational challenges. MCDropout method is chosen for its theoretical elegance and ease of integration into existing DNN models with minimal adjustments~\cite{gagh16}. Our findings show that both MCDropout and Ensemble methods produce comparable uncertainty patterns, but the Ensemble method offers slightly more accurate predictions at the cost of significantly longer training times (Table~\ref{train_inference_time}). Therefore, in scenarios where resources are limited and immediate uncertainty estimation is crucial, MCDropout proves to be a dependable alternative to the Ensemble method.

\rmark{
\textbf{Extension to Time-varying Data.} A natural extension of this work is to perform uncertainty-aware flow analysis on time-varying data. Time-varying data sets will be significantly larger and more challenging to model. Instead of directly modeling the time-varying vector fields, a computationally viable alternative could be to use flowmap-based representations~\cite{neuralflowmap} or directly learning particle tracing reults~\cite{particle_trace_NN}. Additionally, due to the high computational cost, training an Ensemble for uncertainty estimation may not be feasible and MCDropout and, other single-model methods will be preferable.}

\section{Conclusions and Future Work}
This paper emphasizes the importance of understanding uncertainty by applying two  uncertainty estimation methods. \rmark{While in this work, we study the impact of uncertainty on tasks such as vector data prediction, streamline generation, and critical point detection, in the future, we plan to conduct a comprehensive topological analysis and thoroughly study flow map characteristics to evaluate reconstruction quality further using the proposed methods.} Other future research endeavors include exploring alternative deep uncertainty estimation techniques and expanding our method to handle time-varying data. Insights from uncertainty estimates help identify areas needing explicit training and recognize model limitations in specific data regions. In critical scenarios, a model's confidence in its predictions is crucial, fostering greater trust when the model acknowledges its uncertainties.

\section*{Acknowledgments}{
We acknowledge the anonymous reviewers for their insightful comments. We thank Prof. Nairita Pal from IIT Kharagpur, India for providing constructive expert feedback for the proposed work and suggesting potential future research directions. We also extend our gratitude to Dr. Ayan Acharya for the valuable discussions and suggestions, which helped in improving the paper.}

\bibliographystyle{abbrv-doi-hyperref}

\bibliography{template}

\end{document}